%% file: main.tex
\documentclass[a4paper,11pt]{article}
\usepackage{jcappub} 

\usepackage{orcidlink}
\usepackage{graphicx}	
\usepackage{amsmath}	
\usepackage{multirow}
\usepackage{diagbox}
\usepackage{newtxtext,newtxmath}
\usepackage{booktabs}
\usepackage{array}
\usepackage{arydshln}
\usepackage{adjustbox}
\usepackage{rotating,tabularx}
\usepackage{xspace}
\usepackage{afterpage}
\usepackage{hanging} 

\usepackage{aas_macros}

\usepackage[T1]{fontenc} 
\hypersetup{
  colorlinks = true,
  linkcolor = blue,
  urlcolor  = blue,
  citecolor = blue,
  anchorcolor = blue
}
\newcommand{\ChenHowlett}{\cite{KP4s2-Chen})\xspace}
\newcommand{\Chen}{\cite{KP4s3-Chen})\xspace}
\newcommand{\Paillas}{\cite{KP4s4-Paillas})\xspace}
\newcommand{\ForeroSanchez}{\cite{KP4s6-Forero-Sanchez})\xspace}
\newcommand{\Valcin}{\cite{KP4s5-Valcin})\xspace}
\newcommand{\Rashkovetskyi}{\cite{KP4s7-Rashkovetskyi})\xspace}
\newcommand{\Alves}{\cite{KP4s8-Alves})\xspace}
\newcommand{\PerezFernandez}{\cite{KP4s9-Perez-Fernandez})\xspace}
\newcommand{\MenaFernandez}{\cite{KP4s10-Mena-Fernandez})\xspace}

\newcommand{\Findlay}{\cite{KP5s7-Findlay})\xspace}

\title{\textbf{HOD-Dependent Systematics in Emission Line Galaxies for the DESI 2024 BAO analysis}}

\input{DESI-2023-0285_author_list}

\abstract{The Dark Energy Spectroscopic Instrument (DESI) will provide precise measurements of Baryon Acoustic Oscillations (BAO) to constrain the expansion history of the Universe and set stringent constraints on dark energy. Therefore, precise control of the global error budget due to various systematic effects is required for the DESI 2024 BAO analysis. In this work, we focus on the robustness of the BAO analysis against the Halo Occupation Distribution (HOD) modeling for the Emission Line Galaxy (ELG) tracer. Based on a common dark matter simulation, our analysis relies on HOD mocks tuned to early DESI data, namely the One-Percent survey data. To build the mocks, we use several HOD models for the ELG tracer as well as extensions to the baseline HOD models. Among these extensions, we consider distinct recipes for galactic conformity and assembly bias. We perform two independent analyses in the Fourier space and in the configuration space. We recover the BAO signal from two-point measurements after performing reconstruction on our mocks. Additionally, we also apply the control variates technique to reduce sample variance noise. Our BAO analysis can recover the isotropic BAO parameter $\alpha_\text{iso}$ within 0.1\%  and the Alcock Paczynski parameter $\alpha_\text{AP}$ within 0.3\%. Overall, we find that our systematic error due to the HOD dependence is below 0.17\%, with the Fourier space analysis being more robust against the HOD systematics. We conclude that our analysis pipeline is robust enough against the HOD systematics for the ELG tracer in the DESI 2024 BAO analysis.}

\begin{document}
\maketitle
\flushbottom

\section{Introduction}
\label{Section:Introduction}
    \input{BodyText/Introduction}

\section{Simulations and  data}\label{Section:Simulations}

\input{BodyText/Simulations_and_mock_data}

\section{Galaxy-halo connection} \label{Section:Galaxy_halo_connection}
    \input{BodyText/Galaxy_halo_connection_and_mock_data}

\section{Methodology} \label{Section:Methodology}
    \input{BodyText/Methodology}

\section{Results} \label{Section:Results}
    \input{BodyText/Results}
    
\section{Conclusions}\label{Section:Conclusions}
    \input{BodyText/Conclusions}

\section{Data Availability}
The data used in this analysis will be made public along with the DESI Data Release 1 (details in \url{https://data.desi.lbl.gov/doc/releases/}). Also, the data and code to reproduce the figures are available at \url{https://doi.org/10.5281/zenodo.10905805}, as part of DESI’s Data Management Plan.

\acknowledgments
We would like to thank Shun Saito and Lado Samushia for their useful comments on the overall manuscript as well as for serving as internal reviewers. Also, we would like to thank Santiago \'Avila for useful comments on the HOD framework involved in this work.

CGQ gratefully acknowledges a Ph.D. scholarship from the National Council of Humanities, Science and Technology of Mexico (CONAHCYT). H-JS acknowledges support from the U.S. Department of Energy, Office of Science, Office of High Energy Physics under grant No. DE-SC0019091 and No. DE-SC0023241. H-JS also acknowledges support from Lawrence Berkeley National Laboratory and the Director, Office of Science, Office of High Energy Physics of the U.S. Department of Energy under Contract No. DE-AC02-05CH1123 during the sabbatical visit. SN acknowledges support from an STFC Ernest Rutherford Fellowship, grant reference ST/T005009/2. MI acknowledges that this material is based upon work supported in part by the Department of Energy, Office of Science, under Award Number DE-SC0022184 and also in part by the U.S. National Science Foundation under grant AST2327245.

This material is based upon work supported by the U.S. Department of Energy (DOE), Office of Science, Office of High-Energy Physics, under Contract No. DE–AC02–05CH11231, and by the National Energy Research Scientific Computing Center, a DOE Office of Science User Facility under the same contract. Additional support for DESI was provided by the U.S. National Science Foundation (NSF), Division of Astronomical Sciences under Contract No. AST-0950945 to the NSF’s National Optical-Infrared Astronomy Research Laboratory; the Science and Technology Facilities Council of the United Kingdom; the Gordon and Betty Moore Foundation; the Heising-Simons Foundation; the French Alternative Energies and Atomic Energy Commission (CEA); the National Council of Humanities, Science and Technology of Mexico (CONAHCYT); the Ministry of Science and Innovation of Spain (MICINN), and by the DESI Member Institutions: \url{https://www.desi.lbl.gov/collaborating-institutions}. Any opinions, findings, and conclusions or recommendations expressed in this material are those of the author(s) and do not necessarily reflect the views of the U. S. National Science Foundation, the U. S. Department of Energy, or any of the listed funding agencies.

The authors are honored to be permitted to conduct scientific research on Iolkam Du’ag (Kitt Peak), a mountain with particular significance to the Tohono O’odham Nation.

\bibliographystyle{JHEP}
\bibliography{References/references} 

\appendix
\input{Appendix/Consistency_FS_CF}
\input{Appendix/Extended_results}
\newpage
\input{DESI-2023-0285_author_list.affiliations}
\end{document}

%% file: DESI-2023-0285_author_list.tex
\emailAdd{gqcristhian@utdallas.edu}
\affiliation{Affiliations are in Appendix \ref{sec:affiliations}}

\author[1]{{C.~Garcia-Quintero}\orcidlink{0000-0003-1481-4294},}
\author[2]{{J.~Mena-Fern\'andez}\orcidlink{0000-0001-9497-7266},}
\author[3,4]{{A.~Rocher}\orcidlink{0000-0003-4349-6424},}
\author[5]{{S.~Yuan}\orcidlink{0000-0002-5992-7586},}
\author[6,7]{{B.~Hadzhiyska}\orcidlink{0000-0002-2312-3121},}
\author[8]{{O.~Alves},}
\author[9]{{M.~Rashkovetskyi}\orcidlink{0000-0001-7144-2349},}
\author[10]{{H.~Seo}\orcidlink{0000-0002-6588-3508},}
\author[11]{{N.~Padmanabhan},}
\author[12]{{S.~Nadathur}\orcidlink{0000-0001-9070-3102},}
\author[13]{{C.~Howlett}\orcidlink{0000-0002-1081-9410},}
\author[1]{{M.~Ishak}\orcidlink{0000-0002-6024-466X},}
\author[1]{{L.~Medina-Varela},}
\author[6]{{P.~McDonald}\orcidlink{0000-0001-8346-8394},}
\author[14,15,16]{{A.~J.~Ross}\orcidlink{0000-0002-7522-9083},}
\author[1]{{Y.~Xie},}
\author[11]{{X.~Chen},}
\author[1]{{A.~Bera},}
\author[6]{{J.~Aguilar},}
\author[17]{{S.~Ahlen}\orcidlink{0000-0001-6098-7247},}
\author[18,8]{{U.~Andrade}\orcidlink{0000-0002-4118-8236},}
\author[19]{{S.~BenZvi}\orcidlink{0000-0001-5537-4710},}
\author[20]{{D.~Brooks},}
\author[4]{{E.~Burtin},}
\author[21]{{S.~Chen}\orcidlink{0000-0002-5762-6405},}
\author[6]{{T.~Claybaugh},}
\author[22]{{S.~Cole}\orcidlink{0000-0002-5954-7903},}
\author[23]{{A.~de la Macorra}\orcidlink{0000-0002-1769-1640},}
\author[4]{{A.~de~Mattia},}
\author[24]{{A.~Dey}\orcidlink{0000-0002-4928-4003},}
\author[25]{{B.~Dey}\orcidlink{0000-0002-5665-7912},}
\author[26]{{Z.~Ding}\orcidlink{0000-0002-3369-3718},}
\author[20]{{P.~Doel},}
\author[27,5]{{K.~Fanning}\orcidlink{0000-0003-2371-3356},}
\author[28,29]{{J.~E.~Forero-Romero}\orcidlink{0000-0002-2890-3725},}
\author[30,12,31]{{E.~Gaztañaga},}
\author[32,30,33]{{H.~Gil-Mar\'in}\orcidlink{0000-0003-0265-6217},}
\author[6]{{S.~Gontcho A Gontcho}\orcidlink{0000-0003-3142-233X},}
\author[34]{{G.~Gutierrez},}
\author[6]{{J.~Guy}\orcidlink{0000-0001-9822-6793},}
\author[35]{{C.~Hahn}\orcidlink{0000-0003-1197-0902},}
\author[14,36,16]{{K.~Honscheid},}
\author[6]{{A.~Kremin}\orcidlink{0000-0001-6356-7424},}
\author[6]{{M.~Landriau}\orcidlink{0000-0003-1838-8528},}
\author[37]{{L.~Le~Guillou}\orcidlink{0000-0001-7178-8868},}
\author[6]{{M.~E.~Levi}\orcidlink{0000-0003-1887-1018},}
\author[38,39]{{M.~Manera}\orcidlink{0000-0003-4962-8934},}
\author[14,15,16]{{P.~Martini}\orcidlink{0000-0002-4279-4182},}
\author[24]{{A.~Meisner}\orcidlink{0000-0002-1125-7384},}
\author[40,39]{{R.~Miquel},}
\author[41]{{J.~Moustakas}\orcidlink{0000-0002-2733-4559},}
\author[42]{{E.~Mueller},}
\author[23]{{A.~Muñoz-Gutiérrez},}
\author[43]{{A.~D.~Myers},}
\author[25]{{J.~ A.~Newman}\orcidlink{0000-0001-8684-2222},}
\author[44]{{J.~Nie}\orcidlink{0000-0001-6590-8122},}
\author[45,46]{{G.~Niz}\orcidlink{0000-0002-1544-8946},}
\author[47,48]{{E.~Paillas}\orcidlink{0000-0002-4637-2868},}
\author[4,6]{{N.~Palanque-Delabrouille}\orcidlink{0000-0003-3188-784X},}
\author[47,49,48]{{W.~J.~Percival}\orcidlink{0000-0002-0644-5727},}
\author[6,50,7]{{C.~Poppett},}
\author[23,51]{{A.~P\'{e}rez-Fern\'{a}ndez}\orcidlink{0009-0006-1331-4035},}
\author[10]{{A.~Rosado-Marin},}
\author[52]{{G.~Rossi},}
\author[53,13]{{R.~Ruggeri}\orcidlink{0000-0002-0394-0896},}
\author[54]{{E.~Sanchez}\orcidlink{0000-0002-9646-8198},}
\author[6]{{D.~Schlegel},}
\author[55,8]{{M.~Schubnell},}
\author[24]{{D.~Sprayberry},}
\author[8]{{G.~Tarl\'{e}}\orcidlink{0000-0003-1704-0781},}
\author[23]{{M.~Vargas-Maga\~na}\orcidlink{0000-0003-3841-1836},}
\author[24]{{B.~A.~Weaver},}
\author[3]{{J.~Yu},}
\author[47,48]{{H.~Zhang}\orcidlink{0000-0001-6847-5254},}
\author[6]{{R.~Zhou}\orcidlink{0000-0001-5381-4372},}
\author[44]{{H.~Zou}\orcidlink{0000-0002-6684-3997},}

%% file: BodyText/Introduction.tex
The Dark Energy Spectroscopic Instrument (DESI) is designed to conduct a survey of 14,000 square degrees and measure around 40 million galaxy redshifts throughout five years~\cite{2023_DESI_instrument,DESI_2016a}. DESI observation range will span redshifts going from redshift 0.1 up to redshift 2.1 for clustering analysis, extending this range further up to redshift of around 4.1 for Lyman-$\alpha$ forest analyses. DESI target selection program makes a distinction between four tracers in this redshift range, namely Bright Galaxy Survey (BGS)~\cite{BGS_TS} over $0.1<z<0.4$,
Luminous Red Galaxy (LRG)~\cite{LRG_TS} in the range $0.4<z<1.1$, Emission Line Galaxy (ELG)~\cite{ELG_TS} covering $0.8<z<1.6$, and Quasars (QSO)~\cite{QSO_TS}, ranging from $z=0.8$ up to $z=2.1$ for galaxy clustering analyses. Currently, DESI already surpassed its survey validation stage~\cite{2023_SV-overview} and made an early data release publicly available~\cite{2023_EDR_DESI}. So far, various cosmological probes have been explored by previous and ongoing surveys to shed some light on dark energy. Indeed, Stage-III experiments have pushed the boundaries of dark energy parameters up to figures of merit of 92 as pointed out by~\cite{2021_SDSS_twodecades} when combining SDSS with several Stage-III experiments, such as Planck~\cite{Planck:2015fie,Planck:2018vyg}, Pantheon~\cite{Pan-STARRS1:2017jku}, and DES~\cite{DES:2017myr,DESI:2013agm}. The search for answers to the query of dark energy has made efforts to move toward the next generation of Stage-IV experiments, which will push the figure of merit of dark energy even further. One of the key measurements for the determination of dark energy properties is the expansion history of the Universe. By creating a 3D map of the Universe covering $\sim$1/3 of the sky surface area, DESI plans to constrain the expansion history of the Universe and better understand the growth of structure.

As the nature of dark energy remains an open question in cosmology, one of the key probes in cosmology that could offer potential answers is measurements of Baryon Acoustic Oscillations (BAO). DESI will produce precise measurements of BAO using the aforementioned tracers. In particular, the LRG tracer already exhibited a $\sim$5-$\sigma$ detection of BAO with only two months of data~\cite{2023_Moon}, this is just $\sim$2.2 times less precise than the full BOSS and eBOSS BAO measurements for this tracer with $\sim$1/4 less galaxies observed. The precision of this early analysis is a testimony in support of the quality of the DESI data. Since the first detection of BAO~\cite{2005_Eisenstein}, and further BAO measurements~(see for example \cite{2013_Anderson,eBOSS:2020_deMattia,BOSS:2016_Ross,6dF}), not only has the quality of the data improved, but also BAO analyses have become more accurate in terms of modeling. Current BAO analyses estimate the BAO feature from galaxy clustering statistics by measuring an isotropic shift in the BAO scale parameterized by $\alpha_\text{iso}$ and an anisotropic warping parameterized by $\alpha_\text{AP}$ (AP stands for Alcock-Paczynski). These two parameters can be translated into measurements of the angular diameter distance (BAO scale perpendicular to the line-of-sight) and the Hubble parameter (BAO scale along the line-of-sight), after accounting for the AP effect~\cite{1979_Alcock_Paczynski}. 

The results presented in the DESI 2024 BAO analysis \cite{DESI2024.III.KP4} report the constraints on $\alpha_\text{iso}$ and $\alpha_\text{AP}$ for the ELG tracer in the redshift bins $0.8<z<1.1$ and $1.1<z<1.6$. The statistical errors concerning these parameters are presented in Table \ref{Table:DESI_2024_constraints}, and we focus on the aggregated error over both redshift bins. Additionally, internal forecast using early DESI data estimate the expected statistical precision the full survey to be $\sigma^\text{Y5}_\text{stat}(\alpha_\text{iso})=0.0038$ and $\sigma^\text{Y5}_\text{stat}(\alpha_\text{AP})=0.011$. Yet, the statistical precision reported on BAO measurements should be accompanied by the characterization of potential systematics, and it should be ensured that the analysis pipeline is robust against them. Therefore, a precise knowledge of the global error budget due to various systematics is required. There are several systematics to be examined in companion papers to support the DESI 2024 BAO analysis presented at \cite{DESI2024.III.KP4}. The optimal BAO reconstruction settings are investigated in \Chen while \Paillas shows tests on unblinded mocks and catalogs with this configuration. Tests to validate analytical covariance matrices are performed in \ForeroSanchez, based on independent studies about analytical covariance matrices in Fourier space \Alves, and configuration space \Rashkovetskyi. \ChenHowlett presents the systematic error budget for the BAO theoretical modeling. Potential systematic errors due to assumptions regarding the fiducial cosmology assumed in the theoretical template are explored in \PerezFernandez. In particular, this work aims to focus on the robustness of the BAO modeling against the Halo Occupation Distribution (HOD) model assumed in the mocks.

The HOD model is an empirical framework that assumes that galaxies are spatially distributed within dark matter halos. Given the number of mocks needed to perform all the required analyses by DESI, the HOD framework turns out to be a suitable tool for mock mass production. Rather than evolving costly hydrodynamical simulations, the HOD framework allows one to populate a dark matter-only N-body simulation with galaxies while tuning the clustering to match the desired clustering signal to reproduce. Naturally, the dark matter halos have to be identified in advance by employing a halo finder algorithm. However, selecting an HOD model over another to build the mocks could introduce some bias. For this study, rather than trying to improve the HOD modeling itself, we are interested in assessing that our analysis pipeline is robust against different choices of HOD models.

\begin{table}
    \centering
    \resizebox{\textwidth}{!}{
    {\renewcommand{\arraystretch}{1.35}
    \begin{tabular}{c|cccc|cccc}
        \hline\hline
        \multirow{2}{*}{Redshift bin} & \multicolumn{4}{c|}{\textbf{Fourier space}} & \multicolumn{4}{c}{\textbf{Configuration space}} \\\cline{2-9}
        & $\sigma_\text{stat}(\alpha_\text{iso})$ & $\sigma_\text{stat}(\alpha_\text{AP})$ & $\sigma_\text{stat}(\alpha_\parallel)$ & $\sigma_\text{stat}(\alpha_\perp)$ & $\sigma_\text{stat}(\alpha_\text{iso})$ & $\sigma_\text{stat}(\alpha_\text{AP})$ & $\sigma_\text{stat}(\alpha_\parallel)$ & $\sigma_\text{stat}(\alpha_\perp)$ \\\hline
        $0.8<z<1.1$ & 0.020 &  &  &  & 0.018 &  &  &  \\
        $1.1<z<1.6$ & 0.011 & 0.033 & 0.022 & 0.018 & 0.014 & 0.044 & 0.029 & 0.023 \\\hline
        \textbf{Aggregated precision} & \textbf{0.0096} & \textbf{0.033} & 0.012 & 0.008 & \textbf{0.011} & \textbf{0.044} & 0.013 & 0.009 \\
        \hline\hline
    \end{tabular}
    }
    }
    \caption{Table with the aggregated error on the BAO parameters and AP parameter based on the results reported in \cite{DESI2024.III.KP4}, for the ELG tracer. The values reported on this table correspond to post-reconstruction fits, in both Fourier space and configuration space. We take the aggregated errors in bold as our reference for the statistical error to which we compare the systematic error from HOD modeling. Note that the empty values for the first redshift bin represent that the BAO fit was isotropic.}
    \label{Table:DESI_2024_constraints}
\end{table}

Our analysis is directly related to the DESI 2024 BAO results, and we shall focus our attention on the ELG tracer. An analogous HOD systematics analysis for the LRG tracer is described in \MenaFernandez and a further multi-tracer analysis for LRG$\times$ELG is presented in~\Valcin. The objective of this work is to characterize the systematic error due to HOD-dependence in DESI 2024 BAO analysis for the ELG tracer. The approach to be followed in this work is to select representative HOD prescriptions that have been widely used in the literature and include new promising HOD models (see \cite{2023_Antoine-Rocher} and \cite{Yuan2023-conf}) to assess robustness against HOD systematics in support of the DESI 2024 BAO analysis presented at \cite{DESI2024.III.KP4}. A similar systematic error analysis is to be examined in~\Findlay for the DESI 2024 full shape analysis that will be presented in \cite{DESI2024.V.KP5}. Both BAO and full shape DESI 2024 results are based on two-point clustering measurements from the DESI Data Release 1 (DESI DR1) \cite{DESI2024.I.DR1} presented in \cite{DESI2024.II.KP3}. The corresponding cosmological constraints using DESI 2024 BAO measurements (including BAO measurements from the Lyman-$\alpha$ forest \cite{DESI2024.IV.KP6}) are described in \cite{DESI2024.VI.KP7A}. Further cosmological constraints based on full shape measurements will be shown in \cite{DESI2024.VII.KP7B}, as well as constraints on primordial non-gaussianity \cite{DESI2024.VIII.KP7C}.

The structure of this paper is organized as follows: Section~\ref{Section:Simulations} presents the simulations on which we rely on our analysis and the data that we use to build our HOD mocks. Section~\ref{Section:Galaxy_halo_connection} offers a review of the HOD formalism and introduces the HOD models that we use to characterize the systematic error due to the assumed HOD prescription. Section~\ref{Section:Methodology} lays out our general methodology for clustering estimation and BAO analysis. Next, we describe our results from the BAO analysis and the robustness against HOD dependence in Section~\ref{Section:Results}. Finally, we summarize our findings and provide a set of conclusions regarding the DESI 2024 BAO analysis in Section~\ref{Section:Conclusions}.

%% file: BodyText/Simulations_and_mock_data.tex
In this section, we provide a concise overview of the N-body simulations employed for evolving the dark matter halos, which are subsequently populated with galaxies utilizing the HOD formalism. Additionally, we describe the data sample used for tuning the clustering of the HOD models.

\subsection{\texttt{AbacusSummit} simulations} 

We based our analysis on the \texttt{AbacusSummit} suite of cosmological N-body simulations. The \texttt{AbacusSummit} simulations are a massive set of high-accuracy and high-resolution N-body simulations~\cite{2021_Nina-Maksimova} designed to support science analyses in the era of Stage-IV surveys. In particular, they are expected to meet and exceed the requirements of DESI~\cite{DESI:2013agm}. This suite of simulations was produced by running \texttt{Abacus}~\cite{2018_Lehman-Garrison,2021_Lehman-Garrison}, which is an optimized N-body code for GPU architectures based on the static multipole mesh method to compute the gravitational potential. 

The \texttt{AbacusSummit} suite of simulations was run on the Summit supercomputer of the Oak Ridge Leadership Computing Facility. We use base resolution simulations consisting of $\left( 2 \text{Gpc}/h \right)^3$ boxes where each of them has $6912^3$ (around 330 billion) particles with mass $2\times 10^9 h^{-1} M_\odot$. Although this suite of simulations consists of more than 150 simulations that span 97 different cosmological models, we only use 25 different realizations and restrict our analysis to the Planck 2018 baseline $\Lambda$CDM model as our fiducial cosmology\footnote{Specifically, our fiducial cosmology refers to the mean values of base\_plikHM\_TTTEEE\_lowl\_lowE\_lensing.}. We refer to \PerezFernandez for systematic effects concerning fiducial cosmology assumptions, where simulations for different cosmologies were considered. Additionally, we make use of two types of simulation output based on discrete redshift snapshots, these being at $z=0.8$ for the HMQ$^{(3\sigma)}_i$ HOD models ($i=1,2,...,6$) and at $z=1.1$ for the rest of the HOD models, as we will explain in the following section.

Dark matter halos are identified with the \texttt{CompaSO} method~\cite{2021_Boryana-Hadzhiyska}, which is a halo-finding algorithm based on previous spherical overdensity algorithms. Before assigning halo membership to the particles, \texttt{CompaSO} accounts for the tidal radius around all nearby halos, rather than simply truncating the halos at the overdensity threshold. Such an algorithm for halo membership allows for a more effective halo deblending, especially during major mergers. A further cleaning procedure as described in~\cite{2022_Sownak-Bose} is applied to eliminate unphysical halos utilizing information from the merger-tree for every halo. Such `cleaning' has been shown to produce a more effective deblending not only by removing the halos that are over-deblended but also by aggregating halos that are separated at the present but were physically unified in the past. Hence, this method enhances the fidelity of the \texttt{AbacusSummit} halo catalogs.

\subsection{The One-Percent DESI sample}

The first public DESI spectroscopic data sample released was tagged as Early Data Release (EDR)\cite{2023_EDR_DESI}. The EDR is derived from the Survey Validation (SV) campaign in April and May 2021~\cite{2023_SV-overview} before the start of the main survey operations. There were three main phases of operation, the Target Selection Validation phase, the Operations Development stage, and the One-Percent survey. The dataset we use to fit our mocks is the so-called DESI One-Percent sample. The DESI One-Percent sample is a pilot survey of the full DESI program where 140 deg$^2$ were covered. This is, the survey integrates 1\% of the total area of the DESI footprint. 

Given the early nature of this work, we rely on a wide variety of ELG mocks constructed with HOD models fitted to the DESI One-Percent survey data, rather than DESI DR1. As described in more detail in \cite{ELG_TS}, the target selection algorithm for the ELG sample relies on a magnitude cuts in the $grz$-bands. A magnitude cut $g<20$ is imposed to obtain samples for $z>0.6$ and a selection box in the $(g-r)$ versus $(r-z)$ color-color space is used to favor [\textsc{OII}] emitters, typical of ELGs. Also, the measured flux of the emitter needs to be greater than zero in all three bands. We refer to the reader to \cite{ELG_TS} for more specific details about the ELGs target selection. On the other hand, the DESI ELG sample originally covers the redshift range $0.6<z<1.6$ based on two sub-samples. However, information below $z<0.8$ is not considered as signal-to-noise is dominated by the LRG sample which exhibits a higher number density below this redshift. The first sub-sample covers lower redshifts from $z=0.6$ up to $z=1.1$ while having a higher redshift sucess rate. The second sub-sample covers $1.1<z<1.6$ and allows to access information from earlier epochs compared to the LRG sample. The One-Percent clustering measurements of the ELG sample between redshift $0.8<z<1.6$ were used to fit different HOD models. It is worth mentioning that early versions of the One-Percent data were used as well to tune some HOD models such as 1st-Gen and HMQ$_i^{(3\sigma)}$ (see Section~\ref{Section:Galaxy_halo_connection} for a description of such HOD models). However, we still expect such mocks to be useful for the HOD systematics studies as the clustering signal is not too distinct. In the following, we briefly introduce the HOD formalism and the models used in this analysis.

%% file: BodyText/Galaxy_halo_connection_and_mock_data.tex
In this section, we provide an overview of the formalism we use to establish the connection between galaxies and dark matter halos. While we define a baseline model for the satellite galaxies in Section \ref{sec:HOD framework}, we continue to describe the various HOD models used in this work for the central galaxies in Section \ref{sec:standard_HODs}. We extend \ref{sec:extended_HODs}. Later, we present an overview of the HOD models considered in this work and briefly describe how the HOD mocks for the ELG tracer are generated.

\subsection{HOD framework}
\label{sec:HOD framework}
In its basic form, the HOD formalism describes the relation between a typical class of galaxies and dark matter halos, as the probability that a halo with mass $M$ contains $N$ such galaxies. It also specifies how the galaxy positions and velocities are distributed within halos. HOD models have contributions from two galaxy populations, namely centrals and satellites, with $\langle N_{\text{cen}}(M)\rangle$ and $\langle N_{\text{sat}}(M)\rangle$ being their respective mean numbers hosted per halo of a given halo mass. Analytical descriptions of these mean HODs have been derived from either semi-analytical models or hydrodynamical simulations of galaxy formation and evolution (see e.g.~\cite{Berlind03,Zheng05}). The most common mean HOD function~\cite{zheng_galaxy_2007} uses a softened step function for centrals, a power law for satellites, and assumes generally that satellites can only be found in halos that already host a central galaxy. This model has proven to describe well the clustering of luminosity selected~\cite{Zehavi11} or stellar mass limited~\cite{Contreras13} samples like Luminous Red Galaxies (LRGs)~\cite{Zheng09} or quasars~\cite{Smith20}. 

However, a step function cannot represent ELG samples. Strong emission lines in galaxy spectra are strongly correlated with the galaxy star formation rate. In DESI, Emission Line Galaxies (ELGs) are selected spectroscopically using the [\textsc{OII}] \cite{ELG_TS} doublet which strongly correlates with a high star formation rate \citep{Moustakas_2006}. At the relevant redshift, quiescent galaxies are more prevalent than star forming galaxies at high stellar masses, leading to the center of massive halos to be dominated by quiescent central galaxies \cite{eBOSS:2018wvf_GUO,Favole:2015xza,Hadzhiyska:2020iri}. Additionally, ELGs are selected to have strong star formation, which becomes inefficient at the center of massive halos given the absence of cold gas. Therefore, ELG centrals are unlikely to reside in high halo masses, which turns into a quenching of the ELG central occupation in the high halo mass end \cite{2018_Gonzalez-Perez}. This can be reflected in a reduction in the central probability of the ELGs in high-mass halos. From semi-analytical models, the predicted mean HOD for central ELGs can be fitted reasonably well by a Gaussian or an asymmetric Gaussian for centrals, together with a power law for satellites~\cite{Contreras13, 2018_Gonzalez-Perez}. 

As part of the HOD framework, the total number density of the galaxy sample is calculated as

\begin{equation}
    \bar{n}_\text{gal} = \int \frac{dn(M)}{dM} [\langle N_{\text{cen}}(M)\rangle + \langle N_{\text{sat}}(M)\rangle] dM,
\end{equation}
where $\frac{dn(M)}{dM}$ is the differential halo mass function. The contribution of the expected number of centrals and satellites depends on the parameterization of $\langle N_{\text{cen}}(M)\rangle$ and $\langle N_{\text{sat}}(M)\rangle$, which are defined below. We assume the number of central (resp. satellite) galaxies per halo of mass $M$ to follow a Bernoulli (resp. Poisson) distribution with mean equal to $\left\langle N_{\mathrm{cent}}(M) \right\rangle$ (resp. $\left\langle N_\mathrm{sat}(M)\right\rangle$). However, alternative probability distribution functions for the satellite galaxies are considered in \cite{2019_EstebanJimenez,2023_Vos-Gines_nonPoissonian,2020_S-Avila}. In the standard approach, central galaxies are positioned at the center of their halos while satellite galaxy positions sample a Navarro-Frenk-White (NFW) profile~\cite{NFW}. On the other hand, other spatial profiles have been investigated to positionate satellite galaxies within dark matter halos. For example, \cite{2020_S-Avila} used a less concentrated NFW profile for ELG satellites. Similarly, \cite{2017_Orsi_NFW} studied the profile of ELG satellites using the Millennium-XXL simulation \cite{MXXL} and \cite{2023_Reyes-Peraza_NFW} proposed a generalization of the NFW profile to describe ELG satellites. For the purpose of this work, satellite velocities are typically assigned in two different ways: 1) They are normally distributed around their mean halo velocity, with a dispersion equal to that of the halo dark matter particles, rescaled by an extra free parameter denoted $f_{\sigma_v}$ that accounts for velocity biases, as described in \cite{HODGP}. 2) Alternatively, satellites are assigned to DM particles of the halo with equal probabilities. However, other discussions beyond the standard approach for modeling the infall velocity of ELG satellites can be found in \cite{2020_S-Avila,2023_Antoine-Rocher,2017_Orsi_NFW}. When satellites are assigned to particles, two extra parameters are added, namely $\alpha_c$ and $\alpha_s$, allowing for both central and satellite velocity biases, respectively. These parameters modify the velocities as $v_\mathrm{cent} = v_h + \alpha_c\delta_v(\sigma_{vh})$ for centrals, where $\delta_v(\sigma_{vh})$ is the Gaussian scatter of the velocity dispersion of the halo, and $v_\mathrm{sat} = v_\mathrm{particles} + \alpha_s(v_\mathrm{particles} - v_h)$, as described in equations 8 \& 9 in~\cite{yuan_abacushod_2022}. $\alpha_c = 0$ and $\alpha_s = 1$ correspond to ``no velocity bias''. While the velocity of the halo is obtained from the mean velocity of all particles within the halo, the dispersion velocity in \texttt{CompaSO} is output as $\sum_i (v_i-v_h)^2/N_\text{part}$, where $N_\text{part}$ is the total number of particles of the halo. In the following, we proceed to define the explicit functional form for these quantities based on standard HOD models and extensions beyond.

\subsection{Baseline HOD models}
\label{sec:standard_HODs}
\begin{figure*}
\centering\includegraphics[width=0.7\textwidth]{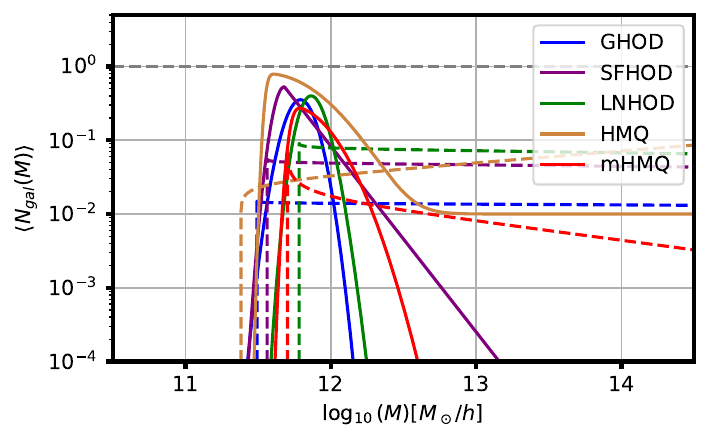}
    \caption{Best-fit baseline HOD models derived from fits to the One-Percent survey ELG data. The values used to plot these HOD models are the ones shown in Table \ref{Table:HOD_parameters}. The plot shows the mean number of galaxies per halo of a given mass $M$. The solid curves represent the contribution from central galaxies while the dashed lines show the contribution from the satellite galaxies.}
    \label{Fig:HOD_models}
\end{figure*}

To study a large variety of ELG mocks, four different HOD models for the expected number of central galaxies are used, one with a Gaussian shape and three different functions with an asymmetric Gaussian shape as described below. On the other hand, we use a standard model for the satellite galaxies as baseline. We show a plot for the baseline HOD models in Fig. \ref{Fig:HOD_models} and we refer to the reader to \cite{2023_Antoine-Rocher} for similar plots for these HOD models and extensions.

\subsubsection{Gaussian HOD}

The Gaussian HOD (GHOD), originally presented in \cite{2020_S-Avila}, is a simple expression for the expected number of centrals used to fit the semi-analytical model of galaxy formation and evolution presented in \citep{2018_Gonzalez-Perez}.
This parameterization is based on a Gaussian distribution around a logarithmic mass mean. Typically, this HOD is written as
\begin{equation}
    \langle N_{\text{cen}}(M) \rangle = \frac{A_c}{\sqrt{2\pi}\sigma_M} \exp\left[-\frac{(\log_{10}{M}-\log_{10}M_c)^2}{2\sigma_M^2}\right].
    \label{eq:GHOD}
\end{equation}
Here, $\log_{10}M_c$ represents the logarithmic mass mean, where $M_c$ is the characteristic mass of the halo. Also, $\sigma_M^2$ corresponds to the variance of the Gaussian functional form for this parameterization. The parameter $A_c$ represents an amplitude. We label this model for the expected number of central galaxies per halo as GHOD. 

\subsubsection{Star forming HOD}

The star-forming HOD (SFHOD) is a combination of a Gaussian distribution for low-mass halos $<M_c$ and a decreasing power law for high-mass halos $>M_c$. This model was used as a baseline model the in eBOSS analysis by \cite{2020_S-Avila} and it is described by 
\begin{equation}
    \langle N_{\text{cen}}(M) \rangle = 
    \begin{cases}
    \frac{A_c}{\sqrt{2\pi}\sigma_M} \exp\left[-\frac{(\log_{10}{M}-\log_{10}M_c)^2}{2\sigma_M^2}\right], & M \leq M_c, \\
    \frac{A_c}{\sqrt{2\pi}\sigma_M} \cdot \left( \frac{M}{M_c} \right)^\gamma,              & M > M_c.
    \end{cases}
    \label{eq:SFHOD}
\end{equation}
The result is an asymmetric shape where the asymmetry is modulated by the $\gamma$ parameter. This model was able to produce a better fit to the semi analytical model from \cite{2018_Gonzalez-Perez}, compared to the GHOD model. We refer to this parameterization as SFHOD.

\subsubsection{High mass quenched}

The third model used is the High Mass Quenched model (HMQ) \citep{2020_Shahab-Alam}, which is described by
\begin{equation}
    \begin{split}
    \langle N_{\text{cen}}(M) \rangle & =  2A\phi(M)\Phi(\gamma M) + \\
    & \frac{1}{2Q}\left[ 1 + \text{erf}\left( \frac{\log_{10} M - \log_{10} M_\text{c}}{0.01} \right) \right],
    \end{split}
    \label{eq:HMQ1}
\end{equation}
where 
\begin{equation}
    \phi(x)=A_c \cdot \mathcal{N}(\log_{10}M_\text{c}, \sigma_M),
    \label{eq:HMQ2}
\end{equation}
\begin{equation}
    \Phi(x) = \frac{1}{2}\left[ 1 + \text{erf} \left( \frac{x}{\sqrt{2}} \right) \right]
    \label{eq:HMQ3}
\end{equation}
and
\begin{equation}
    A=\frac{p_\text{max}-Q^{-1}}{\text{max}(2\phi(x)\Phi(\gamma x))}.
    \label{eq:HMQ4}
\end{equation} 
This model is a combination of a Gaussian function and an error function. The parameter $p_{max}$ controls the amplitude of the low-mass Gaussian part relative to the high-mass plateau, whose level is set by $Q$ that represents the quenching efficiency at high halo masses. The asymmetry of the Gaussian distribution is controlled by the parameter $\gamma$.

Furthermore, a modified version of the HMQ model (labeled mHMQ) is considered in this analysis. In the mHMQ model, the quenching parameter $Q$ goes to infinity and we rather calculate the central occupation as 
\begin{equation}
    \langle N_{\text{cen}}(M) \rangle  =  2p_\text{max}\phi(M)\Phi\left[\gamma (\log_{10}M-\log_{10}M_c)/\sigma_M\right].
    \label{eq:mHMQ}
\end{equation}
Such a limit in the $Q$ parameter is used to suppress the high-mass plateau and only retain the asymmetric shape of the central distribution.
\subsubsection{Lognormal HOD}

The last model we test in our set of standard HODs is a lognormal HOD (LNHOD), which by construction is an asymmetric function with its asymmetry led by the width of the distribution $\sigma_M$. Defining $x = \log_{10}M - (\log_{10}M_c-1)$, the prescription for central galaxies is given by
\begin{equation}
    \langle N_{\text{cen}}(x) \rangle = \frac{A_c}{\sqrt{2\pi}\sigma_M} \cdot \frac{1}{x} \cdot \exp{\left[-\frac{(\ln{x})^2}{2\sigma_M^2}\right]}.
    \label{eq:LNHOD}
\end{equation}
This model was introduced in \cite{2023_Antoine-Rocher}, and it can lead to a sharp asymmetry (close to a staircase function) when $\sigma_M > 1$ (see Figure 10 in \cite{2023_Antoine-Rocher}) which is not a physically motivated model for ELGs but still can reproduce the observed clustering with similar goodness of fit values to other HOD models.

\subsubsection{Standard model for satellite galaxies}
In this work, our baseline model for the expected number of satellites is a standard power law distribution given by 
\begin{equation}
    \langle N_{\text{sat}}(M) \rangle = A_s \left( \frac{M-M_{\text{cut}}}{M_1} \right)^\alpha,
    \label{eq:N_sat}
\end{equation}
where $A_s$ controls the overall normalization of the satellite occupation, $M_\text{cut}$ is the cut-off mass, which defines the minimum mass of the halo to host satellite galaxies. $M_1$ is a normalization parameter that characterizes the halo mass at which we expect to have one satellite per halo. Finally, $\alpha$ represents a power law index parameter. 

\subsection{Extensions for HOD modeling}
\label{sec:extended_HODs}
In addition to the 4 different HOD models described above, ELG mocks were also constructed using extensions to the baseline HOD framework, such as galactic conformity, assembly bias, and using a modified profile for satellites. In the following, such extensions are described.

\subsubsection{1-halo conformity}

The phenomenon in which properties of satellite galaxies exhibit some correlation with those of their central galaxy is called ``galactic conformity''. Galactic conformity was introduced in \cite{Weinmann2006}, which reported that the properties of satellite galaxies in SDSS data are strongly correlated with those of the central galaxy in their halo. Since then, other studies have found a significant trend in favor of galactic conformity
\cite{Kauffmann2013, Knobel2015, Phillibs2014, Robotham2013, Wang_White2012}. 
In a recent study, \cite{2022_Boryana_Hadzhiyska_1halo} used hydrodynamical models to investigate the ELG–halo connection and suggested the inclusion of 1-halo conformity by correlating the central and satellite occupation in the HOD. Similar conclusions about the influence of galactic conformity were derived in \cite{2023_Reyes-Peraza_NFW}. This was further investigated with DESI One-Percent data, where we measured ELG clustering deep into 1-halo scales with high accuracy, for the first time. This measurement exhibits an unexpectedly large amplitude on very small scales $ < 0.2\  h^{-1}$Mpc. This signal was modeled in \cite{2023_Antoine-Rocher} by introducing 1-halo conformity, which generates an excess in small-separation pairs by populating ELG satellites in the same halos as ELG centrals and further studied in \cite{Yuan2023-conf}. In this paper, three different prescriptions for galactic conformity are used to generate mocks. The three prescriptions are defined as follows:
\begin{itemize}
    \item Strict conformity: Halos can host ELG satellites only if they already host an ELG central:
    \begin{equation}
    \left\langle N_\text{sat}(M)\right\rangle = 
    \begin{cases}
     A_s \left( \frac{M-M_{\text{cut}}}{M_1} \right)^\alpha & \textrm{if ELG central},\\
    0 & \textrm{if not}.\\
    \end{cases}
    \label{equ:elgconfhod}
    \end{equation}
    \item Parameterized conformity model: One parameter that modulates the strength of the ELG central-satellite conformity is added to the HOD model. Specifically, the $M_1$ parameter, which controls the overall amplitude of satellite occupation, is modulated by whether the halo hosts an ELG central or not. We have that
    \begin{equation}
    \left\langle N_\text{sat}(M)\right\rangle = 
    \begin{cases}
     A_s \left(\frac{M- M_\mathrm{cut}}{M_\mathrm{1, EE}}\right)^\alpha & \textrm{if ELG central},\\
     A_s \left(\frac{M- M_\mathrm{cut}}{M_\mathrm{1}}\right)^\alpha & \textrm{if not},\\
    \end{cases}
    \label{equ:elg_param_confhod}
    \end{equation}
    where $M_{1,\text{EE}}$ is the new parameter that modulates the ELG-ELG conformity strength. If there is no conformity, then $M_{1,\text{EE}} = M_1$. Also, if there is maximal conformity, i.e. ELG satellites only occupy halos with ELG centrals, then $M_{1,\text{EE}} \ll M_1$. This model is studied in detail in~\cite{Yuan2023-conf}.
    \item Complex conformity model: We also test a model where a more complex form for galactic conformity was attempted. The model is given by
    \begin{equation}
    \left\langle N_\text{sat}(M)\right\rangle = 
    \begin{cases}
     \frac{(M/M_1)^\alpha}{1+(M/\kappa M_\text{cut})^{-A_s}} +\beta \left( \frac{M}{M_1^\prime} \right)^{\alpha_1} & \textrm{if ELG central},\\
     \frac{(M/M_1)^\alpha}{1+(M/\kappa M_\text{cut})^{-A_s}} & \textrm{if not},\\
    \end{cases}
    \label{equ:elg_complex_confhod}
    \end{equation}
    where $M_1^\prime$, $\alpha_1$, $\beta$ are galactic conformity parameters. This piece-wise satellite conformity model allows for different conformity behaviors in two halo mass regimes. This model is effectively a further generalization of Eq.~\ref{equ:elg_param_confhod}. We note that we have yet to find observational evidence for this model, but we include it as an ``overkill'' model in our systematic tests.
\end{itemize}

In our study for HOD systematics, we rely on some HOD models with strict conformity, and on some models that use complex conformity, which is a generalization of the parameterized conformity.

 

\subsubsection{2-halo conformity}

While HOD modeling is primarily a function of halo mass, both semi-analytical models and hydrodynamical simulations predict dependencies in other properties that are referred to as secondary biases in the literature. Evidence for 2-halo galactic conformity was found in \cite{2022_Boryana-Hadzhiyska}, based on the addition of extrinsic halo properties to the halo occupation model to alleviate discrepancies found at scales $r\gtrsim 1 h^{-1}\text{Mpc}$ using hydrodynamical simulations. Deviations from the assumption that the distribution of galaxies depends exclusively on the mass of the host halo are known as ``assembly bias''. We consider three forms of assembly bias that manifest primarily in the 2-halo term. In \cite{2023_Antoine-Rocher}, different halo secondary dependencies have been tested to model the ELG clustering: The halo concentration (referred to as C in the following), the local halo density (referred to as Env) and the local shear/halo density anisotropies (referred as Sh). They modified the probability of occupation following the parameterization suggested in~\cite{2022_Boryana-Hadzhiyska}, where
\begin{align}
 &\left\langle N^\prime_{\text{cen}}(M)\right\rangle &&= 
[1 + a_{\text{cen}} f_a (1-\left\langle N_{\text{cen}}(M)\right\rangle)]
\left\langle N_{\text{cen}}(M)\right\rangle, \\
&\left\langle N^\prime_{\text{sat}}(M)\right\rangle &&= 
[1 + a_{\text{sat}} f_a ] \left\langle N_{\text{sat}}(M)\right\rangle.
\end{align}
Here, $\left\langle N_{\text{cen}}(M)\right\rangle$ and $\left\langle N_{\text{sat}}(M)\right\rangle$ are the standard HODs described above and the parameter $f_a$ is introduced to describe the strength of the secondary property for each halo per mass bin. Within each mass bin, halos are ranked in decreasing order based on the specific property of interest. Subsequently, each halo is assigned a distinct $f_a$ value, assuming that this value decreases linearly from 0.5 to $-0.5$ as one progresses from the highest-ranked halo to the lowest-ranked one.

\subsubsection{Modified satellite profile}

The last extension to the standard HOD framework is a modification of the satellite profile. In the standard HOD framework, satellite galaxies are often positioned following an NFW profile or using dark matter particles. However, this description fails to describe correctly the transition between the 1-halo and 2-halo terms in the clustering of the ELG sample from the One-Percent survey. Inspired by several theoretical works \cite{AnguloOrsi2018, Blanton2007, Wetzel2012},  \cite{2023_Antoine-Rocher} introduced a modified NFW profile (referred to as mNFW in the following) to position ELG satellites using a combination of an NFW profile and an exponential law. Basically, the radial position of a fraction of the satellite galaxies is governed by 
\begin{equation}
\frac{dN(r)}{dr} = \exp{\left(-\frac{r}{\tau\cdot r_s}\right)}.
\label{eq:mNFW_exp}
\end{equation}
Here, Eq.~\ref{eq:mNFW_exp} modifies the expected number of satellites by means of the extra exponential term. The parameter $\tau$ modulates the slope of the exponential and $r_s$ is the scale radius of the profile. This new profile increases the probability of placing satellites in at larger radii and allows satellites to be placed outside the virial radius defined from the simulation (see Figure 16 of  \cite{2023_Antoine-Rocher} for details). This modified positioning of satellites translates into a significant improvement of the agreement of the goodness of fit, with a $\chi^2$ value dropping from 2.38 to 1.44. It is worth mentioning that other modifications to the NFW profile have been proposed (see for example \cite{2023_Reyes-Peraza_NFW}).

\subsection{Overview of the HOD models}\label{subsection:HOD_models_overview}

In summary, we considered 5 different HOD models for the central occupation: Gaussian HOD (Eq.~\ref{eq:GHOD}), Lognormal HOD (Eq.~\ref{eq:LNHOD}), Star Forming HOD (Eq.~\ref{eq:SFHOD}), High-Mass Quenched HOD (Eq.~\ref{eq:HMQ1}) and the  Modified HMQ model (Eq.~\ref{eq:mHMQ}) together with a single prescription for satellite galaxies (Eq.~\ref{eq:N_sat}). Then, extensions to the standard HOD framework are considered: Assembly bias with concentration, environment, and shear (local anisotropy); galactic conformity; and a modified NFW profile for satellite positions. The mocks presented in this analysis are fitted to the ELG sample from the DESI One-Percent survey and are generated using different fitting procedures as described below. A first set of mocks is generated based on the work from \cite{2023_Antoine-Rocher}, using a fitting pipeline based on Gaussian Processes described \cite{HODGP}. In these mocks, satellites are positioned using an NFW profile, if not specified otherwise. Then, following the first approach we mentioned before regarding satellite velocities, these are normally distributed around their mean halo velocity, with a dispersion equal to that of the halo dark matter particles, rescaled by an extra free parameter denoted $f_{\sigma_v}$ that accounts for velocity biases (see \cite{HODGP}). To fit the data, the projected clustering correlation function $w_p$ fitting range lies between $r_p$ 0.04 to 32 Mpc/$h$, while $\pi$ goes up to $\pi_\text{max}=40$ Mpc/$h$. Alongside $w_p$, we jointly fit the first two multipoles $\xi_0$, $\xi_2$ between two different $s$ ranges, being 0.8 to 32 Mpc/$h$ the first choice (1) and 0.17 to 32 Mpc/$h$ the second choice (2). Below is the list of the HOD models from the first set of mocks fitted using the first (1) binning range: 

\begin{itemize}
    \item \textbf{GHOD}: Gaussian HOD.
    \item \textbf{SFHOD}: Star forming HOD.
    \item \textbf{SFHOD+cf}: Star-forming HOD with strict galactic conformity.
    \item \textbf{HMQ}: HMQ model with $Q=100$.
    \item \textbf{1st-Gen}: First generation of DESI mocks based on HMQ model ($Q=0.1$) tuned to preliminar SV3 spectroscopic data from DESI.
    \item \textbf{LNHOD$_1$}: Lognormal HOD (1).
    \item \textbf{LNHOD$_1$-1h}: Same mock as the Lognormal HOD where the 1-halo term was removed, i.e. the satellite galaxies hosted by a halo with a central were removed.
    \item \textbf{LNHOD$_1$+cf}: Lognormal HOD with strict galactic conformity.
\end{itemize}
The following mocks were fitted using the second (2) binning range: 
\begin{itemize}
    \item \textbf{LNHOD$_2$}: Lognormal HOD (2). 
    \item \textbf{LNHOD$_2$+cf}: Lognormal HOD (2) with strict galactic conformity.
    \item \textbf{mHMQ}: Modified HMQ model. This model is HMQ in the limit $Q\rightarrow\infty$.
    \item \textbf{mHMQ+cf}: Modified HMQ model with strict galactic conformity.
    \item \textbf{mHMQ+cf+mNFW}: Modified HMQ model with strict galactic conformity and a modified NFW profile with an exponential function.
    \item \textbf{mHMQ+cf+C}: Modified HMQ with strict galactic conformity and concentration-based assembly bias.
    \item \textbf{mHMQ+cf+Env}: Modified HMQ with strict galactic conformity and environment-based assembly bias.
    \item \textbf{mHMQ+cf+Sh}: Modified HMQ with strict galactic conformity and shear-based assembly bias.
\end{itemize}

Finally, a second set of mocks is generated using the HMQ model and flexible conformity implementations, described in \cite{Yuan2023-conf}. 
These mocks follow the \texttt{AbacusHOD} framework, where satellites are assigned to dark matter particles inside the host halo as was the second approach for velocities assignment that we discussed, but at the same time velocity bias is allowed for both centrals and satellites using two extra parameters $\alpha_c$ and $\alpha_s$ (see Section~\ref{sec:HOD framework} and equations 8 \& 9 in~\cite{yuan_abacushod_2022}). Additionally, the more flexible conformity model defined in Eq.~\ref{equ:elg_complex_confhod} is used in these mocks. The models were fitted to the 2D clustering measurement $\xi(r_p,\pi)$ of the DESI ELG data from the One-Percent survey between $r_p$ from 0.04 to 32 Mpc/$h$ and $\pi$ going up to $\pi_\text{max}=40$ Mpc/$h$. Then, from the best-fit model, 6 mocks were generated with different HOD parameters sampled along the $3\sigma$ contour of the best-fit model, labeled as
\begin{itemize}
    \item \textbf{HMQ$^{(3\sigma)}_i$}: HMQ model spanning the 3-$\sigma$ region for the HOD parameters. The models have velocity bias applied to both centrals and satellites. The models with $i=1,2,3$ have no galactic conformity, while models with $i=4,5,6$ have complex galactic conformity. Overall, these models led to a good fit to the data with the best-fit $\chi^2$ per degree of freedom being close to 1.2.
\end{itemize}

In the end, a total of 22 HOD mocks are used in this analysis. All the models presented in this analysis reproduce the clustering measurement of the ELG sample from the DESI One-Percent survey up to 32 Mpc/$h$ with a similar goodness of fit, except for the model with the modified NFW profile, which obtained a significantly better $\chi^2$. Additionally, the HMQ$_i^{(3\sigma)}$ models also led to a good $\chi^2$ but these were fitted to an early version of the One-Percent survey data.

\begin{table*}
\begin{center}
\setlength{\tabcolsep}{4pt} 
\resizebox{\textwidth}{!}{
{\renewcommand{\arraystretch}{1.2}
\begin{tabular} {l|c|c|c|c|c|c|c|c|c|c|c|c|c|c|c|c|c|c|c|c}\hline\hline
\diagbox[innerwidth=\textwidth*1/5]{HOD}{parameters} & $A_s$ & $A_c$ & $\log_{10} M_c$ & $\alpha$ & $\log_{10}M_\text{cut}$ & $\log_{10}M_1$ & $f_{\sigma_v}$ & $\sigma_M$ & $\gamma$ & $p_\text{max}$ & $a_\text{cen}$ & $a_\text{sat}$ & $\alpha_\text{s}$ & $\alpha_\text{c}$ & $f_\text{exp}$ & $\tau$  & $\lambda_\text{NFW}$ & $\chi^2/$DoF & $\bar{n}_g$ & $z$\\
 \hline
1st-Gen & 1.00 & 1.00 & 13.22 & 0.007 & 11.22 & 12.28 & - & 0.60 & 4.70 & 0.65 & - & - & - & - & - & - & - & - & $3\times 10^{-3}$ & 1.1 \\
GHOD & 0.17 & 1.00 & 11.79 & -0.01 & 11.49 & 13.00 & 1.06 & 0.09 & - & - & - & - & - & - & - & - & - & - & $3\times 10^{-3}$ & 1.1 \\
SFHOD & 0.58 & 1.00 & 11.68 & -0.02 & 11.56 & 13.00 & 1.00 & 0.06 & -2.52 & - & - & - & -& - & - & - & - & - & $3\times 10^{-3}$ & 1.1 \\
SFHOD+cf & 0.47 & 0.50 & 11.99 & 0.55 & 11.17 & 13.00 & 1.31 & 0.24 & -2.69 & - & - & - & - & - & - & - & - & - & $2\times 10^{-3}$ & 1.1 \\
HMQ & 1.00 & 1.00 & 11.55 & 0.16 & 11.38 & 14.30 & 0.83 & 0.32 & 6.71 & 0.79 & - & - & - & - & - & - & - & - & $3\times 10^{-3}$ & 1.1 \\
LNHOD$_1$ & 0.21 & 0.50 & 11.80 & -0.03 & 11.40 & 13.00 & 0.92 & 0.17 & - & -  & - & - & - & - & - & - & - & - & $3\times 10^{-3}$ & 1.1 \\
LNHOD$_1$-1h & 0.21 & 0.50 & 11.80 & -0.03 & 11.40 & 13.00 & 0.92 & 0.17 & - & -  & - & -& -& -&- & - & - & - & $3\times 10^{-3}$ & 1.1 \\
LNHOD$_1$+cf & 0.50 & 0.50 & 12.32 & 0.58 & 11.19 & 13.00 & 1.49 & 1.16 & - & - & - & - & -& -& -& - & - & - & $3\times 10^{-3}$ & 1.1 \\
LNHOD$_2$ & 0.09 & 1.00 & 11.87 & -0.28 & 11.78 & 13.00 & 1.29 & 0.08 & - & - & - & - & - & - & - & - & - & 2.40 & $2\times 10^{-3}$ & 1.1 \\
LNHOD$_2$+cf & 0.26 & 1.00 & 12.55 & 0.80 & 11.23 & 13.00 & 1.25 & 1.71 &  & - & - & - & - & - & - & - & - & 2.37 & $2\times 10^{-3}$ & 1.1 \\
mHMQ  & 0.10 & 0.10 & 11.72 & -0.26 & 11.70 & 13.00 & 1.27 & 0.22 & 7.06 & 1.00 & - & - & - & - & - & - & - & 2.44 & $2\times 10^{-3}$ & 1.1 \\
mHMQ+cf  & 0.31 & 0.10  & 11.64 & 0.91 & 11.19 & 13.00 & 1.34 & 0.39 & 4.50 & 1.00 & -  &  - & - & - & - & - & - & 2.38 & $2\times 10^{-3}$ & 1.1 \\
mHMQ+cf+C & 0.35 & 0.10  & 11.63 & 0.91 & 11.17 & 13.00 & 1.35 & 0.39 & 4.54 & 1.00 & 0.75 & -0.32 & - & - & - & - &  - & 2.34 & $2\times 10^{-3}$ & 1.1 \\
mHMQ+cf+Env & 0.28& 0.10  & 11.61 & 0.86 & 11.19 & 13.00 & 1.34 & 0.44 & 5.76 & 1.00 & -0.02  & 0.02 & - & - & - & - & - & 2.43 & $2\times 10^{-3}$ & 1.1 \\
mHMQ+cf+Sh & 0.27 & 0.10 & 11.66 & 0.92 & 11.19 & 13.00 & 1.31 & 0.41 & 6.05 & 1.00 & 0.10 & 0.00 & - & - & - & - & - & 2.39 & $2\times 10^{-3}$ & 1.1 \\
mHMQ+cf+mNFW & 0.41 & 0.10 & 11.64 & 0.81 & 11.20 & 13.00 & 1.63 & 0.30 & 5.47 & 1.00 & -  & - & -& - & 0.58 & 6.14 & 0.67 & 1.44 & $2\times 10^{-3}$ & 1.1 \\ \hdashline
& $A_s$ & $A_c$ & $\log_{10} M_c$ & $\alpha$ & $\kappa$ & $M_1$ & $f_{\sigma_v}$ & $\sigma_M$ & $\gamma$ & $p_\text{max}$ & $a_\text{cen}$ & $a_\text{sat}$ & $\alpha_\text{s}$ & $\alpha_\text{c}$ & $f_\text{exp}$ & $\tau$  & $\lambda_\text{NFW}$ & $\chi^2/$DoF & $\bar{n}_g$ & $z$\\
 \hline
HMQ$^{(3\sigma)}_1$ & 3.94 & 1.00 & 11.34  & 0.42  & 2.88  & 15.49 & - & 0.16 & 5.52 & 0.03 & -  & - & 0.13 & 0.51 & - & -&- & - & $8\times 10^{-4}$ &   0.8 \\
HMQ$^{(3\sigma)}_2$ & 6.23 & 1.00 & 11.35 & 0.42 & 1.85 & 15.53 & - & 0.12 & 4.94 & 0.03 & -  & - & 0.06 & 0.52 & - & -&- & - & $8\times 10^{-4}$ &   0.8 \\
HMQ$^{(3\sigma)}_3$ & 4.08 & 1.00 & 11.27  & 0.42 & 3.41 & 15.50 & - & 0.22 & 6.79 & 0.02 & -  & - & 0.15 & 0.51 & - & -&- & - & $8\times 10^{-4}$ &   0.8 \\
HMQ$^{(3\sigma)}_4$ & 0.27 & 1.00 & 11.46  & 1.35 & 0.14 & 15.42 & - & 1.10 & 6.07 & 0.11 & -  & - & 0.27 & 0.59 & - & -&- & - & $8\times 10^{-4}$ &  0.8  \\
HMQ$^{(3\sigma)}_5$ & 0.56 & 1.00 & 11.50  & 1.23 & 4.33 & 15.58 & - & 1.16 & 5.19 & 0.12 & -  & - & 0.23 & 0.54 & - & -&- & - & $8\times 10^{-4}$ &   0.8 \\
HMQ$^{(3\sigma)}_6$ & 0.27 & 1.00 & 11.49  & 1.36 & 2.50 & 15.28 & - & 1.03 & 5.08 & 0.10 & -  & - & 0.22 & 0.62 & - & -&- & - & $8\times 10^{-4}$ &  0.8  \\
\hline\hline
\end{tabular}}}
\end{center}
\caption{Summary of the HOD models and HOD parameters used to estimate our systematic error budget. The table shows the HOD parameter values used to generate our mocks and some other characteristics such as number density and central redshift. We have a total of 22 HOD models with 25 mocks based on a common dark matter simulation for each of them. }
\label{Table:HOD_parameters}
\end{table*}

\subsection{HOD mocks for ELGs}
The HOD mocks used for our analysis are based on the \texttt{AbacusSummit} halo catalogs mentioned before in Section~\ref{Section:Simulations} and the aforementioned HOD prescriptions. To generate these mocks, a fitting methodology of two steps was used: The HOD mock generation and HOD parameter fitting. Most of the HOD mocks used in this work were obtained during the effort presented at~\cite{2023_Antoine-Rocher}, while in particular, the mocks for the HMQ$^{(3\sigma)}_i$ models ($i=1,2,..,6$) were generated following~\cite{yuan_abacushod_2022}.

The HOD mock generation for most of the HOD models declared in Section \ref{subsection:HOD_models_overview} were based on a fixed galaxy density of either $2\times 10^{-3} (h/\text{Mpc})^3$ or $3\times 10^{-3}(h/\text{Mpc})^3$. These mocks are built from simulations at $z=1.1$. As described in~\cite{2023_Antoine-Rocher}, central galaxies are positioned at the center of the host halos while, unless stated otherwise, satellites follow an NFW profile using $r_{25}$ (the radius of a sphere enclosing 25\% of the halo particles). Satellite velocities follow a normal distribution around the mean velocity of the host halo with a dispersion equal to that of the dark matter particles in the halo but rescaled by an extra parameter $f_{\sigma_v}$ as done in ~\cite{2021_Shahab-Alam}. In the case of the HMQ$^{(3\sigma)}_i$ models, centered at $z=0.8$, the number density used for the mock generation is of about $0.8\times 10^{-3} (h/\text{Mpc})^3$. Moreover, the HMQ$^{(3\sigma)}_i$ mocks were generated using \texttt{AbacusHOD}~\cite{yuan_abacushod_2022}, which introduces some generalizations based on \cite{2018MNRAS_Yuan, 2021MNRAS_Yuan}, allowing satellite profile to deviate from that of the host halo. Additionally, while the rest of the HOD models use a velocity bias prescription based on the rescaling parameter $f_{\sigma_v}$, the HMQ$^{(3\sigma)}_i$ mocks use velocity bias in both centrals and satellites through two parameters $\alpha_\text{c}$ and $\alpha_\text{s}$, respectively, as shown in~\cite{2023_Sihan-Yuan} (labeled there as $\alpha_\text{vel,c}$ and $\alpha_\text{vel,s}$, respectively). However, the rest of our mocks assume the centrals have the same velocity as that of the dark matter halo. It is worth mentioning that the measured number density (corrected for completeness) of the DESI ELG sample is $\sim 10^{-3} (h/\text{Mpc})^3$ and $\sim 0.75 \times 10^{-3} (h/\text{Mpc})^3$, for $z=0.8$ and $z=1.1$, respectively (see Fig. 1 in \cite{2023_Antoine-Rocher}). However, here we take the freedom to consider mocks at high number density to reduce the shot-noise and focus on HOD effects.

For most of the HOD mocks, the HOD parameter fitting was performed by generating 20 mock catalogs at every point of the parameter space and then comparing the clustering signal to that of the data. Then an average $\chi^2$ from 20 realizations is computed as well as a standard deviation for $\langle \chi^2 \rangle$, and both are fed into a Gaussian Process (GP) fitting pipeline to obtain a surrogate model of the likelihood surface. The computed $\chi^2$ is defined as 
\begin{equation}
    \begin{split}
    \chi^2 = & (\boldsymbol{\xi}_\text{data}-\boldsymbol{\xi}_\text{model})^T [\boldsymbol{C}_\text{data}/(1-D_\text{data})+\boldsymbol{C}_\text{model}/(1-D_\text{model}))]^{-1} \\ & (\boldsymbol{\xi}_\text{data}-\boldsymbol{\xi}_\text{model}),
    \end{split}
    \label{eq:chi2_antoine}
\end{equation}
where $\boldsymbol{\xi}_\text{model}$ is the clustering measurements data vector, $\boldsymbol{C}$ is the covariance matrix calculated by applying the Jackknife method to the One-Percent survey footprint divided by 128 independent regions, as described in~\cite{2023_Antoine-Rocher}. Additionally, $D$ is a Hartlap correction factor~\cite{2007_J-Hartlap}. Thus, Eq. \ref{eq:chi2_antoine} is used as our statistic to fit $\xi_0$, $\xi_2$ and $w_p$ data measurements and generate mocks for our HODs. In the case of the HMQ$^{(3\sigma)}_i$ mocks, \texttt{AbacusHOD} rather samples the HOD parameter space using the \texttt{DYNESTY} nested sampler~\cite{2020_Joshua-Speagle} and assumes a $\chi^2$ statistic given by two contributions. The first one is a simple Gaussian likelihood-based $\chi^2$ given by
\begin{equation}
    \chi^2 = (\boldsymbol{\xi}_\text{model}-\boldsymbol{\xi}_\text{data})^T \boldsymbol{C}^{-1}(\boldsymbol{\xi}_\text{model}-\boldsymbol{\xi}_\text{data}),
\end{equation}
and the second one is a contribution that penalizes the HOD models with insufficient galaxy number density, defined as
\begin{equation}
    \chi_{n_g} = 
    \begin{cases}
    \left( \frac{n_\text{mock}-n_{data}}{\sigma_n} \right)^2, & n_\text{mock} < n_\text{data}, \\
    0,  & n_\text{mock} \geq n_\text{data}.
    \end{cases}
\end{equation}
Here, $n_\text{data}$ is the observed number density of the data, $n_\text{mock}$ is the number density of the mock and $\sigma_n$ is the associated uncertainty of the galaxy number density of the mock. Then, the total $\chi^2$ statistic can be written as
\begin{equation}
    \chi^2 = \chi^2_{\xi}+\chi^2_{n_g}.
\end{equation}
Such a statistic is used to fit the two-point correlation function $\xi(r_p,\pi)$ and get the best fit HOD parameters. From there, we generate the 3-$\sigma$ contours sampled along the best-fit model to define our mocks for the HMQ$^{(3\sigma)}_i$ models.

As we are using 25 different realizations of \texttt{AbacusSummit}, we produce 25 different mocks for every HOD. Additionally, as we will mention later, to have HOD mocks with similar number density we also sub-sample some of our high-density mocks to obtain a low-density version of such HOD mocks, with galaxy number density of around $10^{-3}(h/\text{Mpc})^3$. Finally, we provide in Table~\ref{Table:HOD_parameters} a summary table where we list the values of the HOD parameters used to generate the HOD mocks as well as some properties of interests and variants.

%% file: BodyText/Methodology.tex
In this section, we describe the methodology for the data analysis that we follow in this work. Ultimately, we aim to assess the impact on the BAO fits due to the assumed HOD model. As stated before, to estimate the impact due to the HOD prescription in our BAO analysis we opt to generate mocks tuned with early DESI data for various HOD models and use these mocks to assess the systematics level. In the following, we describe in a more specific way our galaxy clustering estimation and our modeling of the BAO template and fitting scheme.
\subsection{Galaxy clustering estimation}\label{subsection:Galaxy_clustering}
We perform analyses in both configuration and Fourier spaces. In the case of the configuration space analysis, typically one can measure the galaxy two-point correlation function using the~\cite{1993_Landy&Szalay} estimator, 
given by
\begin{equation}
    \xi(s,\mu) = \frac{DD(s,\mu) - 2DR(s,\mu) - RR(s,\mu)}{RR(s,\mu)},
    \label{eq:estimator_cf_pre}
\end{equation}
where $DD$ are the counts for galaxy pairs and $RR$ are the counts for pairs in the randomly generated catalog. Similarly, $DS$ are the cross-pair counts. Here, $s$ is the separation of the galaxies, and $\mu$ is the cosine of the angle formed between the line-of-sight and the galaxy pair. While in principle we could estimate $RR$ analytically, at the time of this work we used Eq.~\ref{eq:estimator_cf_pre} as done in for example ,~\cite {Mohammad:2021aqc}. We use the publicly available DESI package \texttt{pycorr}~\footnote{\url{https://github.com/cosmodesi/pycorr} (version 1.0.0)} to measure the two-point correlation function. 

Additionally, the BAO measurement is enhanced by a process called reconstruction~\cite{Eisestein2007-reconstruction, 2012_Nikhil_Padmanabhan, 2015_Seo}, where the displacement field of the galaxies due to the peculiar velocities (which arise from bulk flows due to non-linear structure growth) is reconstructed from the observed density field. Then, The position of each galaxy can be reversed to where each galaxy would reside if such non-linear effects had not occurred. We estimate the shift in the galaxies due to reconstruction using \texttt{pyrecon}~\footnote{\url{https://github.com/cosmodesi/pyrecon} (version 1.0.0)}. We follow the details and methodology for optimal reconstruction presented at \cite{KP4s3-Chen} and applied it to the unblinded data in \cite{KP4s4-Paillas}. For the reconstruction procedure, we assume a galaxy bias $b=1.2$ for the ELG tracer and use a MultiGrid~\footnote{\url{https://github.com/martinjameswhite/recon_code}} reconstruction algorithm. We only apply the rec-sym reconstruction~\cite{2019_Chen}, which preserves the linear kaiser factor and so the quadrupole term, which accounts for redshift space distortions.

The two-point correlation functions for post-reconstruction measurements were run by taking into account the pair counts $SS$ from the shifted galaxy catalog. In this case, the estimator can be written as
\begin{equation}
    \xi(s,\mu) = \frac{DD(s,\mu) - 2DS(s,\mu) - SS(s,\mu)}{RR(s,\mu)}.
    \label{eq:estimator_cf_post}
\end{equation}
In the case of the two estimators, we project the two-point correlation function into correlation function multipoles by computing
\begin{equation}
    \xi_\ell(s) = \frac{2\ell+1}{2} \int_1^1 d\mu \xi(s,\mu)\mathcal{L}_\ell(\mu),
    \label{eq:cf_multipoles}
\end{equation}
where $\ell=0,2,4$. However, we do not consider the hexadecapole term since it can introduce noise from the measurements to the fits beyond the HOD-dependent systematics. Then, we simplify our analysis by considering only the monopole and quadrupole terms. We compute $\xi(s,\mu)$ for evenly spaced bins of $1\text{Mpc}/h$ for $s$ going from $0\text{Mpc}/h$ up to $200\text{Mpc}/h$ and $200$ bins for $\mu$ between $-1$ and $1$. However, ultimately we rebin our results for $\xi_\ell(s)$ to bin steps of $4\text{Mpc}/h$.

In the case of the Fourier space measurements, we use the implementation of the periodic box power spectrum estimator as shown in~\cite{2017_Nick-Hand} into the DESI package \texttt{pypower}~\footnote{\url{https://github.com/cosmodesi/pypower} (version 1.0.0)}. If the density contrast is given by $\delta_g(\boldsymbol{r})=\frac{n_g(\boldsymbol{r})}{\Bar{n}_g}-1$ then the power spectrum multipoles can be calculated as
\begin{equation}
    P_\ell(k) = \frac{2\ell+1}{V} \int \frac{d\Omega_k}{4\pi}\delta_g(\boldsymbol{k})\delta_g(-\boldsymbol{k})\mathcal{L}_\ell(\hat{\boldsymbol{k}}\cdot\hat{\eta})-P_\ell^{\text{noise}}(k),
    \label{eq:ps_multipoles}
\end{equation}
where $V$ is the volume of the box, $\boldsymbol{k}$ is the wavenumber vector and $\eta$ is the line-of-sight vector. The shot-noise term is only considered for the monopole term. We interpolate the density field using meshes of $512^3$ based on a Triangular-Shaped Cloud (TSC) prescription. We use steps of $\Delta k=0.001 h/\text{Mpc}$ starting at $k=0 h/\text{Mpc}$ and then we rebin the measurements to $\Delta k=0.005 h/\text{Mpc}$. For both configuration space and Fourier space analyses, the procedures described above match the method adopted for the final DESI 2024 results.

\subsection{Zeldovich control variates technique }\label{subsection:Control_variates}
We go beyond the typical two-point measurements and take advantage of the Control Variates (CV) technique to produce additional measurements where statistical noise is reduced. Given a random variable $X$ and a correlated random variable $C$ with mean $\mu_c$, the CV technique is based on building a new estimator for $X$ based on
\begin{equation}
    Y = X - \beta (C-\mu_c),
    \label{eq:cv_estimator}
\end{equation}
where $Y$ is the new random variable and $\beta$ is some arbitrary coefficient that can be set to minimize the variance of $Y$. It can then be shown that the optimal choice for such coefficient is $\beta^{\star}=\text{Cov}[X,C]/\text{Var}[C]$ (see~\cite{2022_N-Chartier}). In the case where $\mu_c$ can be obtained analytically, the variance of the new random variable can be written just as
\begin{equation}
    \text{Var}[Y] = \text{Var}[X] (1-\rho_{xc}^2),
    \label{eq:cv_variance}
\end{equation}
where $\rho_{xc}$ is the Pearson correlation coefficient between $X$ and $C$. Therefore, we observe that there is a reduction in the variance of the new random variable $Y$ coming from the extra information provided by the correlation between $X$ and $C$. Here, we use the CV technique as described in~\cite{2023_Boryana-Hadzhiyska} to produce measurements in both Fourier space and configuration space. We explain below the CV technique applied to the power spectrum measurements as an example case.

Recently,~\cite{2022_N-Kokron} provided a recipe to combine Lagrangian perturbation theory models with N-body simulations to reduce the effects of finite volume in calculating ensemble average properties. More precisely, using the Zeldovich approximation~\cite{1970_Y-Zeldovich} they noticed that the Zeldovich displacements calculated during the initial conditions are strongly correlated with the final density field. Therefore, one can get a reduced noise version of the biased tracer power spectrum if we use the fact that this one has a correlation with the Zeldovich approximation version of the power spectrum. This is referred to as Zeldovich Control Variates (ZCV). While~\cite{2022_N-Kokron} worked out the real space version,~\cite{2023_J-DeRose} developed the redshift space version of the formalism. In the last one, we can write
\begin{equation}
    \hat{P}^{\ast, tt}_{\ell}(k) = \hat{P}^{tt}_{\ell}(k) - \beta_{\ell}(k) \left(\hat{P}^{ZZ}_{\ell}(k) - P^{ZZ}_{\ell}(k)\right) ,
    \label{eq:zcv_power_spectrum}
\end{equation}
where $\hat{P}^{tt}_{\ell}(k)$ is the power spectrum measured at late times in the N-body simulation, $\hat{P}^{ZZ}_{\ell}(k)$ is the power spectrum measured in the Zeldovich approximation and $P^{ZZ}_{\ell}(k)$ is the ensemble-average power spectrum in the Zeldovich approximation. It can be shown that  
\begin{equation}
    \beta_{\ell}(k) = \left[ \frac{\hat{P}^{tZ}_{\ell}(k)}{\hat{P}^{ZZ}_{\ell}(k)} \right]^2,
    \label{eq:zcv_beta}
\end{equation}
where $\hat{P}^{tZ}_{\ell}(k)$ is the cross-power spectrum  and $\hat{P}^{ZZ}_{\ell}(k)$ is the auto-power spectrum of the ZCV. We apply the ZCV technique to all our HOD galaxy catalogs to get a set of noise-reduced measurements. For more details on ZCV or derivations for the correlation function multipoles we refer to the reader to~\cite{2023_Boryana-Hadzhiyska}.

We can also produce post-reconstruction CV measurements using a linear model. This formalism is called Linear Control Variates (LCV) and is described in detail in~\cite{2023_Boryana-Hadzhiyska}. The LCV equation is given by
\begin{equation}
    \hat{P}^{\ast, rr}_{\ell}(k) = \hat{P}^{rr}_{\ell}(k) - \beta_{\ell}(k) \left(\hat{P}^{LL}_{\ell}(k) - P^{LL}_{\ell}(k)\right) ,
    \label{eq:lcv_power_spectrum}
\end{equation}
where $\hat{P}^{rr}_{\ell}(k)$ is the measured power spectrum for a given tracer, but now $\hat{P}^{LL}_{\ell}(k)$ and $P^{LL}_{\ell}(k)$ are the measured and analytical reconstructed power spectrum using linear theory. Similarly, we also have that
\begin{equation}
    \beta_{\ell}(k) = \left[ \frac{\hat{P}^{rL}_{\ell}(k)}{\hat{P}^{LL}_{\ell}(k)} \right]^2,
\end{equation}
where $\hat{P}^{rL}_{\ell}$ is the measured cross-power spectrum between the true and the linear modeled reconstructed fields. Finally, for both ZCV and LCV, we can obtain the correlation function multipoles by performing an inverse Fourier transform, followed by some appropriate treatment for remnant ringing effects. In sum, we can use the fact that we have a reliable analytic approximation at large scales such as the Zeldovich approximation to remove sample variance noise using the CV technique.

\subsection{Baryon acoustic oscillations modeling}\label{subsection:BAO_fitting}
We perform an anisotropic analysis to extract the BAO feature information. We use a slightly modified version of~\cite{2017_Beutler} for the modeling of our BAO template. We adopt a BAO modeling version close to the one presented in \ChenHowlett and explain the differences further below. We start by splitting the linear matter power spectrum into `wiggle' and `no-wiggle' parts, labeled as $P_\text{w}(k)$ and $P_\text{nw}(k)$ respectively, using the method as described in~\cite{2016_Hinton}. From there we define an oscillatory term $\mathcal{O}(k) = 1 + P_\text{w}(k)/P_\text{nw}(k)$, which contains the BAO information. We proceed to fit the BAO feature by constructing a template of the galaxy power spectrum. The random peculiar velocities at small scales cause an elongation of the positions in the redshift space. This produces damping due to the non-linear velocity field that can be parameterized with
\begin{equation}
    \mathcal{D}_\text{FoG}(k,\mu,\Sigma_\text{s}) = \frac{1}{(1+k^2\mu^2\Sigma_s^2/2)^2},
\end{equation}
where $\Sigma_\text{s}$ is the damping scale parameter for the Fingers-of-God. Additionally, we have to introduce a term to account for the coherent infall of galaxies at large scales, given by $(1+\beta\mu^2)^2$~\cite{1987_Kaiser}, where $\beta$ is a free parameter to fit. These two terms allow us to define a smoothed power spectrum that accounts for the effect of galaxy bias and peculiar velocities, as
\begin{equation}
    P_\text{sm}(k,\mu) = B^2 (1+\beta\mu^2)^2 \mathcal{D}_\text{FoG}(k,\mu,\Sigma_\text{s}) P_\text{nw}(k).
    \label{eq:smooth_power}
\end{equation}
We also have added a factor $B$ that acts as a linear galaxy bias. Now, the growth of non-linear structure can also wash out the BAO feature. Then, we introduce an extra damping factor given by
\begin{equation} 
\mathcal{C}(k,\mu) = \exp\left[-k^2\left( \frac{(1-\mu^2)\Sigma_\perp^2}{2} + \frac{\mu^2\Sigma_\parallel^2}{2} \right)\right].
\label{eq:damping_factor}
\end{equation}
The first mode affects the BAO signal perpendicular to the line-of-sight and is parameterized by $\Sigma_\perp$, while the second mode acts along the line-of-sight and is represented by $\Sigma_\parallel$. We now add Eq.~\ref{eq:damping_factor} to Eq.~\ref{eq:smooth_power} and calculate the power spectrum multipoles as
\begin{equation}
    \begin{split}
    P_\ell(k) &= \frac{2\ell+1}{2} \int_1^1 d\mu\mathcal{L}_\ell(\mu) P_\text{sm}[k^\prime(k,\mu),\mu^\prime(\mu)] \\
    & \times [1+(\mathcal{O}(k^\prime)-1)\mathcal{C}(k^\prime,\mu^\prime)]  + \sum_{i=-1}^{4} A_\ell^{(i+1)} k^i.
    \end{split}
    \label{eq:pow_spec_multipoles}
\end{equation}
Here, $\mathcal{L}_\ell$ represents the Legendre multipoles and $A_\ell^{i}$ are a set of polynomials that allows us to fit the broadband of the power spectrum. The number and power of polynomial terms were chosen based on the work of \cite{KP4s2-Chen}. We perform analytic marginalization of both the broadband terms and the linear galaxy bias as described in \cite{KP4s2-Chen}. The prime ($\prime$) represents the use of the true wave numbers compared to the observed wave numbers (without $\prime$). This rescaling is added to our template to account for the radial dilation of BAO and the anisotropic warping. The radial dilation of BAO is parameterized by $\alpha_\text{iso}$, given by
\begin{equation}
    \alpha_\text{iso}(z) = 
    \frac{D_V(z)r^\text{fid}_s(z_d)}{D_V^\text{fid}(z)r_s(z_d)}.
    \label{eq:alpha_iso}
\end{equation}
Here, the spherically-averaged distance $D_V(z)$ is divided by the sound horizon $r_s$ evaluated at the drag epoch redshift $z_d$. This ratio $D_V(z)/r_s(z_d)$ is then divided by the fiducial value used to construct our template. On the other hand, the fact that we measure redshifts and these are converted to physical distances by assuming a concrete cosmology can lead to bias since the assumed cosmology can be different from the true one. Such a bias can introduce an anisotropic warping effect deviating us from having a purely isotropic scale (besides non-linear structure evolution). We account for this effect (referred to as AP, \cite{1979_Alcock_Paczynski}) by using the scaling parameter
\begin{equation}
    \alpha_\text{AP}(z) = \frac{H(z)^\text{fid}D_A(z)^\text{fid}}{H(z)D_A(z)}.
    \label{eq:alpha_ap}
\end{equation}
Hence, the parameters defined in Eq.~\ref{eq:alpha_iso} and Eq.~\ref{eq:alpha_ap} account for the dilation of the coordinates according to (compare to ~\cite{1996_Ballinger})
\begin{equation}
    k^\prime = k\frac{\alpha_\text{AP}^{1/3}}{\alpha_\text{iso}}\left[ 1+\mu^2 \left( \frac{1}{\alpha_\text{AP}^2}-1 \right) \right]^{1/2}
\end{equation}
and
\begin{equation}
    \mu^\prime = \frac{\mu}{\alpha_\text{AP}}\left[ 1+\mu^2 \left( \frac{1}{\alpha_\text{AP}^2}-1 \right) \right]^{-1/2}.
\end{equation}
If we work rather in configuration space, we just transform the power spectrum multipoles from Eq.~\ref{eq:pow_spec_multipoles}, \textit{without including any contribution from the polynomial terms}, according to
\begin{equation}
    \xi_\ell(s) = i^\ell \int_0^\infty \frac{k^2}{2\pi^2} P_\ell(k) j_\ell(ks)dk + \sum_{i=-2}^{1} A_\ell^{(i+2)} s^i,
    \label{eq:corr_func_multipoles}
\end{equation}
where $j_\ell$ are the spherical Bessel functions and 4 polynomial terms were added in this case. Notice that the polynomial terms to characterize the broadband are added after the spherical Hankel transformation of the power spectrum multipoles. Thus, our fits in Fourier space are performed using Eq.~\ref{eq:pow_spec_multipoles}, while the fits in configuration space are computed by using Eq.~\ref{eq:corr_func_multipoles}. 

We note that, due to the parallel nature in which the two studies were performed, our BAO modeling procedure differs slightly from that recommended by \ChenHowlett, which is the BAO modeling method adopted for the DESI 2024 results, in that 1) we allow the FoG damping term to affect both the wiggle and no-wiggle terms; 2) we use a `polynomial'-based broadband method instead of their preferred `spline'-based method; 3) we allow the BAO dilation parameters to affect both the wiggle and no-wiggle model components and 4) we use $k$ and $\mu$ for the redshift space distortion factor within $C(k,\mu)$ instead of the dilated coordinates. However, they conclude that the BAO fitting methodology is robust to any of these choices, at the $0.1\%$ level for $\alpha_\text{iso}$ and $0.2\%$ level for $\alpha_\text{AP}$. In any case, as demonstrated later, our method is also able to recover unbiased BAO constraints, and because we focus on comparative differences between different simulations in this work our results are immune to these choices.

A summary table with the parameters used in our modeling of the BAO template, along with some derived parameters and statistical definitions are provided in Table~\ref{Table:BAO_parameter_description}. In the following, we describe our parameter estimation methodology.

\begin{table*}
\begin{center}
\setlength{\tabcolsep}{4pt} 
\resizebox{\textwidth}{!}{
{\renewcommand{\arraystretch}{1.2}
\begin{tabular}{c|l|l}
\hline\hline
Parameter & Description & Prior \\ \hline
\multicolumn{3}{l}{1. BAO template MCMC parameters}  \\ \hline
$\alpha_{\text{iso}}$ & Isotropic shift in the BAO scale. & $(0.8,1.0)$ \\
$\epsilon$ & Anisotropic warping of the BAO signal. & $(-0.2,0.2)$ \\ 
\multirow{2}{*}{$\Sigma_{\parallel}$} & \multirow{2}{*}{Non-linear damping of the BAO feature mode in the line-of-sight.} & Pre, CS: $\mathcal{N}(8.75,2.0)$, Post, CS: $\mathcal{N}(5.42,2.0)$ \\
 & & Pre, FS: $\mathcal{N}(8.94,2.0)$, Post, FS: $\mathcal{N}(5.35,2.0)$\\
\multirow{2}{*}{$\Sigma_{\perp}$} & \multirow{2}{*}{Non-linear damping of the BAO feature mode perpendicular to the line-of-sight.} & Pre, CS: $\mathcal{N}(4.23,2.0)$, Post, CS: $\mathcal{N}(1.92,2.0)$\\
 & & Pre, FS: $\mathcal{N}(3.98,2.0)$, Pre, FS: $\mathcal{N}(1.40,2.0)$ \\
\multirow{2}{*}{$\Sigma_{\text{s}}$} & \multirow{2}{*}{Fingers of god parameter for the velocity dispersion in the Lorentzian form.} & Post, CS: $\mathcal{N}(5.36,4.0)$, Post, CS: $\mathcal{N}(1.70,4.0)$ \\ 
 & & Pre, FS: $\mathcal{N}(2.0,2.0)$, Post, FS: $\mathcal{N}(0.0,2.0)$ \\
$\beta$ & Kaiser term parameter equal to $f/b$. & $(0.01,4.0)$ \\
$b$ & Linear galaxy bias. Obtained by analytic marginalization. & (0.1,10.0) \\
$A_{\ell,i}$ & Polynomial coefficients for broad band terms. Obtained by analytic marginalization. & (-20000.0, 20000.0) \\ \hline
\multicolumn{3}{l}{2. Derived BAO parameters}  \\ \hline
$\alpha_{\text{AP}}$ & \multicolumn{2}{l}{Alcock-Paczynski scale distortion parameter. Derived from $\epsilon$.} \\ 
$\alpha_{\parallel}$ & \multicolumn{2}{l}{BAO scaling parameter along the line-of-sight. Derived from $\alpha_\text{iso}$ and $\epsilon$.} \\
$\alpha_{\perp}$ & \multicolumn{2}{l}{BAO scaling parameter perpendicular to the line-of-sight. Derived from $\alpha_\text{iso}$ and $\epsilon$.} \\ \hline
\multicolumn{3}{l}{3. Statistical definitions}  \\ \hline
$\langle\Delta X\rangle$ & \multicolumn{2}{l}{Mean difference between the measured value of X and the fiducial value of X.} \\
$\sigma_X$ & \multicolumn{2}{l}{Statistical uncertainty associated with variable X given the assumed covariance matrix.}  \\
$\sigma(\overline{X})$ & \multicolumn{2}{l}{Standard deviation for the mean of X, obtained from 25 independent measurements.} \\
\hline\hline
\end{tabular}}}
\end{center}
\caption{Description of the BAO template parameters used throughout this work. The BAO template parameters are described in the first section of the table along with the priors used during the fitting stage. The parameters shown in the next section of the table are considered derived parameters after the MCMC parameter estimation. Finally, a brief description of the statistical notation we use in follow-up sections is described at the bottom of the table. As explained in the text, our results are unaffected compared to the BAO modeling methodology used for the DESI 2024 results.}
\label{Table:BAO_parameter_description}
\end{table*}

\begin{figure*}
	\includegraphics[width=\textwidth]{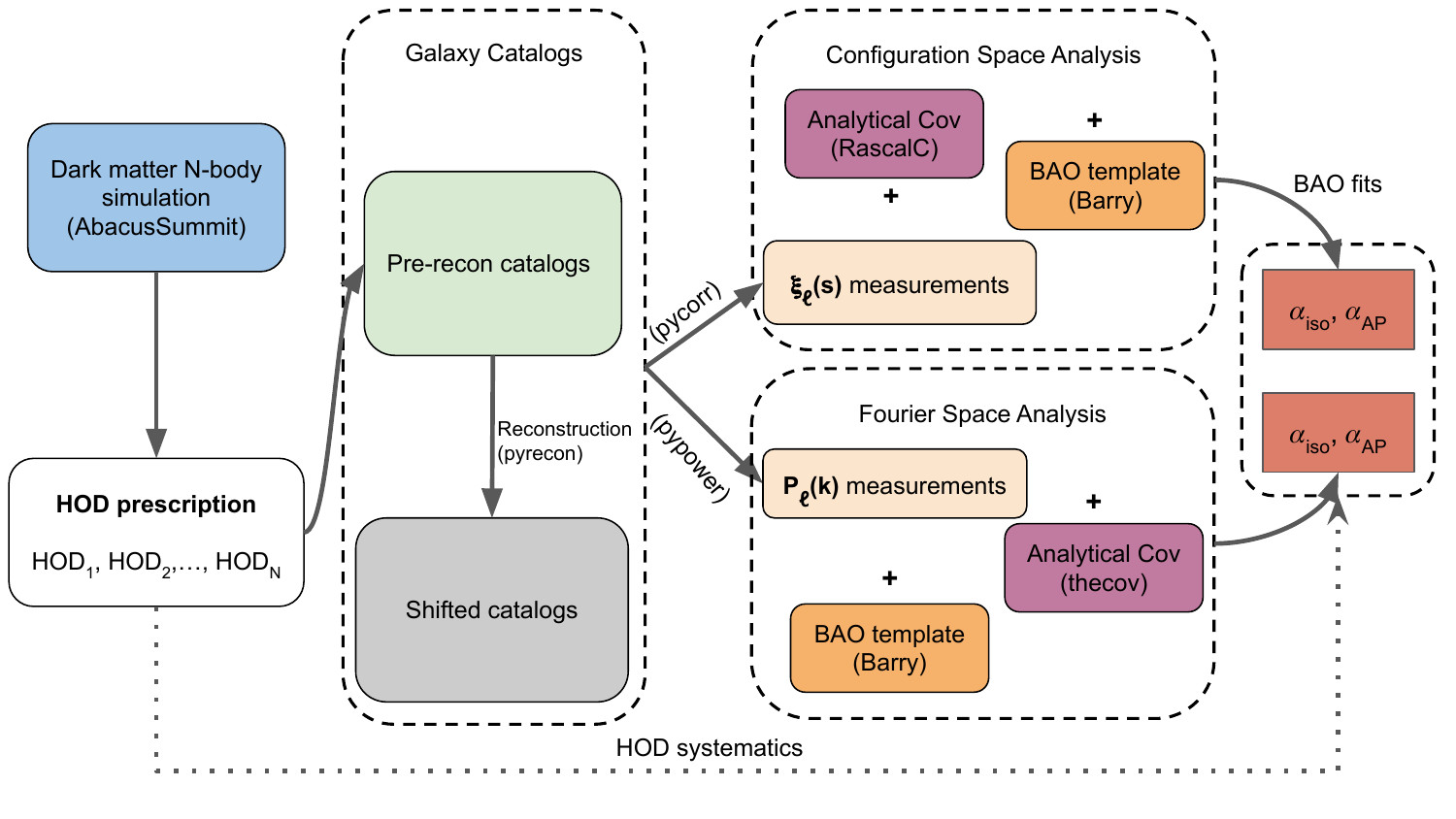}
    \caption{Flowchart that summarizes the algorithm followed in deriving the results of this work. While the HOD prescription is used at the top of the dark matter simulation to produce pre-reconstruction catalogs, we follow a series of calculations to obtain the impact on the BAO parameters. The different codes or simulations used in each step are written between parentheses.}
    \label{Fig:Flowchart}
\end{figure*}

\subsection{Parameter fitting scheme}\label{subsection:fitting}
For both Fourier space fits and configuration space fits, we perform MCMC analysis to fit the BAO template described before. We use \texttt{Barry}~\footnote{\url{https://github.com/Samreay/Barry} (version April 2023)} ~\cite{2019_Hinton} for our parameter estimation\footnote{Note that \texttt{Barry} uses $\epsilon=\alpha_\text{AP}^{1/3}-1$ as a base parameter for MCMC sampling.}. As stated in \cite{DESI2024.III.KP4}, this BAO fitting pipeline is consistent with the \texttt{desilike}\footnote{\url{https://github.com/cosmodesi/desilike}} code used for the DESI 2024 results, but given the early nature of our analysis we use \texttt{Barry}. We explore the parameter space by using the nested sampling algorithm in \texttt{DYNESTY}~\cite{DYNESTY}, while we minimize the $\chi^2$ as defined by
\begin{equation}
    \chi^2 = \left( P_\text{model}-P_\text{data} \right)^T W \left( P_\text{model}-P_\text{data} \right).
\end{equation}
Here, $P_\text{model}$ is given by Eq.~\ref{eq:pow_spec_multipoles} if we perform a Fourier space analysis or Eq.~\ref{eq:corr_func_multipoles} if we rather work with the correlation function. Similarly, $P_\text{data}$ represents the corresponding data vector, which can be either the power spectrum multipoles or the correlation function multipoles. The matrix $W$ is given by 
\begin{equation}
    W = \begin{cases}
C^{-1} \left(\frac{N_\text{m}-N_\text{d}-2}{N_\text{m}-1}\right) &\text{, if EZMocks covariance available (only for 1st-Gen),}\\
C^{-1} &\text{, if analytical covariance is used.}
\end{cases}
\end{equation}
Hence, the form we adopt for $W$ depends if we are fitting a particular HOD model using an analytical covariance matrix or rather by a covariance matrix coming from several mock realizations. Overall, we use analytical covariance matrices for all of our HOD models except by 1st-Gen, where we use a covariance matrix generated from 1000 EZMocks simulations~\cite{2015_Chuang_EZMocks}. In the case of the EZMocks covariance matrix, we apply the Hartlap correction factor~\cite{2007_J-Hartlap} given the fact that we are inverting a covariance matrix generated from a finite amount of mocks, which can lead to biased parameter estimation. This correction is enough since we use the mean value obtained from the fits to calculate the systematic error budget as described in Section \ref{subsection:robustness_against_HODs}, and the errors we report come from the dispersion among 25 realizations. However, as shown in \cite{Percival:2021cuq}, including extra correction factors could slightly broad the posterior distribution of the parameters. For the rest of the HOD models, we generate analytical covariance matrices tuned to the clustering of 25 mocks for every HOD. In the case of Fourier space analysis, we generate Gaussian analytical covariances by using \texttt{thecov}~\footnote{\url{https://github.com/cosmodesi/thecov} (version 0.1.0)}, which is based on \cite{2019_Wadekar} and is validated in the context of DESI 2024 in \Alves. For configuration space analysis, we use \texttt{RascalC}\footnote{\url{https://github.com/oliverphilcox/RascalC} (version 2.2)}. This code was applied to early DESI data in \cite{2023_Rashkovetskyi} and it was originally presented in \cite{2019_Philcox}. The validation of \texttt{RascalC} towards DESI 2024 results will be detailed in \Rashkovetskyi. In both cases, a validation against covariance matrices derived from EZMocks simulations was carried out in \ForeroSanchez. We do not apply any correction factor when using analytical covariance matrices as little inversion bias is expected. The choices used in our analysis match the methods adopted for the DESI 2024 results, as presented in \cite{DESI2024.III.KP4}. The overall methodology used in this work related to HOD-dependent systematics can be summarized by the flowchart given in Fig.~\ref{Fig:Flowchart}, where the various performed calculations and pipeline codes used in our analysis are highlighted.

\begin{figure*}
	\includegraphics[width=\textwidth]{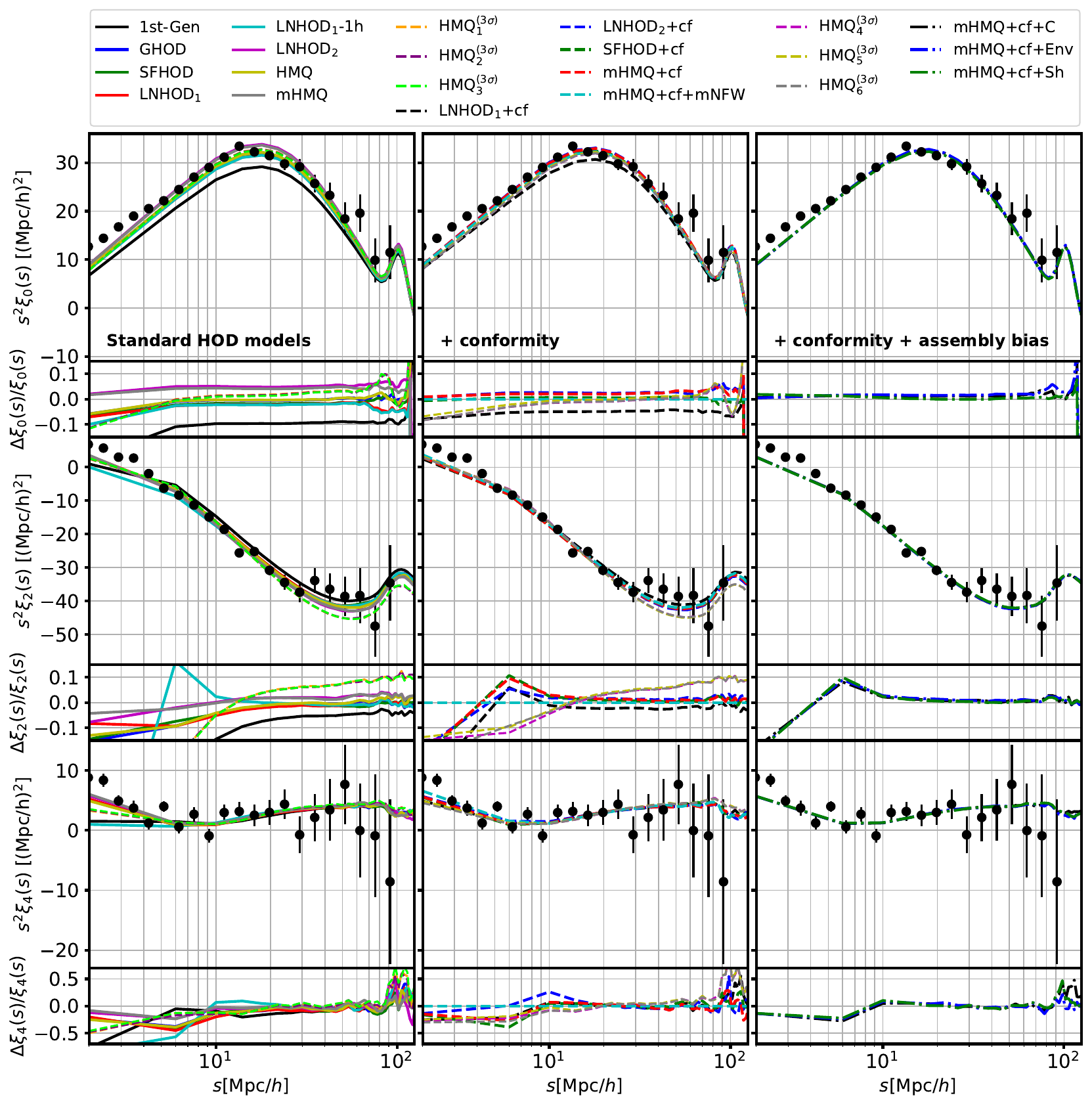}
    \caption{HOD measurements calculated from catalogs prior BAO reconstruction using mocks that were tuned against the One-Percent survey data (black data points). The top panels represent the measurements obtained for the monopole term, the middle panels show the results for the quadrupole term and the bottom panels correspond to the hexadecapole term measurements. We do not use the hexadecapole term neither in the HOD fits nor in the BAO fits, but still, we show it for completeness. While the HODs on the left panel correspond to the standard HOD models used in our work, the HODs shown in the central panels assume galactic conformity and the HOD models in the right-hand side panel assume not only galactic conformity but also assembly bias. Additionally, each correlation function multipole measurement has a smaller panel below where the relative error in \% is shown with respect to the mHMQ+cf+mNFW model, which is the fiducial HOD model for our analysis of systematics. The color legend for each HOD curve is shown at the top of the plot. Note that while the black data points correspond to actual data, the curves correspond to measurements from galaxy catalogs that were further rebinned going from $s=2$Mpc/$h$ up to $s=198$Mpc/$h$ with steps of $4$Mpc/$h$.}
    \label{Fig:HOD_measurements}
\end{figure*}

%% file: BodyText/Results.tex
In this section, we use the mock data based on the HOD models shown in Section \ref{Section:Galaxy_halo_connection} and apply the methodology described in Section \ref{Section:Methodology} to perform comparisons between different HOD prescriptions at the level of the BAO scaling parameters. We then use the BAO fits derived from our HOD mocks to quantify a systematic error budget and compare it to the statistical error from DESI 2024 results.

\subsection{Results from BAO analysis}
To perform a BAO fitting analysis, we calculate two-point measurements in both configuration space and Fourier space. As mentioned before, our measurements in configuration space are performed using \texttt{pycorr}, while our analysis in Fourier space uses \texttt{pypower}. First of all, we focus on analyzing the data without applying reconstruction. We use the catalogs for all of our 22 HOD models, which were generated by populating the \texttt{AbacusSummit} N-body dark matter simulation with galaxies using different HOD prescriptions. We then estimate two-point measurements and produce 25 measurements for each HOD model in both configuration and Fourier Space.

The average correlation function measurements before BAO reconstruction are shown in Fig.~\ref{Fig:HOD_measurements}, where we include the correlation function measurements for the final version of the EDA used to construct the vast majority of our HOD mocks. This helps us to see how similar the clustering of the various HOD models is, just after the HOD fitting stage and without any BAO reconstruction applied yet. We observe that most of the two-point correlation function measurements overlap slightly for $s>60 \text{Mpc}/h$ when considering HOD models without galactic conformity or assembly bias. However, if we look at smaller scales, we observe that the mHMQ model and the LNHOD$_2$ model start to prefer a higher clustering amplitude for the monopole term. This can be explained by the fact that these two models were obtained by fitting to smaller scales compared to the rest of the HODs, therefore they match better the EDA clustering at small scales. Nevertheless, these differences at small scales are not expected to have a big impact on the BAO fits since we fit only scales over $s=50\text{Mpc}/h$. On the other hand, there is a low clustering signal for the 1st-Gen mocks, which is simply explained by the fact that such a model was fitted to a preliminary version of the EDA. Yet, the clustering signal is not too different above $s=50\text{Mpc}/h$ and we should then be still able to recover the BAO scale correctly after reconstruction and include this HOD in our systematic error budget. When we add galactic conformity to some of the HODs (second column in Fig.~\ref{Fig:HOD_measurements}), we see that overall there is a good match in the observed clustering among all HODs, except by the LNHOD$_1$+cf. This model was indeed not fitted to such small scales compared to the others. Let us recall that LNHOD$_1$+cf and LNHOD$_2$+cf are the same model with the exception that LNHOD$_1$+cf is fitted down to 0.8 Mpc$/h$ while LNHOD$_2$+cf goes down to just 0.17 Mpc$/h$. Therefore, the low clustering signal of the LNHOD$_1$+cf model observed in Fig.~\ref{Fig:HOD_measurements} can be explained by the fitting range choices. On the other hand, we observe that adding galactic conformity to the HOD models does not produce a meaningful deviation in the two-point correlation function measurements. Finally, including assembly bias at the top of the galactic conformity, does not seem to have a significant impact on the observed clustering of the mHMQ+cf when adding concentration bias, environment bias, or shear bias.

Next, after looking at the clustering signal for the various HOD models we focus on enhancing the BAO measurement. We apply BAO reconstruction on all of our galaxy catalogs as described in Section~\ref{Section:Methodology}. We then again perform two-point measurements in both configuration space and Fourier space on the catalogs after BAO reconstruction. The measurements after applying BAO reconstruction and CV are shown in Appendix \ref{Appendix:Full_results} in Fig. \ref{Fig:post_recon_data}. These measurements were subtracted with the smoothing component after performing the BAO fits to highlight the BAO feature explicitly. As described in Section~\ref{Section:Methodology}, we use a BAO template based on 11 (13) parameters to fit the monopole and quadrupole in configuration (Fourier) space. To test the HOD dependence on the BAO fits and reduce the effect of systematics due to the BAO template, we apply Gaussian priors for $\Sigma_\parallel$, $\Sigma_\perp$, $\Sigma_\text{s}$. Since we want to test the impact on the BAO measurement due to HOD dependence when choosing a particular HOD for the analysis, we use the mHMQ+cf+mNFW as our fiducial HOD. We tune the priors based on this fiducial HOD to perform an optimal fit for this HOD model and analyze the systematic effect this can have when changing the HOD prescription but holding the same priors. We show the priors we assumed in our fits in Table~\ref{Table:BAO_parameter_description}, along with some useful statistical notation that will be used in the following. To get such prior choices, we tune the optimal non-linear damping parameters for this model by setting $\alpha_\text{iso}=\alpha_\text{AP}=1$ while fitting the nuisance parameters to the average of 25 realizations. We also tested other prior choices such as using 1st-Gen as our fiducial HOD or taking an average prior from all the HODs. However, our results are not significantly sensitive to this choice. In the following, we proceed to describe the results we obtain from the BAO analysis for both Fourier space and configuration space, independently. Even though we use the CV measurements as our final choice for quoting systematics, we also describe both CV and non-CV BAO fits for completeness and to show the gain from the CV approach.

\begin{figure*}
	\includegraphics[width=\textwidth]{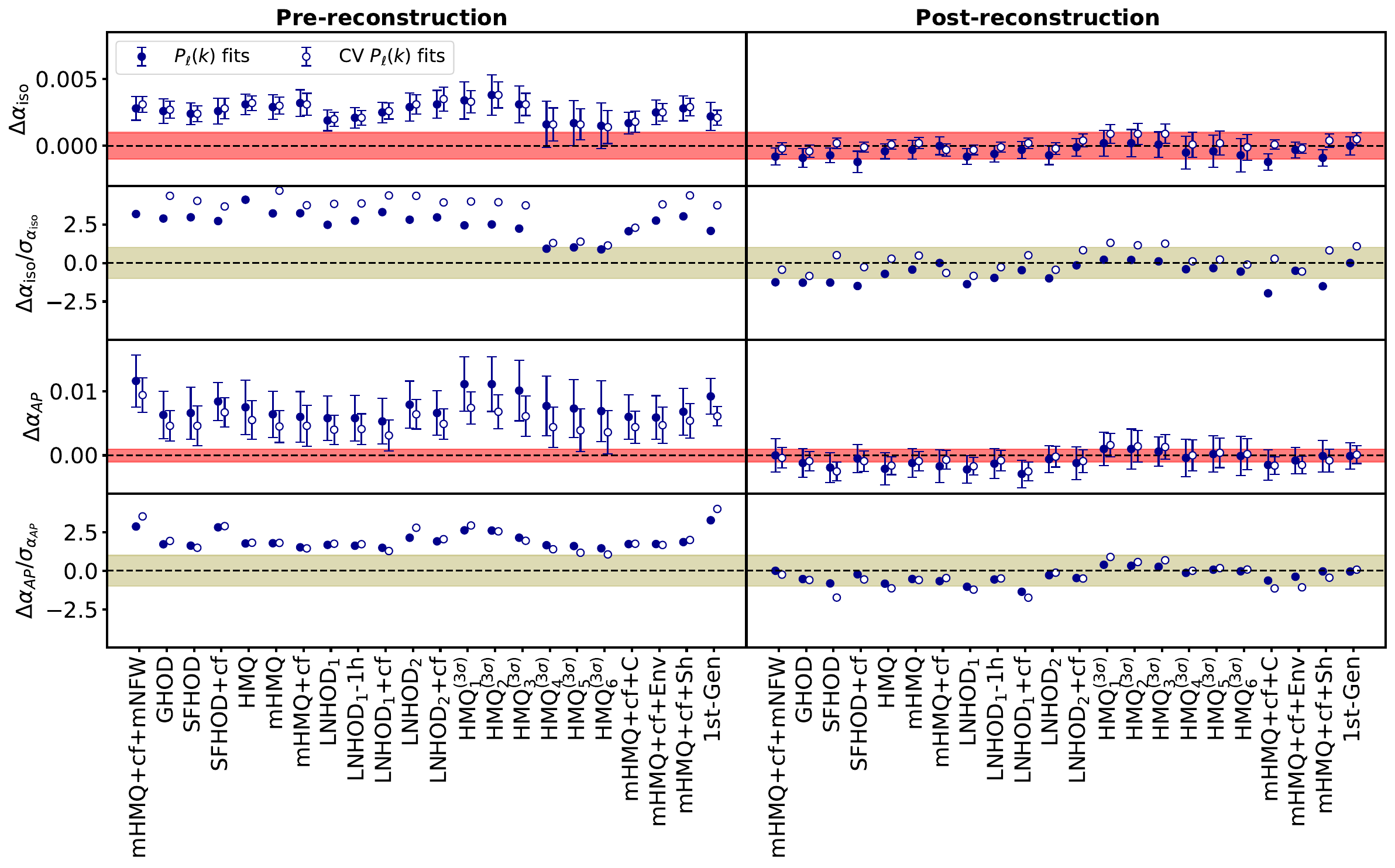}
    \caption{Results for the BAO scaling parameters when fitting the power spectrum monopole and quadrupole. The fits are obtained by averaging the fits of 25 realizations and rescaling accordingly. We show the difference in the scaling parameters with respect to their fiducial value and depict the 0.1\% regions in a red band. We also show the $\sigma$-error level along with a yellow region that corresponds to the interval where the error is below 1-$\sigma$. The results in the left panel are obtained before reconstruction while the right panel fits represent the post-reconstruction results. The blue circles correspond to the fits to $P_\ell(k)$ while the white circles are obtained when fitting the CV noise-reduced measurements. Each index in the $x$-axis corresponds to a particular HOD model.}
    \label{Fig:Performance_BAO_fits_PS}
\end{figure*}


\subsubsection{Fourier Space Analysis}
We shall focus first on the results from the Fourier space analysis. We perform BAO fits in the range $0.02\text{Mpc}^{-1}h<k<0.30\text{Mpc}^{-1}h$, using Gaussian priors for the non-linear damping parameters and $\Sigma_\text{s}$, as described Table~\ref{Table:BAO_parameter_description}. We run fits for 25 realizations for each HOD model and present the average BAO fit in Fig.~\ref{Fig:Performance_BAO_fits_PS}. Here, the error bars correspond to the dispersion on $\alpha_\text{iso}$ (alternatively, $\alpha_\text{AP}$) over 25 fits rather than the statistical uncertainty from the covariance matrix used in the fit. The blue data points represent the fits to the power spectrum multipoles and the white data points correspond to the BAO fits to the CV version of the measurements. We shall focus first on describing the non-CV BAO fits. We observe (left-hand side of the figure) that the pre-reconstruction fits for $\alpha_\text{iso}$ are slightly above (around 2\%) of the fiducial value while estimations of $\alpha_\text{AP}$ also show a bias for most HOD models. Such bias in the fits before BAO reconstruction is expected as our BAO damping model is not sufficient to capture the nonlinear physics included in the HOD mocks. We observe that the bias in such fits is overall around 2-$\sigma$. Looking at the right-hand side of the figure we observe that the post-reconstruction fits are successful in two ways, as follows: First, we find a decrease in the error bars of both $\alpha_\text{iso}$ and $\alpha_\text{AP}$, with respect to the pre-reconstruction BAO fits. This effect can be observed by comparing the left-hand side panel and right-hand side panel error bars in the figure. Second, we observe a shift in the measured values of the BAO scaling parameters which diminishes the bias of the fit. This can be seen by looking at the plots for $\Delta\alpha/\sigma_\alpha$ where most of the data points now lie within the 1-$\sigma$ confidence interval. These effects are expected as reconstruction reverses the positions of galaxies based on the displacement field, to reduce the impact of the non-linear structure growth on the BAO feature. Then, reverting the positions of the galaxies enhances the extra clustering signal coming from the BAO shells at the BAO scale, reducing the bias on the BAO measurement. We can also see from Fig.~\ref{Fig:Performance_BAO_fits_PS} that $\alpha_\text{iso}$ is the best measured BAO statistic for DESI. Furthermore, we found that the dispersion between BAO fits corresponding to different HOD models is way below the aggregated statistical error for DESI 2024 (0.96\% for $\alpha_\text{iso}$). Indeed, we can see that the bias in the BAO fits is overall within the sub-sub-percent level, as highlighted by the red band in Fig.~\ref{Fig:Performance_BAO_fits_PS}. 

If we now draw our attention to the BAO fits corresponding to CV noise-reduced measurements, we can see that overall they are not only consistent with the non-CV BAO fits before reconstruction, but they also show a better performance in terms of the recovery of the BAO feature for the post-reconstruction fits. The last point can be reflected in the panel dedicated $\Delta\alpha_\text{iso}/\sigma_{\alpha_\text{iso}}$ in Fig.~\ref{Fig:Performance_BAO_fits_PS} (second panel on the right-hand side), where overall all white dots stay inside the yellow region. We indeed observe that the maximum bias observed in the BAO scale recovery for $\alpha_\text{iso}$ goes from 2.0-$\sigma$ down to 1.3-$\sigma$ when using CV measurements. Additionally, while the average bias observed for $\alpha_\text{iso}$ is 0.7-$\sigma$ for non-CV BAO fits, the CV BAO fits shows a 0.2-$\sigma$ bias on average. On the other hand, we notice that CV BAO fits are bit more bias compared to non-CV BAO fits for few HODs, such as HMQ$_1^{(3\sigma)}$, HMQ$_3^{(3\sigma)}$ and HMQ$_3^{(3\sigma)}$ since BAO fits moves to a higher value of $\alpha_\text{iso}$. This effect however can be attributed to the trend that overall, all the CV BAO fits seem to lead to a higher value of $\alpha_\text{iso}$. Such an effect can be associated with the fact that all HOD models are correlated to some degree since they belong to the same dark matter simulation and are therefore populating the same halos. Since the quoted error bar is given by the dispersion on $\alpha_\text{iso}$ (alternatively, $\alpha_\text{AP}$) given 25 independent measurements of them, we observe that the CV results show a smaller error compared to the non-CV results. Such reduction in the variance is due to the fact that statistical noise is removed in the CV two-point measurements. 



\begin{figure*}
{\includegraphics[width=\textwidth]{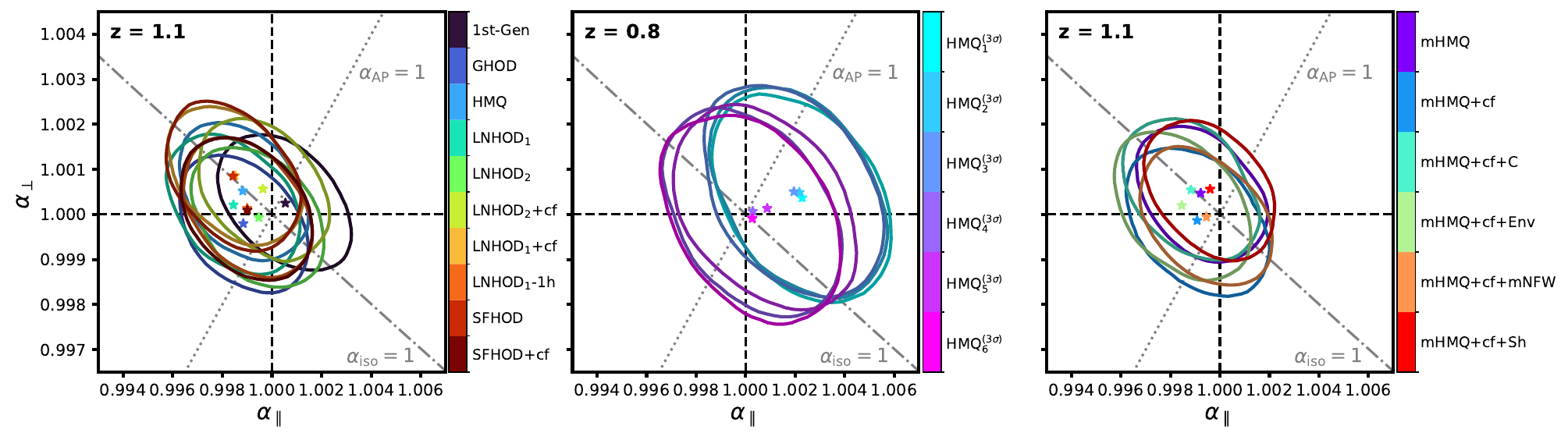}}
\caption{68\% confidence regions for $\alpha_\parallel$ and $\alpha_\perp$. The fit was performed in Fourier space on the averaged data vector over 25 realizations for each HOD mock rather than individual realizations. The star represents the best-fit value. The dashed black lines represent the fiducial value for both $\alpha_\parallel$ and $\alpha_\perp$. The fiducial value of $\alpha_\text{iso}$ is shown as a gray dashed dotted line. Similarly, the fiducial value for $\alpha_\text{AP}$ is shown in a dotted gray line. While HMQ$_i^{(3\sigma)}$ mocks are constructed at $z=0.8$, the rest of the HOD models are applied for simulations at $z=1.1$.}
\label{Fig:BAO_2D_contours}
\end{figure*}


We show the quantitative results for our BAO fits in Table~\ref{Table:BAO_fits_PS} in Appendix~\ref{Appendix:Full_results}, using both the standard two-point measurements and the CV noise-reduced two-point measurements. The table also includes the results of the derived BAO scaling parameters $\alpha_\parallel$ (scaling along the line-of-sight) and $\alpha_\perp$ (scaling transversal to the line-of-sight). These parameters are defined from the relations $\alpha_\text{iso} = \alpha_\parallel^{1/3} \alpha_\perp^{2/3}$ and $\alpha_\text{AP} = \alpha_\parallel/\alpha_\perp$. We notice that the mean error coming from the covariance matrix $\langle \sigma_{\alpha} \rangle$ is of the same order for all HODs except for the HMQ$_i^{(3\sigma)}$ mocks at $z=0.8$, which have the smallest number density. Apart from this, we observe that overall, the mean error on a given BAO scaling parameter $\langle \sigma_{\alpha} \rangle$ is similar to the error coming from the dispersion of 25 fits $\sigma(\overline{\alpha})$, for the standard BAO fits. However, the CV BAO fits show that $\sigma(\overline{\alpha})<\langle \sigma_{\alpha} \rangle$, being the average error improvement factor due to CV around 1.50 for $\alpha_\text{iso}$ and 1.45 for $\alpha_\text{AP}$ (or equivalently, $\epsilon$). This is in agreement with what had been found in~\cite{2023_Boryana-Hadzhiyska}. This can be also seen better from Fig.~\ref{Fig:CV_reduction_factor} in Appendix~\ref{Appendix:Configuration_vs_Fourier}, where we show the CV error improvement factor per HOD model, for various BAO scaling parameters and $\epsilon$. The aforementioned values correspond to the average factors shown in blue and red dashed lines, respectively, in Fig.~\ref{Fig:CV_reduction_factor}. In the case of $\alpha_\perp$ we observe that the error improvement factor is a bit less than 1.3. However, since $\alpha_\perp$ is correlated with $\alpha_\parallel$, this low factor is compensated by a 1.6 factor in the case of $\alpha_\parallel$. While we do not find a clear trend about the error improvement comparing $\alpha_\text{iso}$ and $\epsilon$, we observe that comparing $\alpha_\perp$ improvement with respect to $\alpha_\parallel$ shows a clear tendency, being CV more efficient for the latter. This might be due to the fact that CV is more accurate on linear scales where $\alpha_\parallel$ is impacted by linear redshift space distortion effects. Now, looking at the goodness of the fit, we find that the $\langle \chi^2 \rangle$ seems reasonably close to the degrees of freedom (DoF) of the BAO template (DoF$=93$), as shown in Table~\ref{Table:BAO_fits_PS}. Similarly, we also observe a decrease in $\langle \chi^2 \rangle$ for the CV BAO fits. This drop in the $\langle \chi^2\rangle$ is expected since we are using the same covariance matrices for both CV and non-CV BAO fits. However, the CV BAO fits show a decrease in the dispersion, which ends up impacting the $\chi^2$ computation.

As shown in Fig.~\ref{Fig:Performance_BAO_fits_PS}, our BAO analysis pipeline can recover the BAO feature with fluctuations at a sub-sub-percent level. Some fluctuations in the BAO parameters are present for some HOD models, but they are below 2-$\sigma$. The situation improves in general when we perform BAO fits after using the CV technique. Focusing on our CV BAO fits, we found that the biggest deviation for $\alpha_\text{iso}$ among all our HOD models, is about 0.1\% for the HMQ$_2^{(3\sigma)}$ model, with a bias of 1.3-$\sigma$ with respect to the fiducial value. A similar case is found for HMQ$_1^{(3\sigma)}$ and HMQ$_3^{(3\sigma)}$ as these HOD models correspond to 3-$\sigma$ variations of the best-fit model. We found a maximum shift of 0.24\% in $\alpha_\text{AP}$ when looking at the SFHOD model, with a bias with respect to $\alpha_\text{AP}=1$ of about 1.7-$\sigma$. Again, we stress that $\alpha_\text{iso}$ is our best measured and most robust parameter for the DESI 2024 BAO analysis.

We now analyze the BAO scaling parameters when fitting the average of 25 mocks for each HOD model. We leverage on $\alpha_\parallel$ and $\alpha_\perp$ to investigate the degeneracy direction followed by the BAO fits, as shown in Fig.~\ref{Fig:BAO_2D_contours}, based on the CV measurements. We observe that all the fits are consistent with the fiducial BAO scale within the 1-$\sigma$ confidence regions. We point out that in this case, the error on the parameters is directly coming from the covariance matrix rather than from the dispersion over 25 BAO fits since we are directly fitting the mean. We observe that the conventional HOD models (contours on the left panel in Fig.~\ref{Fig:BAO_2D_contours}) tend to scatter well within $\alpha_\text{iso}=1$ but some of them show slight shifts away from $\alpha_\text{AP}=1$. For example, LNHOD$_1$+cf shows a shift towards a low value of $\alpha_\text{AP}$. Similarly, if we focus on the central panel, we observe that the HMQ$_i^{(3\sigma)}$ best-fit values ($i=1,2,3$) exhibit slight shifts in both $\alpha_\text{iso}$ and $\alpha_\text{AP}$ towards high values. However, we get an optimal recovery of the BAO scale for HMQ$_i^{(3\sigma)}$ models ($i=1,2,3$), where complex galactic conformity is added at the top of the velocity bias. Thus, it is expected that the LNHOD$_1$+cf model and the HMQ$_i^{(3\sigma)}$ models without galactic conformity will end up driving the systematic error for $\alpha_\text{AP}$. Also, that the HMQ$_i^{(3\sigma)}$ models ($i=1,2,3$) will drive the systematics for $\alpha_\text{iso}$. Finally, the mHMQ model along with its extensions, such as strict galactic conformity and assembly bias, show not much scattering from the fiducial BAO scale.


\begin{figure*}
  {\includegraphics[width=\textwidth]{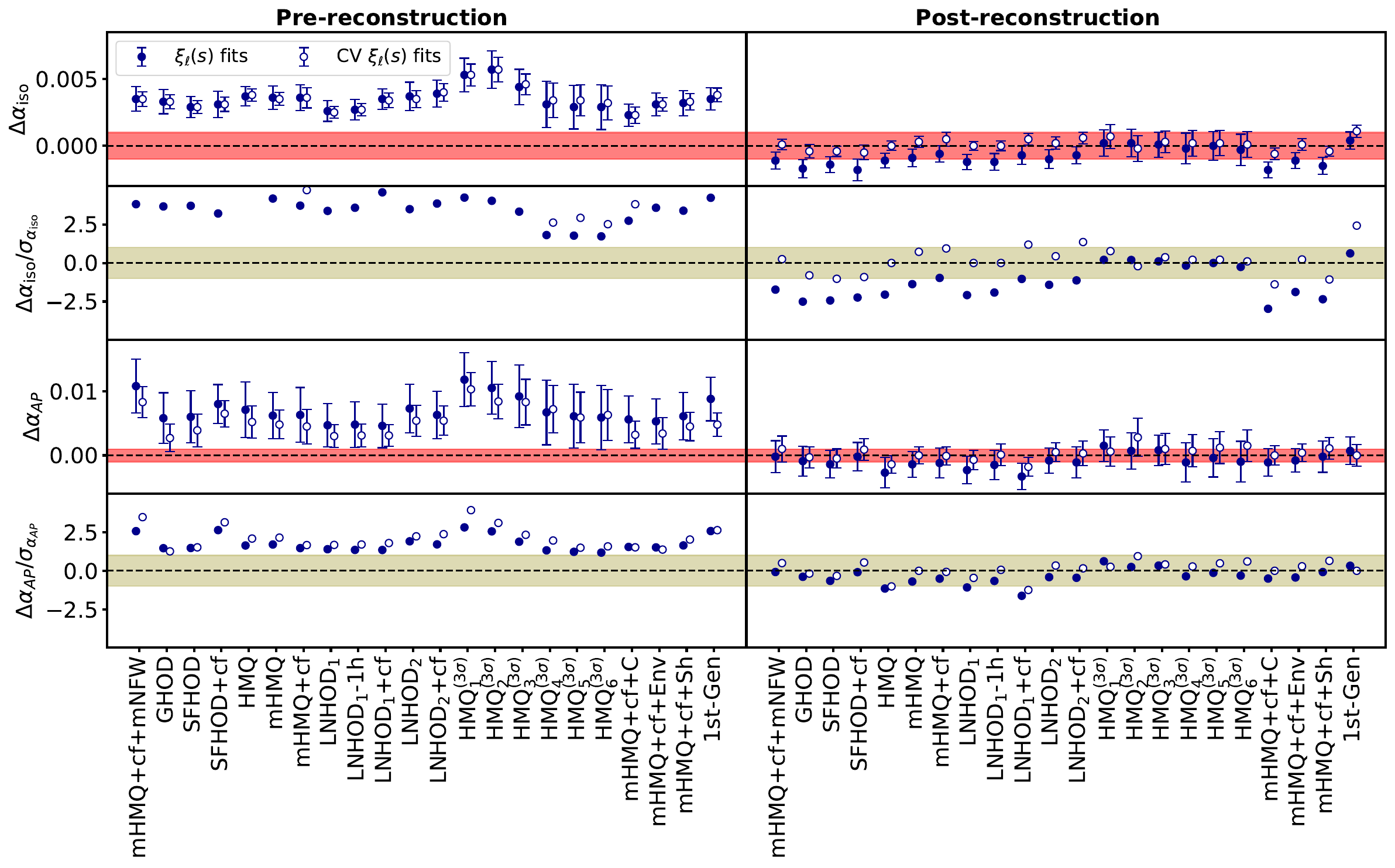}} 
    \caption{Results for the BAO scaling parameters when fitting the correlation function monopole and quadrupole. The fits are obtained by averaging the fits of 25  realizations and rescaling accordingly. We show the difference in the scaling parameters with respect to their fiducial value and depict the 0.1\% region in a red band. We also show the $\sigma$-error level along with a yellow region that corresponds to the interval where the error is below 1-$\sigma$. The results in the left panel are obtained before reconstruction while the right panel fits represent the post-reconstruction results. The blue circles correspond to the fits to $\xi_\ell(s)$ while the white circles are obtained when fitting the CV noise-reduced measurements. Each index in the $x$-axis corresponds to a particular HOD model.}    \label{Fig:Performance_BAO_fits_CS}
\end{figure*}


\subsubsection{Configuration Space Analysis}
For the configuration space BAO analysis we fit $\xi_0(s)$ and $\xi_2(s)$ in the range $50\text{Mpc}/h<s<150\text{Mpc}/h$. The modeling is similar to the Fourier space BAO fits, but here we only use 4 broadband terms compared to the 6 coefficients used in the Fourier space analysis. The fits to the correlation function multipoles in redshift space are shown in Fig.~\ref{Fig:Performance_BAO_fits_CS}. We observe that reconstruction behaves as expected for all the HOD models, being especially efficient for $\alpha_\text{AP}$, by bringing most HOD models below the 1-$\sigma$ threshold. The performance of the BAO fits for $\alpha_\text{AP}$ looks even more encouraging for CV measurements, where the expected value of $\alpha_\text{AP}$ is recovered within 1-$\sigma$ for all HODs. For the isotropic BAO parameter, we found that reconstruction pulls down the over-optimistic values of $\alpha_\text{iso}$. However, some HODs still show some bias up to 2.5-$\sigma$. Nevertheless, the BAO fits for the CV measurements restore the overall consistency of the fits. Except for the 1st-Gen mocks, the rest of the HOD models show consistent BAO fits with $\alpha_\text{iso}=1$ up to 1.3-$\sigma$. We also observe that the $\alpha_\text{iso}$ shift differences in the HOD models for CV BAO fits lie within the 0.1\% precision region (red horizontal band in Fig.~\ref{Fig:Performance_BAO_fits_CS}). These shifts are below the DESI 2024 aggregate precision threshold (1.1\% for $\alpha_\text{iso}$) by a factor of 10. However, this is the maximum difference found between a pair of HOD models, and could be a very pessimistic value to quote for the HOD systematics, as we discuss in Section~\ref{subsection:robustness_against_HODs}.

We observe that the error improvement factors coming from CV show, on average, lower values than those found in the Fourier space analysis (see Fig.~\ref{Fig:CV_reduction_factor}). This is due to the low-density HMQ$_i^{(3\sigma)}$ models with $(i=1,2,3)$, which do not manifest an efficient noise reduction. Overall, we observe the CV technique to work better for the high-density HOD mocks providing a larger improvement in the error, which is consistent with what was found in~\cite{2023_Boryana-Hadzhiyska}. This is potentially due to the fact that high-density mocks have larger satellite fractions and lower shot-noise, which play an important role in the efficacy of the CV method. On the other hand, qualitatively we found that the hierarchy on the error improvement factor is the same in both Fourier space and configuration space. The highest improvement is shown in $\alpha_\parallel$ (a 1.5 error improvement factor) while $\alpha_\perp$ shows just a 1.2 error improvement factor. 


\begin{figure*}
{\includegraphics[width=\textwidth]{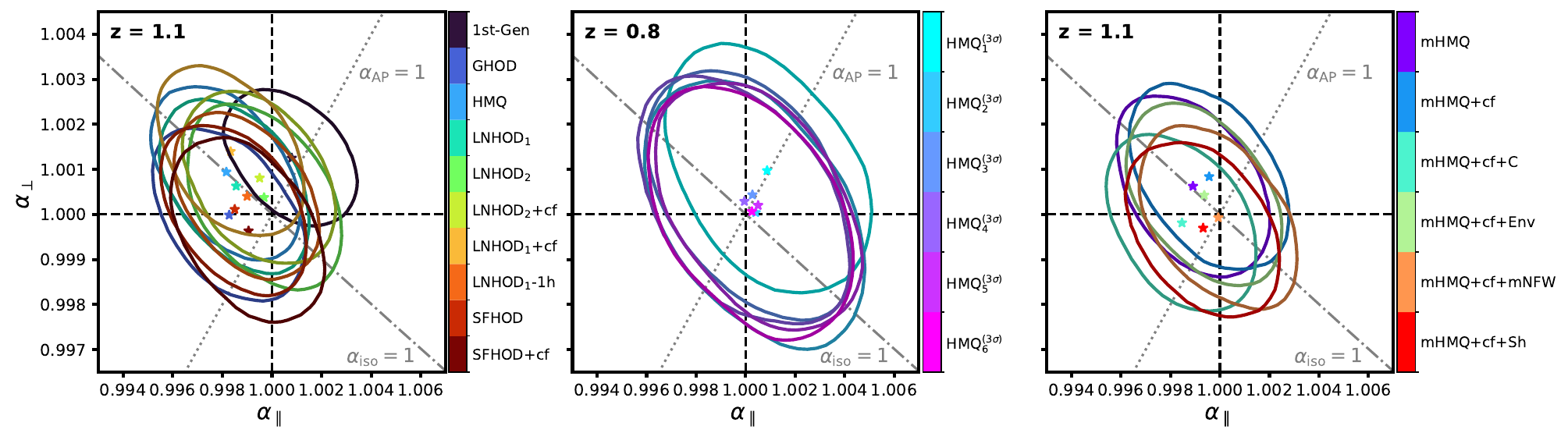}}
\caption{68\% confidence regions for $\alpha_\parallel$ and $\alpha_\perp$. The fit was performed in configuration space on the averaged data vector over 25 realizations for each HOD mock rather than individual realizations. The star represents the best-fit value. The dashed black lines represent the fiducial value for both $\alpha_\parallel$ and $\alpha_\perp$. The fiducial value of $\alpha_\text{iso}$ is shown in a gray dashed-dotted line. Similarly, the fiducial value for $\alpha_\text{AP}$ is shown in a dotted gray line. While HMQ$_i^{(3\sigma)}$ mocks are constructed at $z=0.8$, the rest of the HOD models are applied for simulations at $z=1.1$.}
\label{Fig:BAO_2D_contours_xi}
\end{figure*}

The 2D contours for the BAO fits to the average of 25 measurements are shown in Fig.~\ref{Fig:BAO_2D_contours_xi}. In this case, we observe that the contours are inflated in comparison to the Fourier space results. This comes from the fact that the errors reported in Fig.~\ref{Fig:BAO_2D_contours_xi} come from analytical covariance matrices for the correlation function multipoles. Yet, our error budget should not depend much on these differences in the covariance matrices between both spaces since our systematic error comes from the difference between the mean values of the fit. In general, while we observe similar trends compared to the Fourier space results, yet we observe more scattering in the best-fit values. Similarly as before, the LNHOD$_1$+cf model will drive the systematics for $\alpha_\text{AP}$ in configuration space. However, we observe that the LNHOD$_2$+cf model, which is the same model as LNHOD$_1$+cf but fitted to smaller scales, shows a better performance on the BAO fit. On the other hand, we observe that the CV BAO fit for the 1st-Gen mocks is biased within 1-$\sigma$, opposite to what is found in the Fourier space analysis. The shift affects exclusively $\alpha_\text{iso}$ and will drive the systematics for this parameter. 

In general, we found the results from the configuration space analysis to be slightly more scattered compared to the Fourier space analysis, but overall consistent. A discussion about the consistency between the two analyses is presented in Appendix~\ref{Appendix:Configuration_vs_Fourier}.


\subsection{Robustness against HOD modeling} \label{subsection:robustness_against_HODs}
Considering different prescriptions for sampling a given underlying cosmological field with galaxies can provide consistent results when compared to actual measurements. However, even in the absence of errors in the measurements, such sampling can lead to different results for the BAO scale. This fact turns into an unavoidable systematic error floor on any BAO measurement. In this work, we are interested in estimating such systematic error.

After achieving good performance in our BAO fits, we need to establish a methodology to test the robustness of these fits against the underlying HOD model. This will help us to quantify the systematic error budget due to HOD-dependence in modeling BAO. Our strategy to quantify the level of systematics is the following. Previously, we made some general conclusions about the results from the BAO fits, based on the average BAO fits over 25 realizations for each HOD. Indeed, we can calculate the maximum shift between pairs of HOD models derived from averaging the BAO fits over all realizations as a starting point. However, not all the shifts found between HOD models are statistically significant, and a more conservative strategy needs to be drawn. Then, we rather focus on comparing the BAO fits from individual realizations (rather than the averaged fits) across all HODs to test the robustness of the BAO analysis against the HOD prescription. We rely on two statistics to analyze the HOD systematics on the BAO fits. First, we define 
\begin{equation}
    \langle\Delta\alpha_{ij}\rangle=\langle \alpha_i-\alpha_j\rangle,
    \label{eq:Delta_alpha}
\end{equation}
where $\alpha$ can correspond to either $\alpha_\text{iso}$ or $\alpha_\text{AP}$. Eq.~\ref{eq:Delta_alpha} represents the average over 25 realizations on the BAO fits between different pairs of HODs. We also define
\begin{equation}
    N_\sigma(\alpha_{ij})=\frac{\langle\Delta\alpha_{ij}\rangle}{{\sigma}(\Delta\alpha_{ij})/\sqrt{N}},
    \label{eq:N_sigma}
\end{equation}
as the associated significance of the shifts calculated from Eq.~\ref{eq:Delta_alpha}. Here, $\sigma(\Delta\alpha_{ij})$ is the corresponding dispersion of the shifts in $\alpha_\text{iso}$ (or alternatively, $\alpha_\text{AP}$). We notice that the standard deviation associated with the mean difference in $\alpha_{ij}$ is divided by $\sqrt{N}$, where $N=25$ is the number of mocks. As $\langle \Delta\alpha_{ij} \rangle \rightarrow 0$ for a given pair of HOD models, $N_\sigma(\alpha_{ij})\rightarrow 0$ and there is no HOD systematics detected in our BAO fits. Similarly, if the dispersion in the values of the BAO scaling parameters is too large, $\sigma(\alpha_{ij})\rightarrow \infty$, and the measured average shift between HODs becomes significant. Conversely, if we had an infinite number of simulations and $\sigma(\alpha_{ij})/\sqrt(N)\rightarrow 0$, $N_\sigma(\alpha_{ij})\rightarrow \infty$ and we would claim an HOD systematics detection. To consider a systematics detection due to the HOD modeling, we consider a threshold of 3-$\sigma$ for such a claim when Eq.~\ref{eq:N_sigma} is calculated. We chose this threshold after testing sub-sampled versions of our mocks with higher shot-noise, which led to higher values of $N_\sigma$. Then, we opt for a conservative 3-$\sigma$ threshold to consider differences in number density between our mocks.


\begin{figure*}
\begin{tabular}{c c}
\Large{\textbf{Fourier space analysis}} & \Large{\textbf{Configuration space analysis}}\par\bigskip \\
{\includegraphics[width=0.5\textwidth]{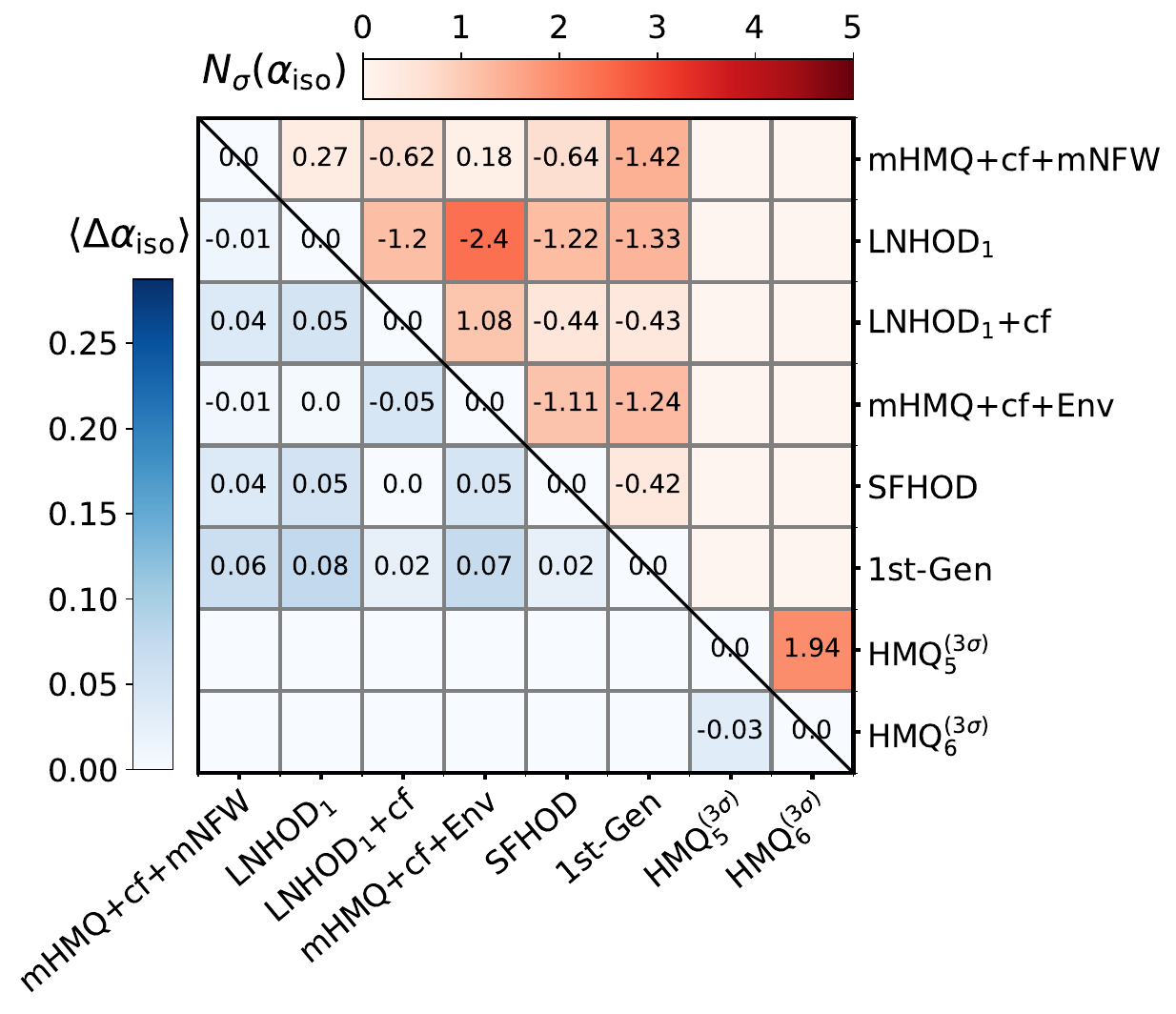}} & {\includegraphics[width=0.5\textwidth]{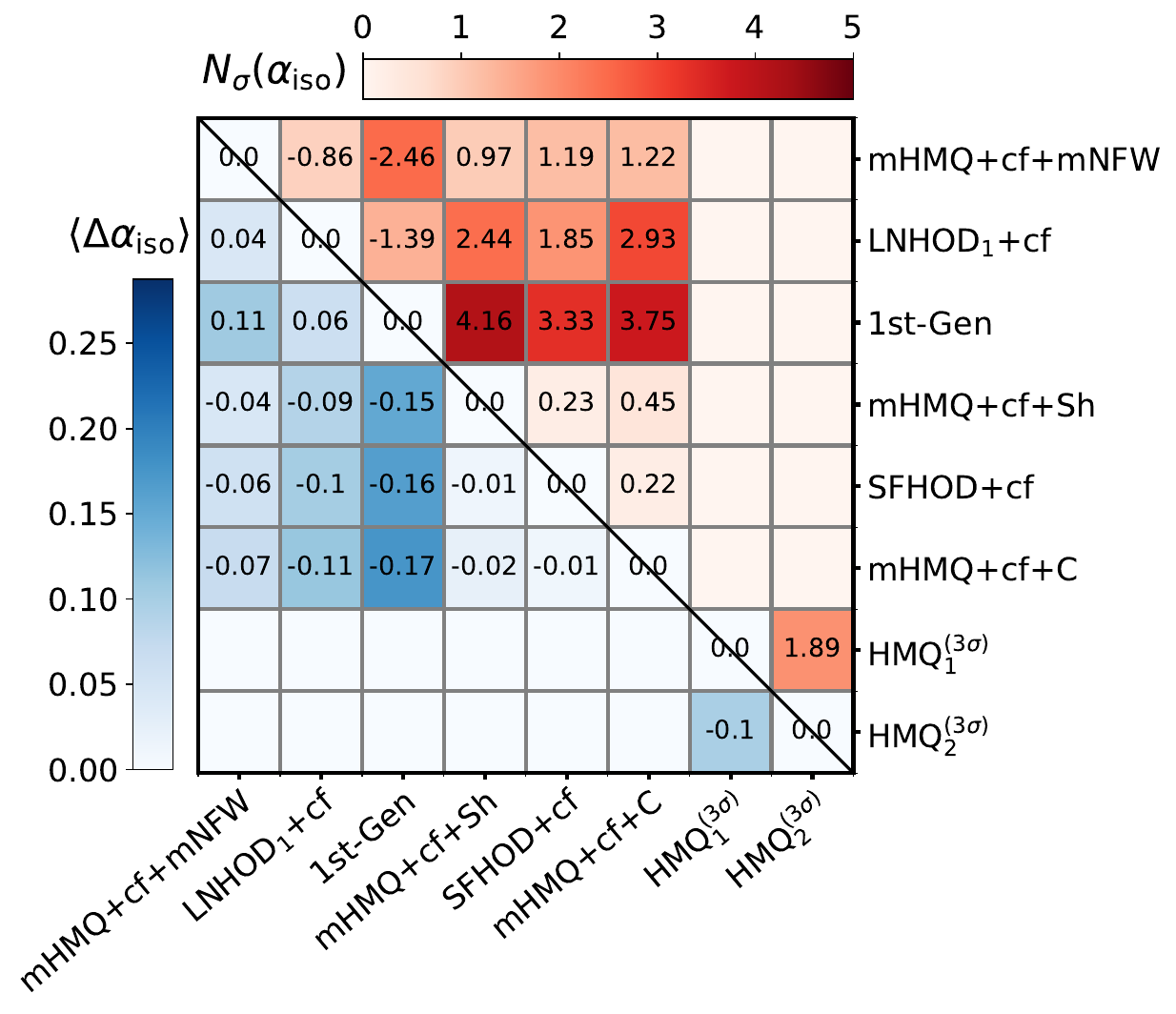}}  \\
{\includegraphics[width=0.5\textwidth]{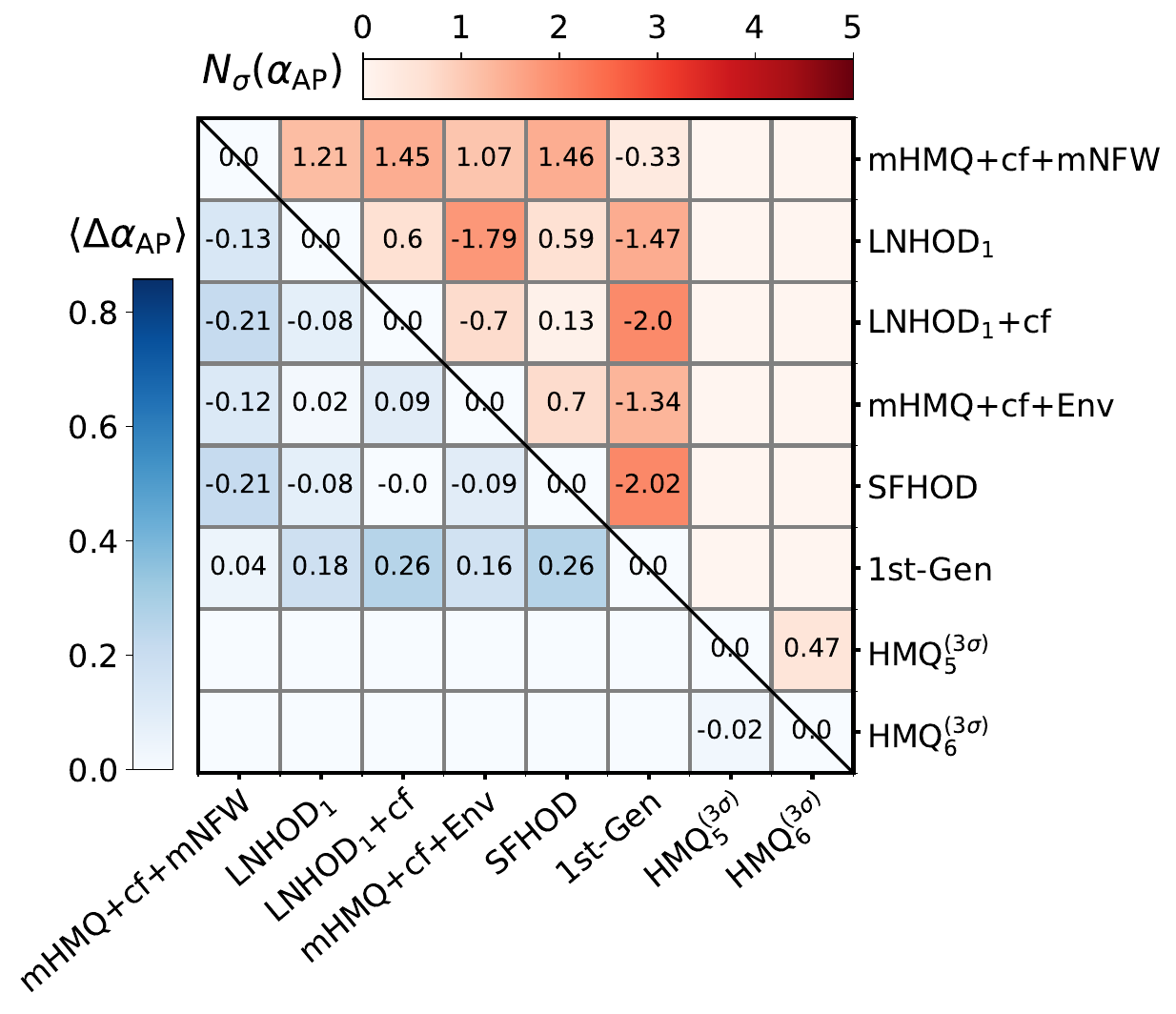}} & {\includegraphics[width=0.5\textwidth]{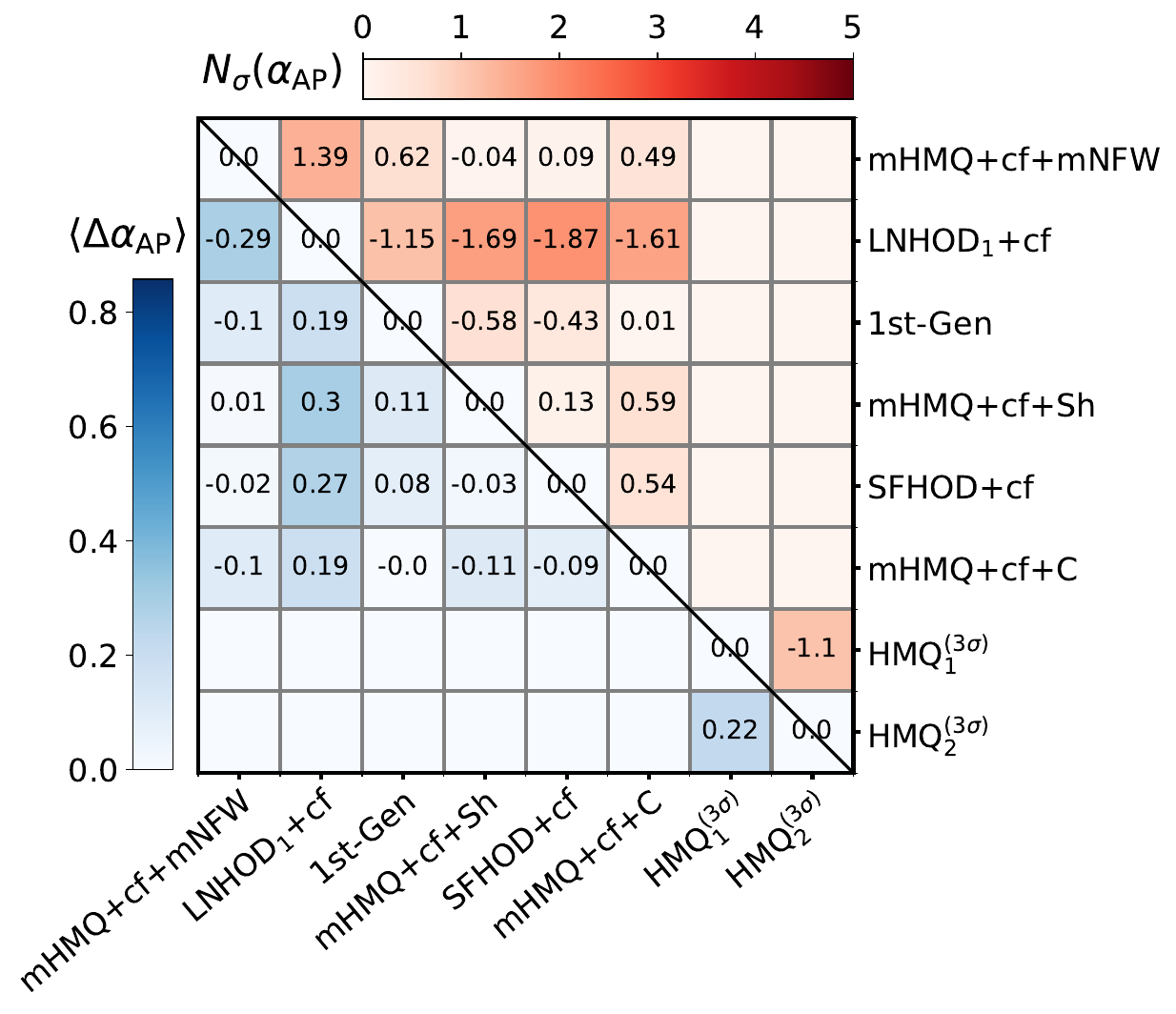}}
\end{tabular}
\caption{BAO systematics heatmap for selected HOD models in Fourier space (left-hand side) and configuration space (right-hand side). The heatmaps at the top correspond to systematics related to $\alpha_\text{iso}$ while the heatmaps at the bottom correspond to $\alpha_\text{AP}$. For each heatmap, the values corresponding to the  blue color scale represent the mean differences in $\alpha_\text{iso}$ (or $\alpha_\text{AP}$, respectively) when comparing different pairs of HOD models. The values of these shifts are multiplied by 100 to show $\langle \Delta\alpha_\text{iso} \rangle$ in \%. Similarly, the upper triangle matrix in red scale shows the corresponding value of $N_\sigma$ as defined in Eq.~\ref{eq:N_sigma}, where the color scale in red is set by the 5-$\sigma$ detection level. Notice that empty values in each heatmap mean that we do not compare HOD mocks not centered at the same redshift to avoid introducing extra systematics not exclusively due to the HOD model itself. The complete versions of these heatmaps are shown in Appendix \ref{Appendix:Full_results}.}
\label{Fig:Heatmaps_CV_simplified}
\end{figure*}

\begin{figure*}
\begin{center}
\begin{tabular}{c c}
{\includegraphics[width=0.48\textwidth]{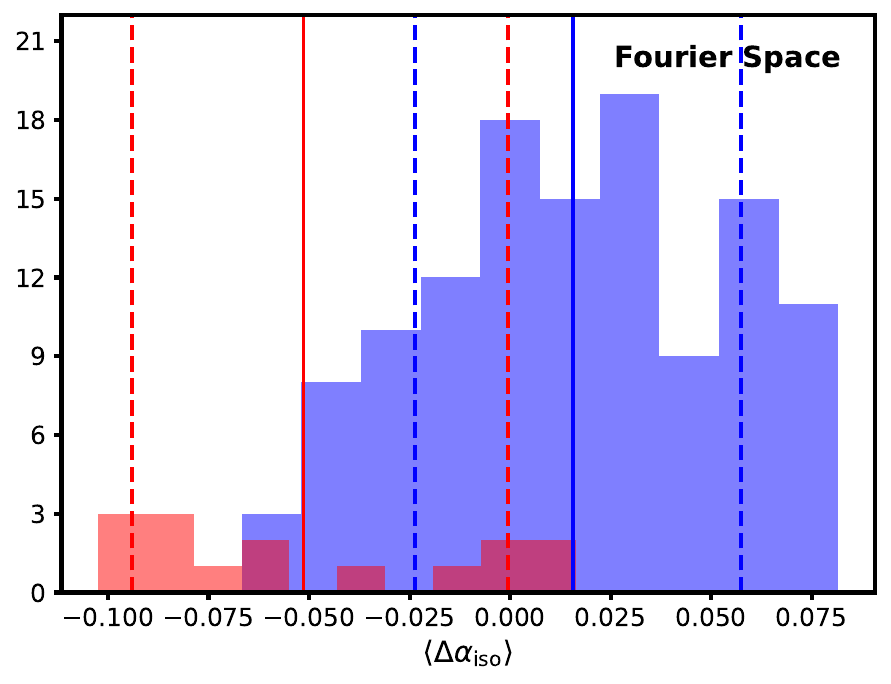}} & {\includegraphics[width=0.47\textwidth]{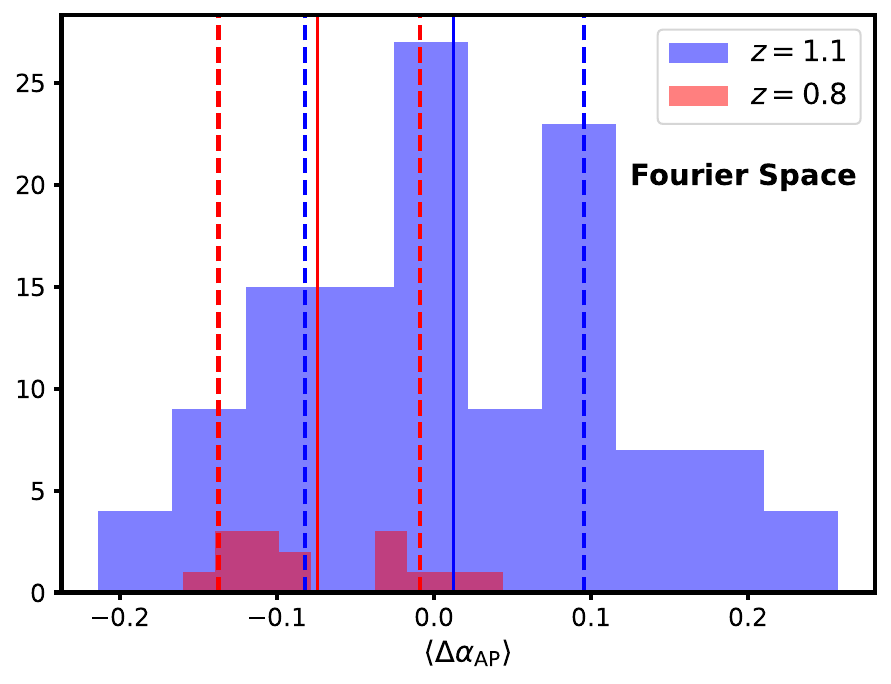}}  \\
{\includegraphics[width=0.48\textwidth]{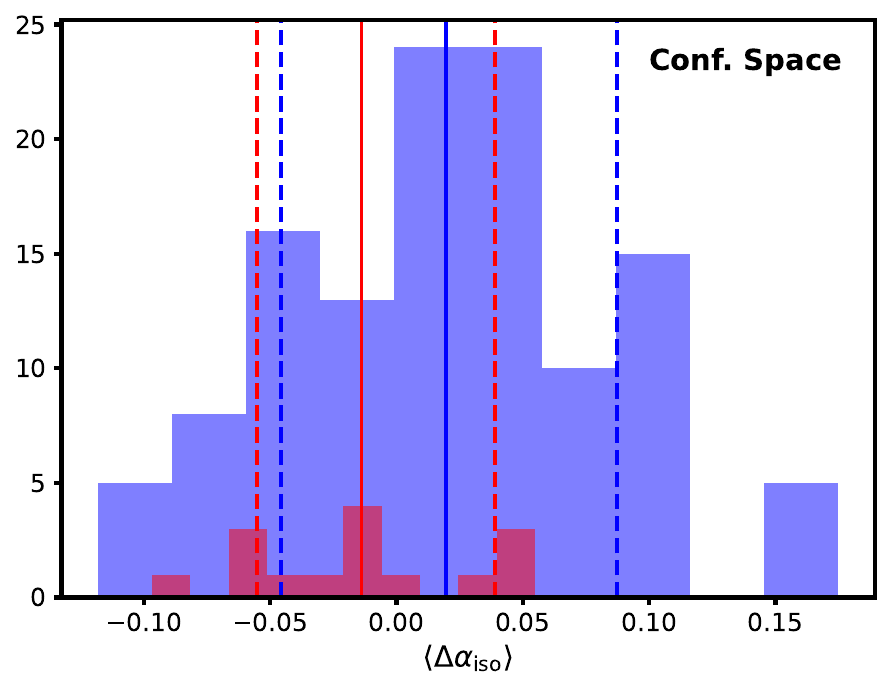}} & {\includegraphics[width=0.47\textwidth]{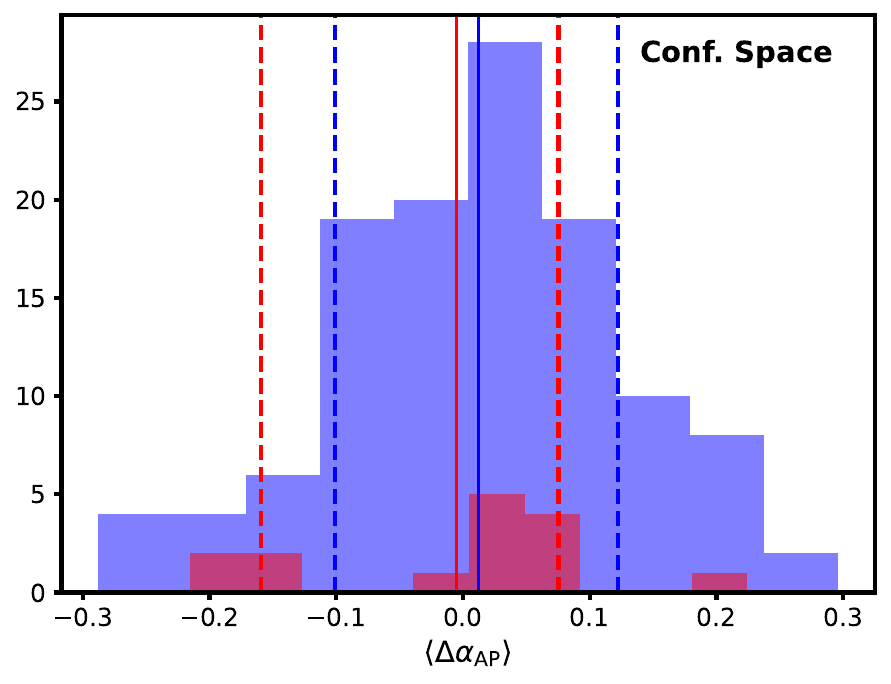}}
\end{tabular}
\end{center}
\caption{Histogram built from the shifts found in the heatmaps from Fig.~\ref{Fig:Heatmaps_CV_simplified}. The values depicted in blue represent the HOD models at $z=1.1$ while the red histogram corresponds to results from the HOD models at $z=0.8$. The solid vertical line represents the mean value from the histogram while the 68\% region is enclosed between vertical dashed lines.}
\label{Fig:Histograms}
\end{figure*}

\begin{figure}
\begin{center}
  {\includegraphics[width=0.6\textwidth]{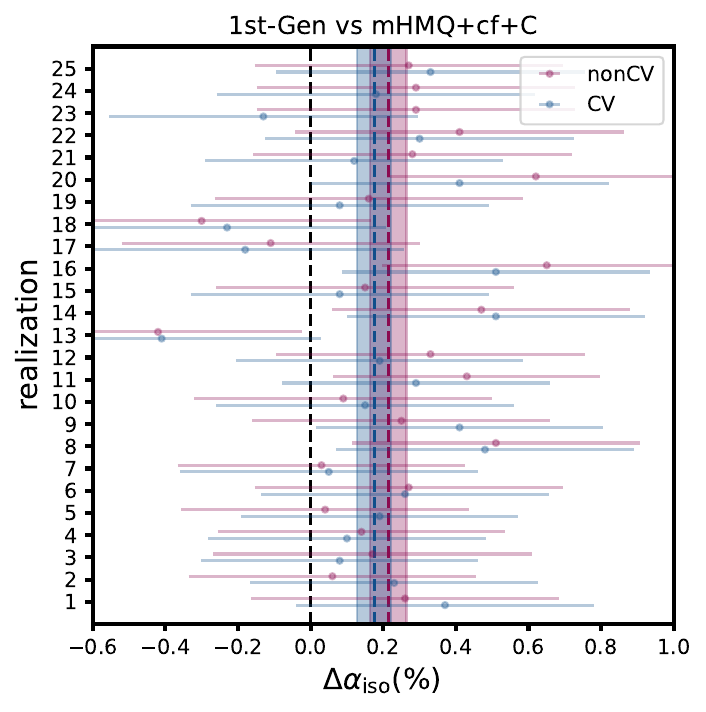}}
    \caption{Differences in $\alpha_\text{iso}$ between 1st-Gen and mHMQ+cf+C models. The purple data represents the results from fitting the two-point measurements while the blue data points correspond to the results for the CV measurements. Similarly, the colored bands represent the 68\% region of the mean value of $\Delta\alpha_\text{iso}$. The final result quoted in this work corresponds to the blue band which indicates a 3.75 HOD systematics detection on Fig.~\ref{Fig:Heatmaps_CV_simplified}.}
    \label{Fig:HOD_detection}
\end{center}
\end{figure}

We compare all the possible permutations of pairs of HOD models and use the 25 mock realizations in each case to calculate Eq.~\ref{eq:Delta_alpha} and Eq.~\ref{eq:N_sigma}. To properly visualize our results, we express the comparisons in terms of heatmaps. We shall focus first on our results in Fourier space. We show in Fig.~\ref{Fig:Heatmaps_CV_simplified} the heatmaps for some selected HOD models which are representative of the biggest shifts found across all permutations of HOD models. While the values in the blue color scale represent the shift in the BAO scaling parameters, the values associated with the red color scale show the significance of such shifts. If we compare the heatmap at the top (for $\alpha_\text{iso}$) with the heatmap at the bottom (for $\alpha_\text{AP}$), we observe that $\alpha_\text{iso}$ turned out to be not only a more precise variable to measure the BAO scale but also the most robust in terms of HOD dependence. As mentioned in Section~\ref{Section:Introduction}, while the statistical error related to $\alpha_\text{iso}$ is around 1\%, the maximum shift in $\alpha_\text{iso}$ as seen from the heatmap is 0.08\% (0.17\%) in Fourier (configuration) Space. This is, the maximum difference in $\alpha_\text{iso}$ due to the HOD prescription found is around 1/12 (1/6) of the statistical error for DESI 2024 statistical error. However, this number is the maximum fluctuation found across all HODs and it should be regarded as a too aggressive number to quote for the systematic error. In the case of $\alpha_\text{AP}$, the maximum shift found from the heatmaps is just 1/13 of the aggregated error for ELGs in Fourier space, and 1/16 in configuration space. At the top of this, we have to remember that these numbers will end up being added in quadrature, which makes the effect smaller.

Overall, the heatmaps show the differences between the measured BAO scaling parameters given a pair of HOD models that are under comparison. These HOD models were constructed from a common dark matter simulation and were populated using the same random seed. Therefore, $\Delta \alpha_\text{iso}$ (alternatively, $\Delta \alpha_\text{AP}$) tells us the impact on $\alpha_\text{iso}$ (alternatively, $\alpha_\text{AP}$) due to differences in the HOD model, given our fiducial choice of mHMQ+cf+mNFW as our default HOD. We can construct a histogram based on all these differences to analyze the dispersion between HOD models for a given BAO scaling parameter. A meaningful quantity to quote from such a histogram is the 68\% region around the mean, which tells us the dispersion given the changes in the underlying HOD model. We show these histograms separately for both sets of HOD models at $z=1.1$ and $z=0.8$ in Fig.~\ref{Fig:Histograms}. These were obtained by using the full heatmaps shown in Appendix~\ref{Appendix:Full_results}. The complete heatmaps as well as the full set of results are shown in Appendix~\ref{Appendix:Full_results}. In general, we observe that the 68\% region covered by both the $z=1.1$ and the $z=0.8$ histograms have nearly the same width. However, we choose to quote a larger value between the two as our HOD systematic error if there is no HOD detection. 

In case the heatmaps directly point towards systematics due to HOD modeling being detected above the 3-$\sigma$ threshold, we quote the corresponding shift found. The top right heatmap in Fig.~\ref{Fig:Heatmaps_CV_simplified} shows various $N_\sigma(\alpha_\text{iso})$ values above the 3-$\sigma$ threshold, pointing out to detected systematics due to HOD modeling. In this scenario we quote the highest average shift $\langle \Delta \alpha_\text{iso} \rangle$ found among all detected cases as our $\sigma_\text{HOD}$ systematic error. Yet, the shift quoted (0.17\%) in the detected systematics case is still below the statistical error for DESI 2024. Fig.~\ref{Fig:HOD_detection} shows the results of $\Delta\alpha_\text{iso}$ for each mock realization along with the mean value obtained when comparing 1st-Gen against mHQM+cf+C, which is the pair that provides the highest shift where a detected systematic due to HOD have occurred. We observe that the results without applying CV show a 4.24-$\sigma$ deviation from $\Delta\alpha_\text{iso}=0$ while the CV results only diminish this down to 3.75-$\sigma$. In comparison, the Fourier space analysis for this combination (not shown in the figure) went down from 2.17-$\sigma$ to 0.62-$\sigma$. This highlights the fact that even though CV is able to reduce the sample variance noise in $\alpha_\text{iso}$, the systematics due to HOD modeling in configuration space remained present between these two models for our BAO template modeling choices. Hence, we conclude that our analysis in Fourier space is slightly more robust against HOD systematics than the analysis in configuration space. A similar conclusion holds for $\alpha_\text{AP}$. On the other hand, the systematic error found in $\alpha_\text{iso}$ is $\sim$3.6 times larger in configuration space compared to Fourier space, where no HOD systematics was detected given our threshold for such a claim. However, we still point out that a systematic error of 0.17\% for $\alpha_\text{iso}$ is 1/6 of the statistical error and, since the error is added quadratically, this would just cause a difference of $\sim$1.2\% in the total error. In comparison, the systematic error quoted in Fourier space is 0.047\%, which would lead to an increase of $\sim$0.1\% in the total error.

In summary, the values we decide to quote as our systematic error due to HOD dependence are given by the following expression,
\begin{equation}
    \sigma_\text{HOD} = \begin{cases}
    \text{max}(\langle\Delta\alpha_{ij}\rangle) &\text{, if $N_\sigma(\alpha_{ij})\geq 3$, (HOD detection)} \\
    \sigma_{68\%}(\langle\Delta\alpha_{ij}\rangle) &\text{, if $N_\sigma(\alpha_{ij})<3$. (No HOD detection)}
\end{cases}
    \label{eq:HOD_systematics}
\end{equation}
Here $\sigma_{68\%}(\langle\Delta\alpha_{ij}\rangle)$ represents the interval which covers 68\% of the histogram region around the mean. Since the histograms are not exactly Gaussian we opt to use this notation. However, the value found for $\sigma_{68\%}$ is close to the standard deviation obtained from assuming a Gaussian histogram. We summarize our HOD systematic errors quoted for the DESI 2024 BAO analysis in Table~\ref{Table:Systematics_Summary}. We observe that the errors found in Fourier space are consistent with those of configuration space. Additionally, the error due to HOD systematics found for $\alpha_\text{iso}$ is around 1/20 of the statistical error for DESI 2024 in Fourier space, while $\alpha_\text{AP}$ is about 1/37 of it. In the case of the configuration space analysis, these factors are 1/6 and 1/40, respectively. Since $\sigma_\text{tot}^2 = \sigma_\text{stat}^2 + \sigma_\text{syst}^2$, we can calculate the increase in $\sigma_\text{tot}$ given our added HOD systematics. We found that adding our values found for $\sigma_\text{HOD}$ leads to a change of 0.12\% on the statistical error for $\alpha_\text{iso}$ and 0.04\% on the statistical error for $\alpha_\text{AP}$ in the Fourier space. In configuration space, we found an increase of 1.19\% for $\alpha_\text{iso}$ and 0.03\% for $\alpha_\text{AP}$ in the statistical error. However, since this is the percentage increase on the statistical error itself, we conclude that this is still a sub-dominant effect for DESI 2024.

Finally, it is worth mentioning that we also tested other effects during our tests related to HOD systematics. For example, we tested the effect of having different number densities across our HOD models. We diminished the number density of our HOD mocks at high redshift by sub-sampling our mocks to have a common number density of $10^{-3}$($h$/Mpc)$^3$. We found that sub-sampling our HOD mocks to have the same number density across all HOD models led to an increase in the shifts between pairs of HODs, but these were coming mostly from a higher shot-noise, which causes CV to also be less efficient. Nevertheless, even though these shifts were small enough for the DESI 2024 statistical precision, we opted to use the high number density version of our mocks as our final choice, since the larger shifts we observed across our HODs were mostly coming from statistical noise due to a higher shot-noise. Additionally, we also tested the effect of stochasticity into our HOD modeling, specifically for the LNHOD$_1$ model. We produced several mocks with alternative random seeds and concluded that such effects were small enough to propagate in a significant way compared to the statistical error for DESI 2024.

\begin{table}
    \centering
    {\renewcommand{\arraystretch}{1.3}
    \begin{tabular}{c | c | c | c | c}
    \hline\hline
    Space & Parameter & $\sigma_\text{HOD}$ & $\Delta\sigma/\sigma_\text{stat}$ & HOD Systematics detected \\ \hline
    \multirow{2}{6em}{Fourier} & $\alpha_\text{iso}$ & 0.047\% & 0.12\% & No \\ \cline{2-5} 
     & $\alpha_\text{AP}$ & 0.089\% & 0.04\% & No \\ \hline
    \multirow{2}{6em}{Configuration} & $\alpha_\text{iso}$ & 0.17\% & 1.19\% & Yes \\ \cline{2-5} 
     & $\alpha_\text{AP}$ & 0.11\% & 0.03\% & No \\ \hline \hline
    \end{tabular}}
    \caption{Summary table with the systematic error budget due to HOD quoted from $\sigma_\text{HOD}$. In the non-detection case, we follow the procedure described in Section \ref{subsection:robustness_against_HODs} where $\sigma_\text{HOD}$ is obtained from the 68\% region around the mean value of a histogram constructed from the shifts between all the permutations of pairs of HODs. If systematics due to HOD modeling is detected, we directly quote $\sigma_\text{HOD}$ as the highest shift found for the pair of HOD models for which an HOD systematics has been detected. In the next column, we use $\Delta\sigma=\sigma_\text{tot}-\sigma_\text{stat}$, once $\sigma_\text{HOD}$ has been added quadratically as a systematic error according to $\sigma_\text{tot}^2=\sigma_\text{stat}^2+\sigma_\text{HOD}^2$. The last column shows whether a systematics due to HOD was detected or not.}
    \label{Table:Systematics_Summary}
\end{table}

%% file: BodyText/Conclusions.tex
In this analysis, our objective was to test the robustness of our BAO modeling against the dependence on the assumed HOD model. Furthermore, our aim was to quantify the error budget due to this particular systematics in order to support the DESI 2024 BAO analysis presented in \cite{DESI2024.III.KP4}. A companion paper presented in \cite{KP4s10-Mena-Fernandez} shows analogous results for the LRG tracer. While the methodologies of these two analyses are consistent, each analysis employs different HOD models for the respective tracers under consideration. To test the sensitivity to the HOD prescription, we based our analysis on the most representative HOD models present in the literature for the ELG tracer. Even though this analysis occurred earlier compared to other DESI 2024 BAO systematics analyses, our pipeline is close enough and analogous to the final theoretical modeling choices for DESI 2024 as described in \ChenHowlett.

In this analysis, we included various HOD models as well as extensions to go beyond the standard modeling, such as galactic conformity and assembly bias. We produced mocks for each of these models based on the \texttt{AbacusSummit} simulations as our underlying dark matter simulation. Such simulations are well resolved for the scales we intended to test (below k=0.3$h$/Mpc). Our HOD mocks were tuned to the early One-Percent DESI data, prior to the DESI 2024 BAO analysis with DESI-DR1. We calculated two-point measurements before and after BAO reconstruction using a common galaxy bias value for all the mocks. This makes our final systematic error to also include effects from the fiducial bias assumptions during BAO reconstruction and this will be included as well in the DESI 2024 BAO systematic error budget for the ELG tracer. To support the fidelity of our analysis when producing BAO fits, we estimated analytical covariance matrices in both Fourier space and configuration space based on the clustering of the 25 mocks per HOD model. To mitigate the sample variance noise coming from the fact that we have a limited number of simulations, we used the CV technique to produce noise-reduced versions of the two-point measurements. We found that the CV technique can reduce the bias with respect to the fiducial value in our BAO parameters. This enhancement in the performance of the fits for the BAO parameters led us to consider the CV two-point measurements after BAO reconstruction as our choice for the final systematic error. We also tested the effect of number density disparity in our mocks and found that sub-sampling over the HOD mocks leads to higher shifts but these are just due to statistical noise.

While the statistical error for the ELGs is approaching the sub-percent level, as seen in the case of $\alpha_\text{AP}$, or has already reached it, as observed in the case of $\alpha_\text{iso}$, within DESI 2024, we found that the variations on $\alpha_\text{iso}$ due to HOD modeling are close or below the sub-sub-percent level. In the case of $\alpha_\text{AP}$, we found that variations due to HOD prescription are small compared to the current statistical error. These conclusions were reached after successfully recovering the isotropic BAO parameter $\alpha_\text{iso}$ within 0.1\% accuracy and the Alcock Paczynski parameter $\alpha_\text{AP}$ within 0.3\% accuracy in our BAO fits. Furthermore, we established a methodology to define the systematic error budget by comparing the BAO fits for every pair of HODs. We built heatmaps for the pairwise differences between HODs that summarize the differences in the fits as well as the statistical significance of such shifts. Then, when no significant shift between a pair of HOD models is detected below a 3-$\sigma$ threshold, we defined our systematic error from the spread among all the shifts between pairs of HOD models. More precisely, for a given BAO parameter (or for the AP parameter), our systematic error is defined by the 68\% percentile around the mean of the histogram constructed from all the shifts in $\alpha_\text{iso}$ (or $\alpha_\text{AP}$) between pairs of HODs. In the case of a histogram very close to a Gaussian distribution, our quoted systematic error would coincide with the standard deviation of such a histogram. If there is a shift or more than one shift between a pair of HOD models with a significance above 3-$\sigma$ (i.e. systematics due to HOD modeling detected) we then directly quote the maximum shift found. We only found an HOD detection in the case of $\alpha_\text{iso}$ for the configuration space analysis, while the rest of the systematic errors were derived from the heatmap approach. 

As a result from our analysis, while the statistical errors derived from aggregated precision for the ELG tracer in the DESI 2024 BAO analysis are 0.96\% (1.1\%) for $\alpha_\text{iso}$ and 3.3\% (4.4\%) for $\alpha_\text{AP}$ in Fourier space (configuration space), the systematic errors attributed to HOD dependence for the ELG tracer can be expressed as follows: In Fourier space, we obtained 0.047\% as our systematic error for $\alpha_\text{iso}$ and 0.089\% as our systematic error for $\alpha_\text{AP}$. These values would lead to an increase of the order 0.12\% and 0.04\%, respectively, in the statistical error. For the configuration space analysis, we found the systematic error for $\alpha_\text{iso}$ to be 0.17\% and 0.11\% for $\alpha_\text{AP}$. Similarly, these errors would produce an increase in the total error of about 1.19\% and 0.03\% with respect to the statistical error. Therefore, we found that the changes due to HOD modeling are small enough for DESI 2024 BAO analysis. On the other hand, it is worth remembering that in addition to the error budget quoted in this work, other systematic errors will be added.

As the next generation of Stage-IV surveys is preparing to upgrade our scientific knowledge of the cosmos, there is no doubt that DESI, as the first Stage-IV experiment to analyze data, will deliver very important results to the scientific community in the BAO domain and beyond. Therefore, it has become crucial to assess systematic effects to assure the precision and accuracy of the results. Based on all the evidence presented in this work, we conclude that our analysis pipeline is robust enough against HOD-dependent systematics for the DESI 2024 BAO results. Moreover, the systematic error budget derived from this analysis seems reasonable and it is close to or below the sub-sub-percent level, depending on whether we refer to the configuration space or Fourier space, respectively. Therefore, the error that is added to the DESI 2024 BAO analysis is small enough for DESI-DR1. Yet, this analysis shall be revisited in the future with more mocks and new HOD models for the DESI result after five years of observations.

%% file: Appendix/Consistency_FS_CF.tex
\section{Consistency between Fourier space and configuration space across HOD models}\label{Appendix:Configuration_vs_Fourier}

The BAO analysis for quantifying the systematic error due to HOD dependence has been conducted for both configuration and Fourier spaces. Hence, it is reasonable not only to perform each analysis separately but also to provide some comparison of both results to assess whether they are consistent with each other or not. It is important as well, to evaluate if the consistency/inconsistency of the results has something to do with a particular HOD model.

We focus first on the efficiency of CV on our BAO fits results. Fig.~\ref{Fig:CV_reduction_factor} shows the CV improvement factor of the dispersion on the BAO parameters obtained from the BAO fits with respect to the standard non-CV fits. We observe that the improvement in the errors is higher for mocks with higher number density compared to mocks with lower number density such as HMQ$_i^{(3\sigma)}$ ($i=1,2,...,6$). This is in agreement with what was argued in~\cite{2023_Boryana-Hadzhiyska}. We also found that on average, the dispersion in the CV results is a factor of $\sim$1.5 times lower for $\alpha_\text{iso}$ and $\epsilon$, compared to the non-CV results. Here we use $\epsilon$ for direct comparison with the original work presented in \cite{2023_Boryana-Hadzhiyska} and due to the fact that the improvement is not so visible in $\alpha_\text{AP}$ compared to $\epsilon$. Indeed, our findings in terms of the values of the improvement factors are consistent with the results from~\cite{2023_Boryana-Hadzhiyska}. We observe that other BAO-related parameters give different improvement factors. For example, while it has been found that $\alpha_\parallel$ gives the highest improvement rate ($\sim$1.6 times in Fourier space, and $\sim$1.5 in configuration space), $\alpha_\perp$ shows no major improvement after CV, with a dispersion of around 1.2 times lower compared to the non-CV results. In principle, we expect these factors as such given the fact that the Zeldovich approximation is more accurate at larger scales, where $\alpha_\parallel$ is more sensitive to, due to the Kaiser effect that causes the coherent infall of galaxies onto still collapsing structures. Overall, we found consistent results between both Fourier space and configuration space in terms of the error improvement factor in the dispersion of the BAO parameters due to the CV technique.

We show scatter plots for both $\alpha_\text{iso}$ and $\alpha_\text{AP}$ for some HOD models in Fig.~\ref{fig:scatter_plots_CS_vs_FS}. We can draw several conclusions from these scatter plots. First of all, we found that the CV results were less scattered with respect to the fiducial values, compared to the non-CV BAO fits. Some exceptions can be observed as in the case of the 1st-Gen mocks, where the BAO fits from the correlation function multipoles prefer a higher value of $\alpha_\text{iso}$ after CV, which turns out to drive our final systematics error budget for configuration space. We observe that while the mean values are consistent between the two analyses, sometimes the results for individual realizations are not, as is the case for the HMQ$_1^{(3\sigma)}$ model. This in principle is due to sample variance noise which is reduced after considering several realizations. Scatter plots such as the ones for $\alpha_\text{iso}$ in the case of the SFHOD, HMQ, and mHMQ+cf+Sh models show us that while CV does not particularly improve the consistency of the results between Fourier space and configuration space, it helps to correctly shift the BAO fits towards consistent values of $\alpha_\text{iso}$ and $\alpha_\text{AP}$ by reducing sample variance noise. Thus, in general, whereas CV shows to reduce the dispersion between the $\alpha_\text{iso}$ and $\alpha_\text{AP}$ measurements there is no clear trend about CV improving the consistency between configuration space and Fourier space. Part of this can be due to the fact that we are using analytical covariance matrices tuned to the clustering of the non-CV two-point measurements for both CV and non-CV BAO fits. Then, this mismatch might introduce some degree of inconsistency between the results of both spaces, even though the dispersion between the BAO scaling parameters is reduced. Nevertheless, overall we found consistency between the results from Fourier space and configuration space analyses. We found no extreme outliers that point out inconsistencies between the two spaces. At most, models with very low number density show more scatter but the mean values found over the 25 realizations are still consistent.

\afterpage{
\begin{figure}
{\includegraphics[width=\textwidth]{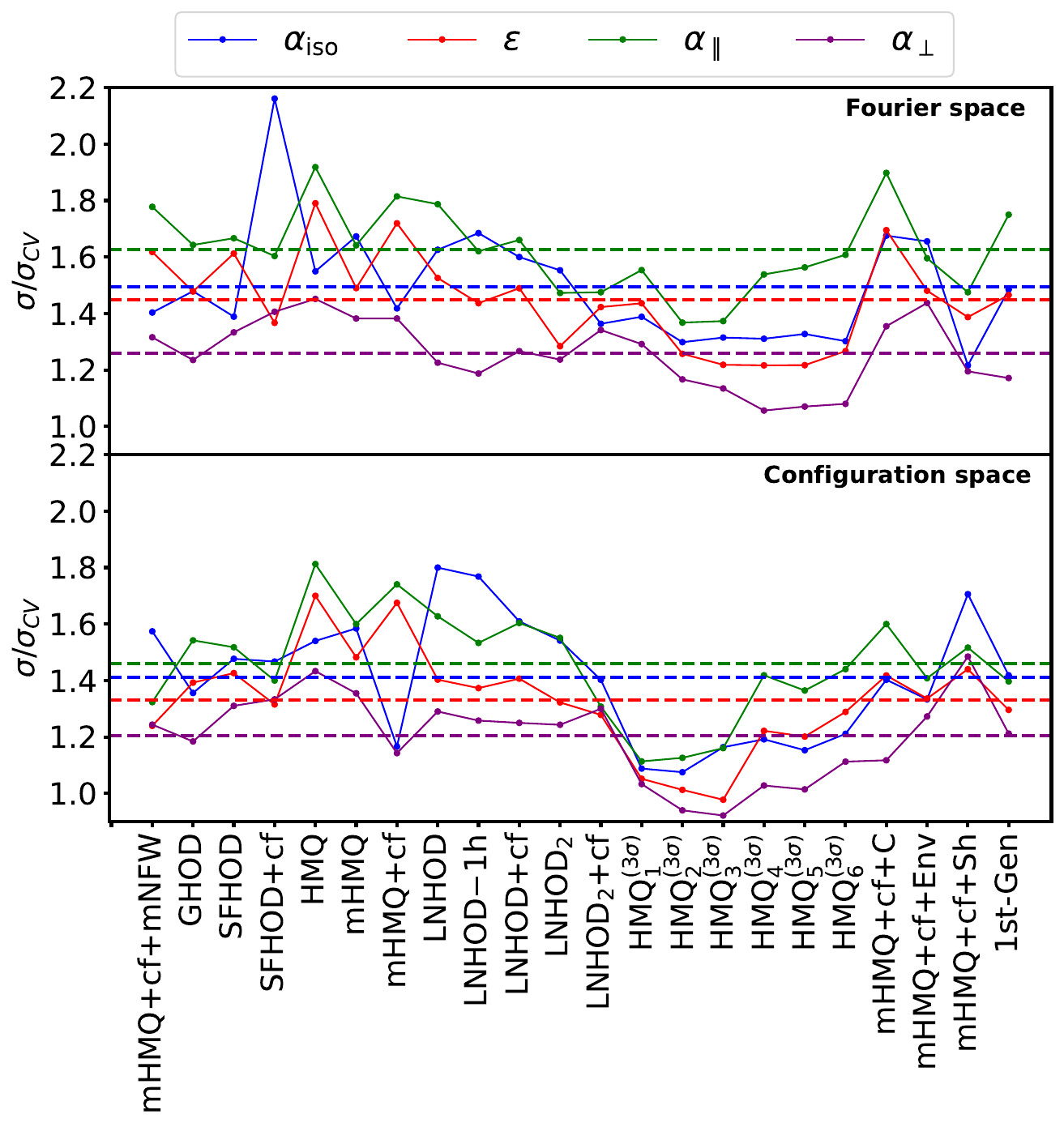}}
\caption{Error improvement factor calculated when comparing the dispersion over 25 mocks using the standard two-point measurements against the dispersion when updating to CV noise-reduced measurements. The top panel shows the error improvement factors in the Fourier space analysis and the bottom panel shows the analogous results in configuration space. The $x$-axis represents the HOD model and the $y$-axis shows the error improvement factor. We show such factors for several BAO scaling parameters. The mean values for the error improvement factors are shown in horizontal dashed lines.}
\label{Fig:CV_reduction_factor}
\end{figure}
\clearpage}

\afterpage{
\begin{figure*}
\begin{center}
\begin{tabular}{|c|c|c|c|}
\hline
{\includegraphics[width=0.22\textwidth]{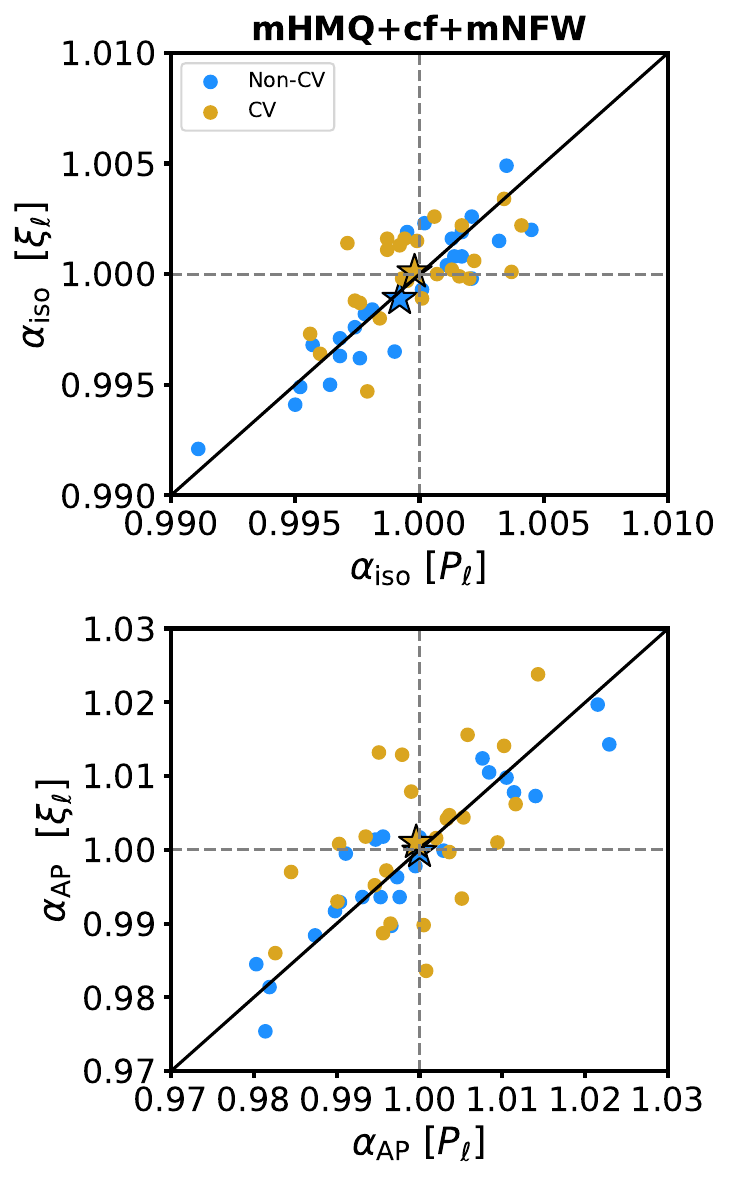}} & {\includegraphics[width=0.22\textwidth]{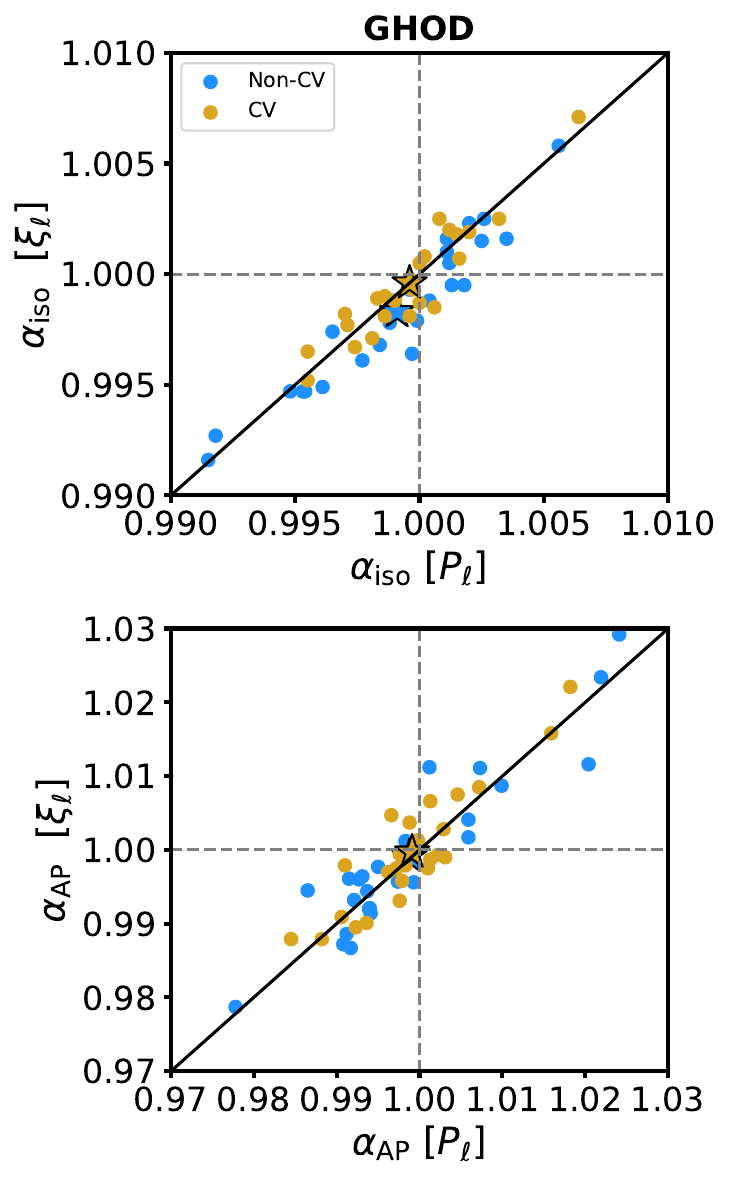}} & {\includegraphics[width=0.22\textwidth]{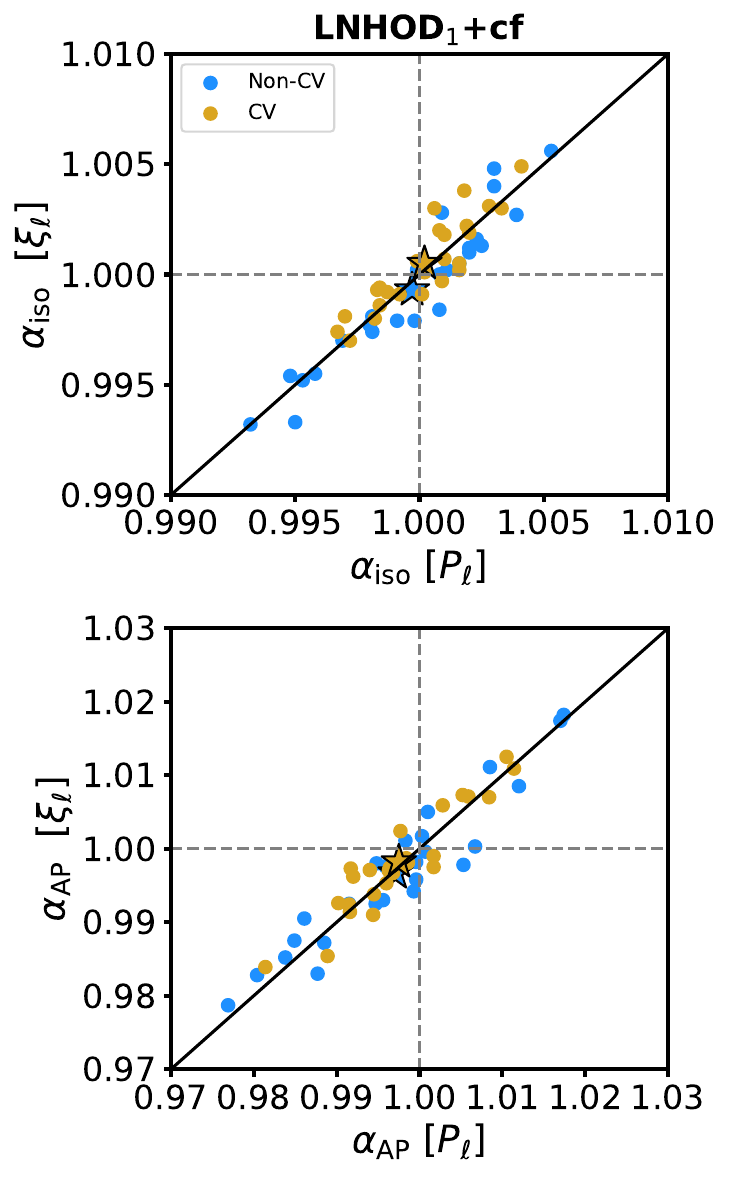}} & {\includegraphics[width=0.22\textwidth]{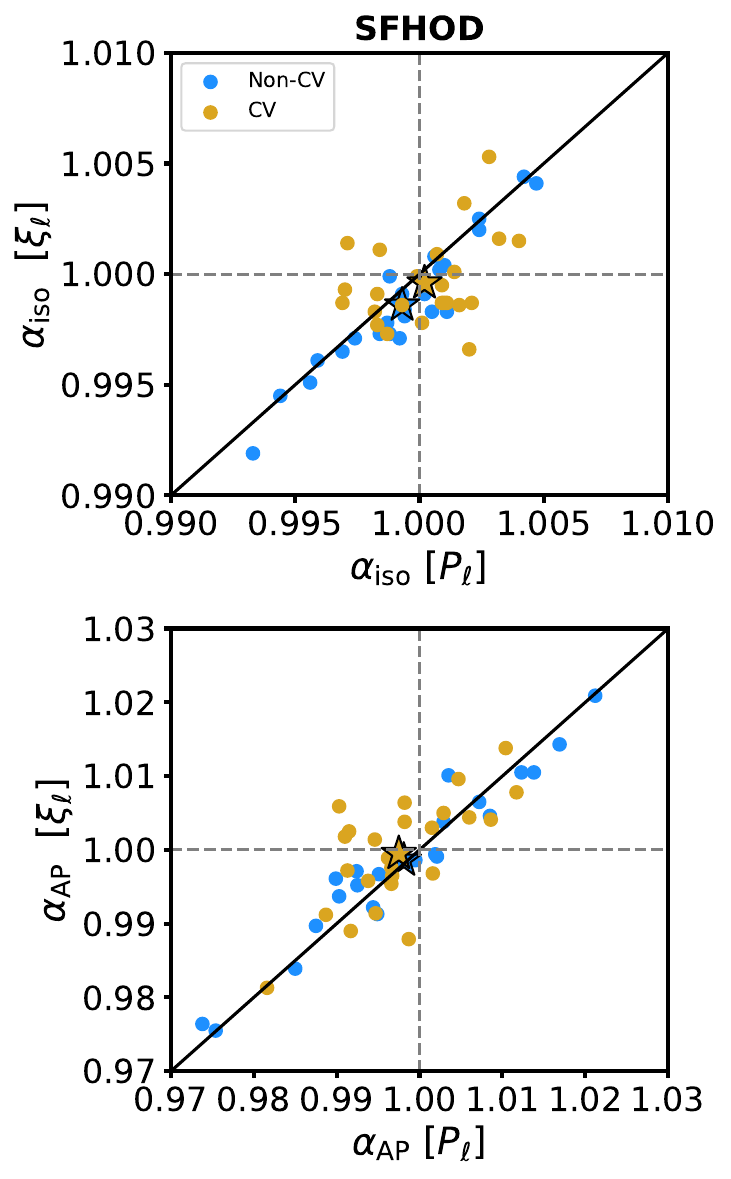}} \\ \hline
{\includegraphics[width=0.22\textwidth]{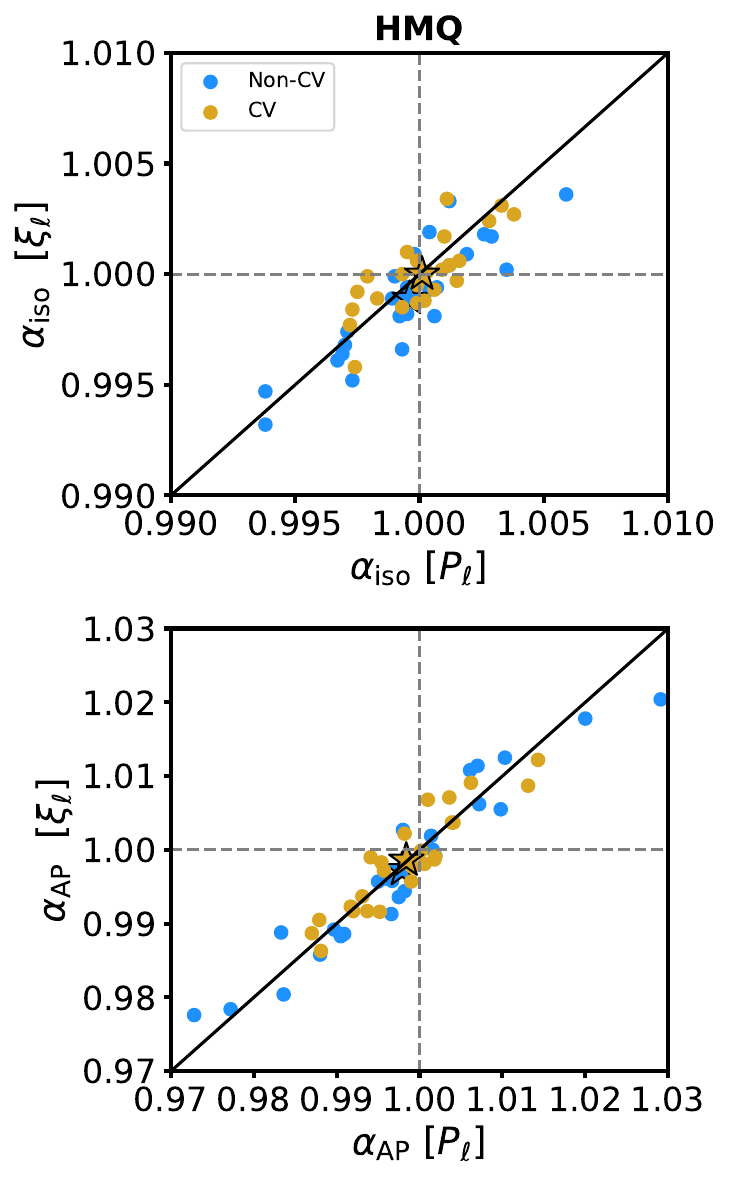}} & {\includegraphics[width=0.22\textwidth]{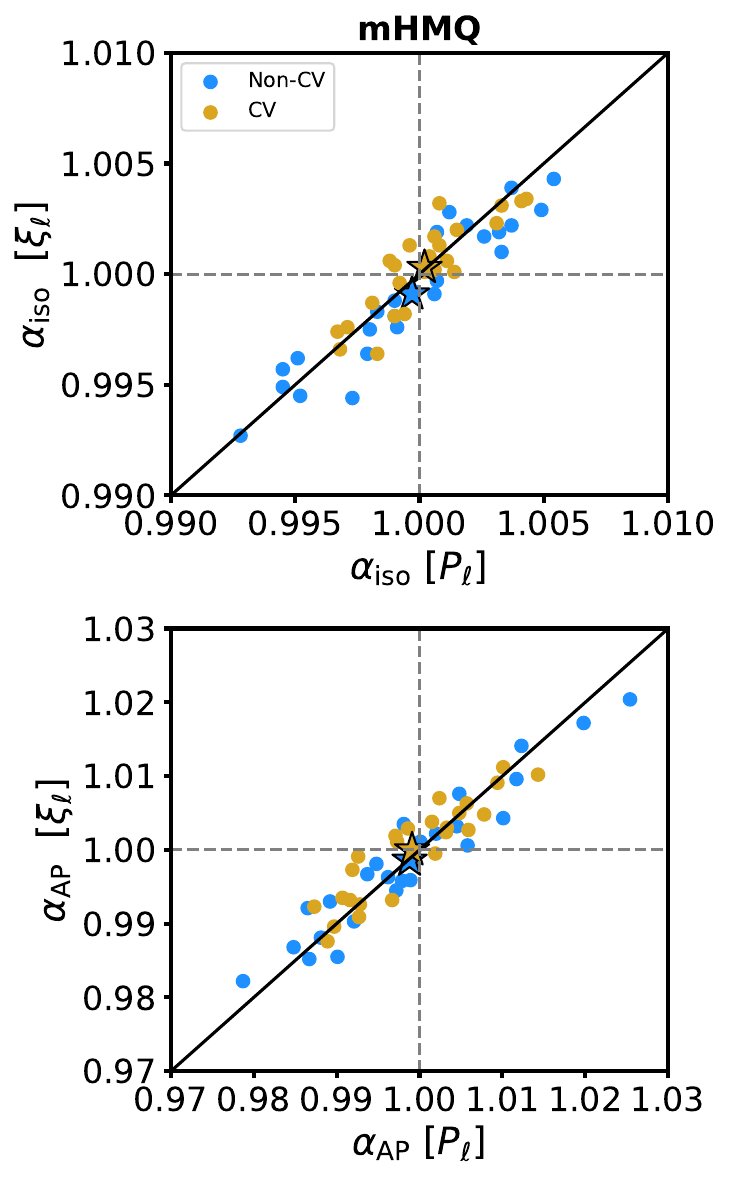}} & {\includegraphics[width=0.22\textwidth]{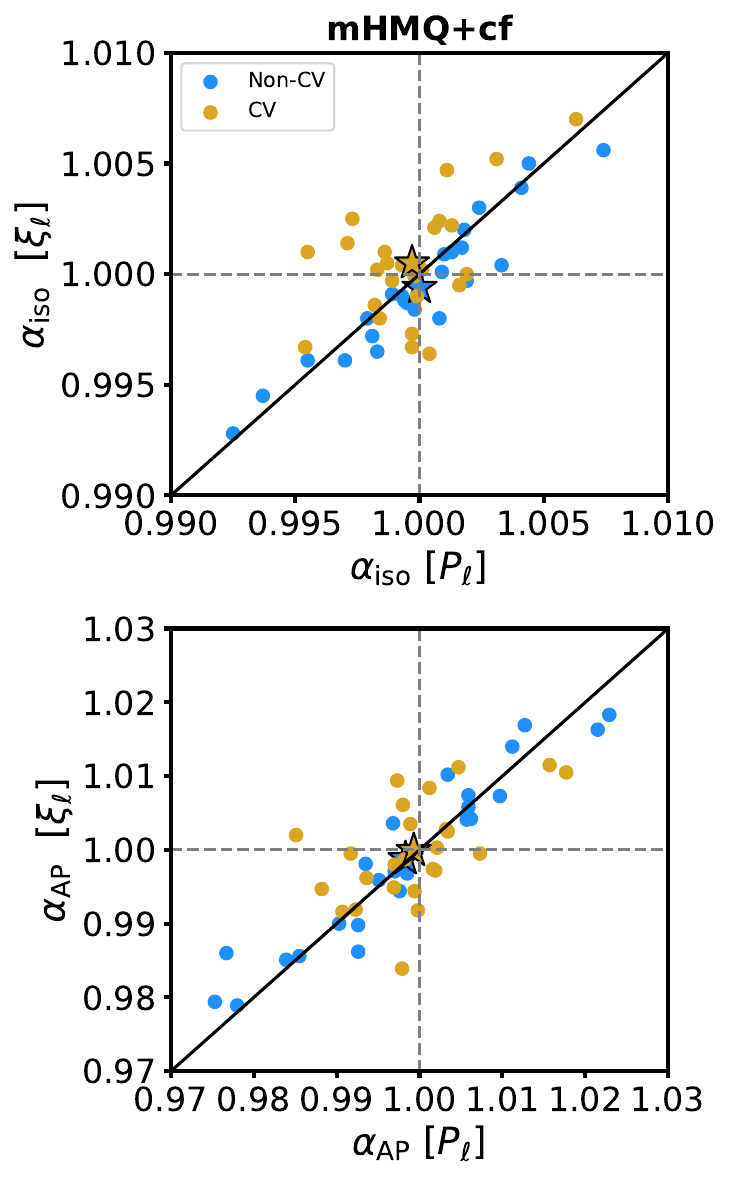}} & {\includegraphics[width=0.22\textwidth]{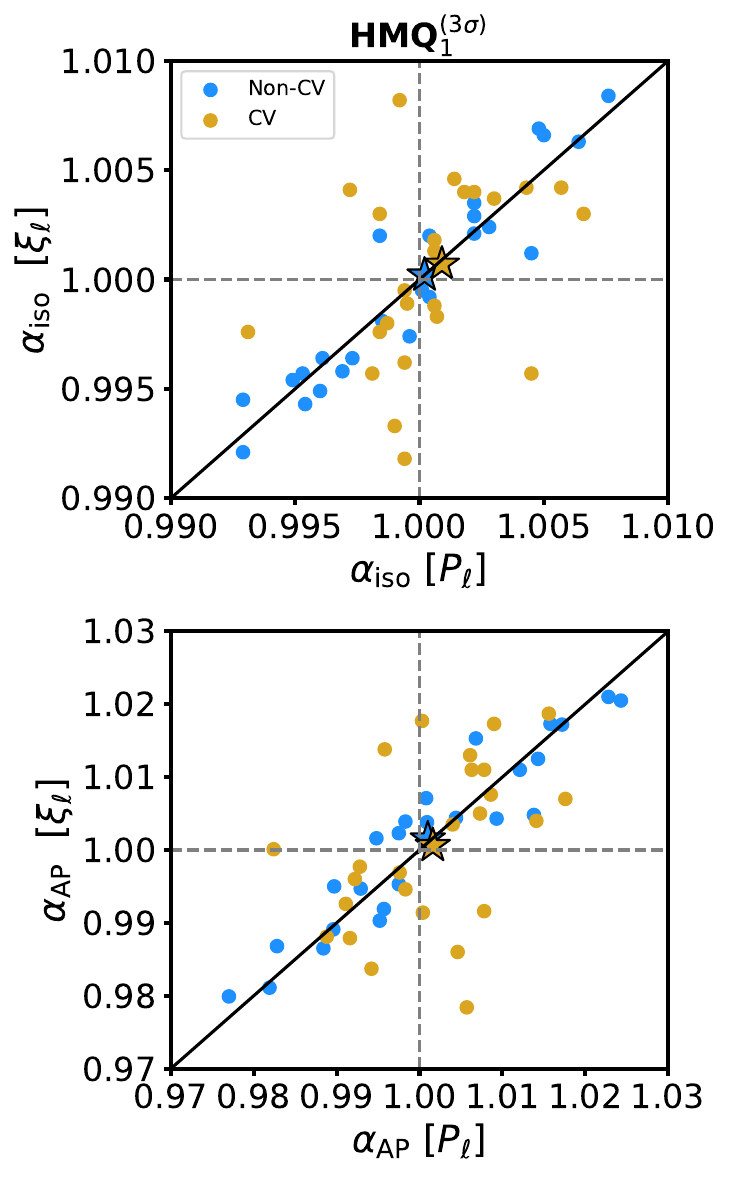}}\\ \hline
{\includegraphics[width=0.22\textwidth]{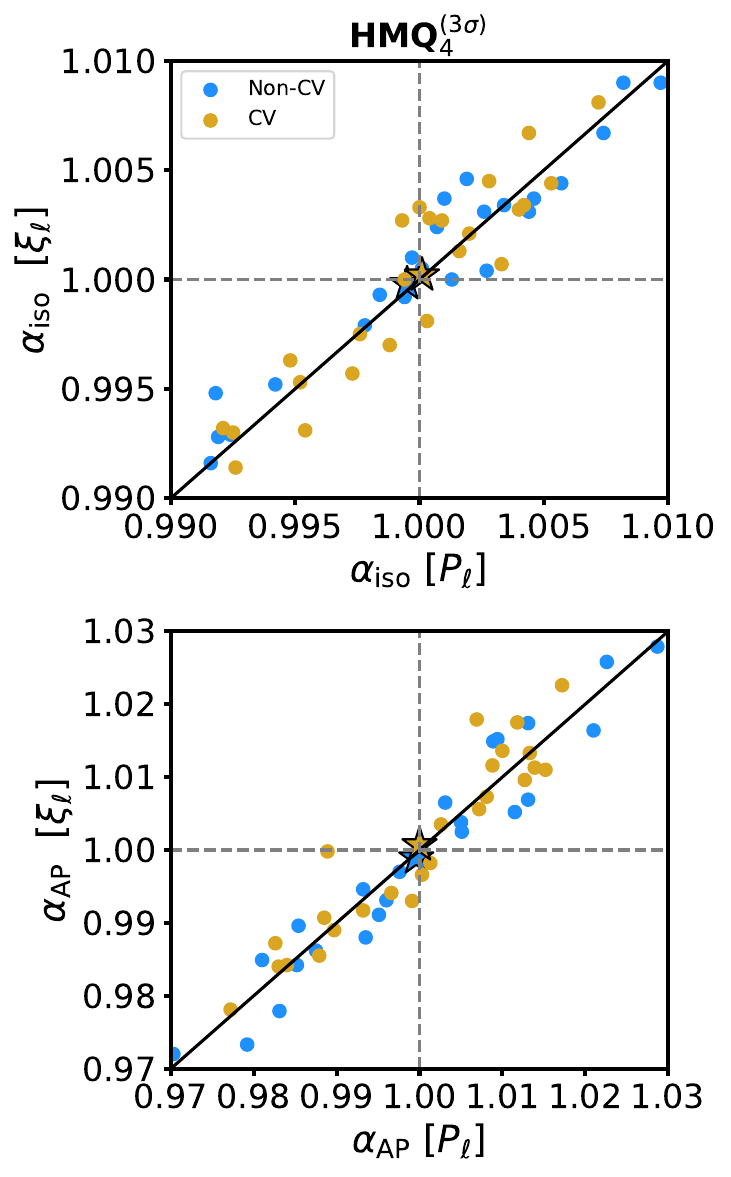}} & {\includegraphics[width=0.22\textwidth]{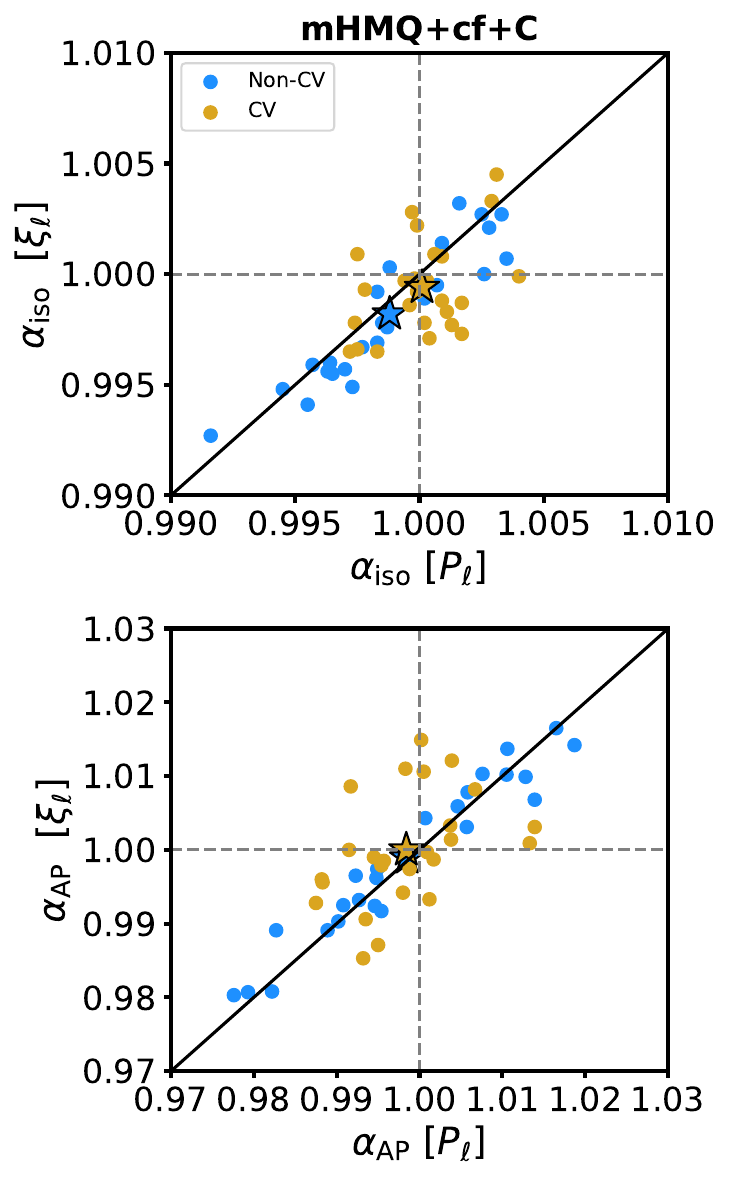}} & {\includegraphics[width=0.22\textwidth]{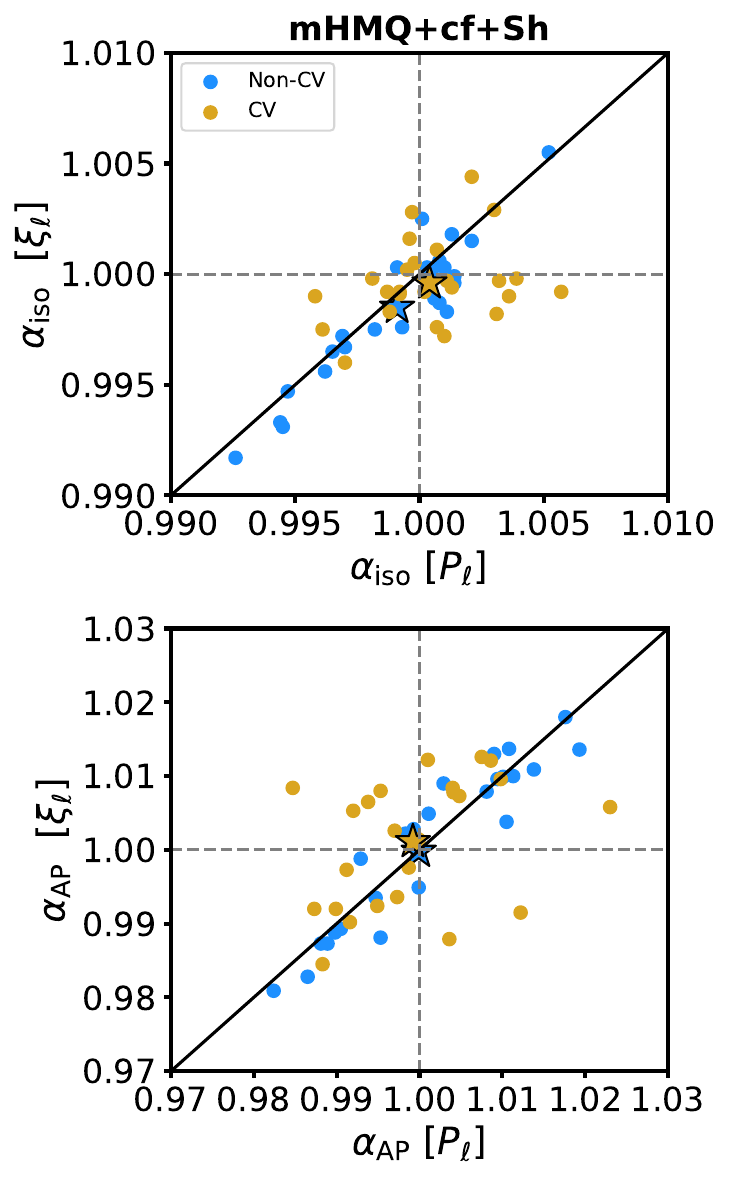}} & {\includegraphics[width=0.22\textwidth]{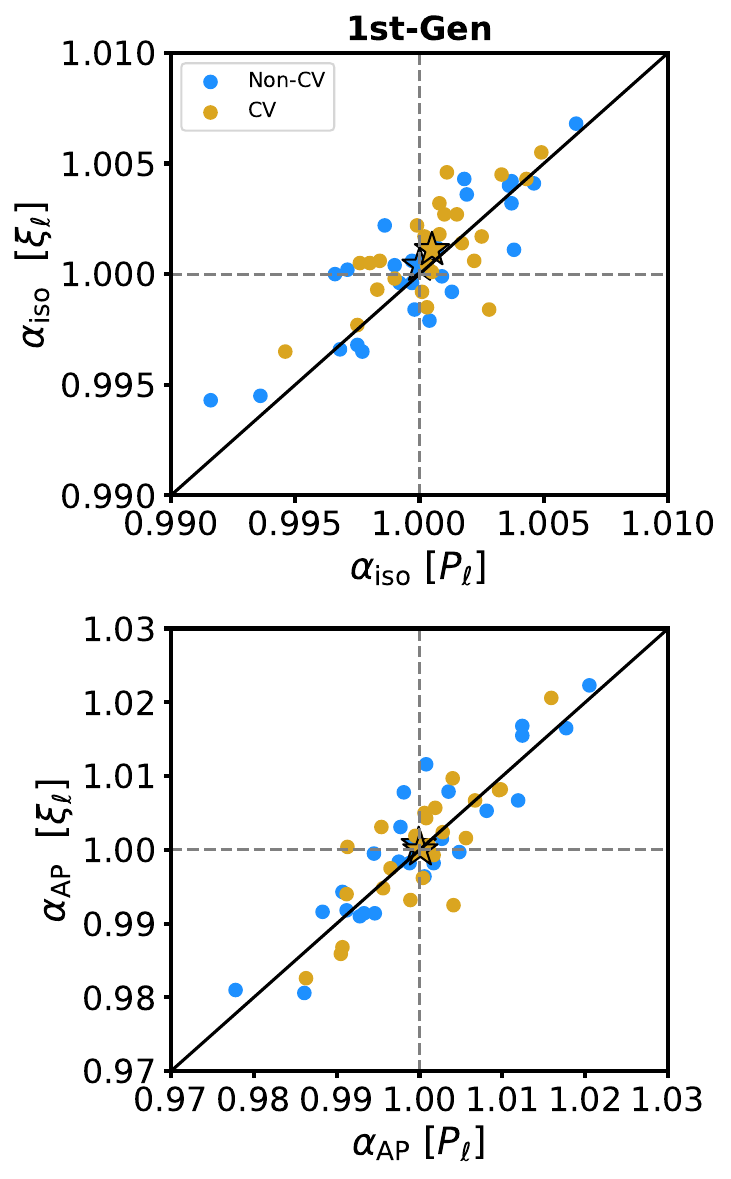}} \\ \hline
\end{tabular}
\end{center}
\caption{Post-reconstruction scatter plots for $\alpha_\text{iso}$ and $\alpha_\text{AP}$ for various HOD models. The scatter plots for $\alpha_\text{iso}$ are shown in the upper plots while the lower plots correspond to $\alpha_\text{AP}$. Each box corresponds to a particular HOD model denoted in boldface. The $x$-axis of each scatter plot represents the measurement using the power spectrum multipoles (Fourier space). On the other hand, the results from the fits on the correlation function multipoles (configuration space) are depicted over the $y$-axis. The blue dots show the results for the non-CV two-point measurements while the CV results are represented by orange points. The stars represent the mean values after averaging the results from the fits of 25 realizations.}
\label{fig:scatter_plots_CS_vs_FS}
\end{figure*}
\clearpage}

%% file: Appendix/Extended_results.tex
\section{Extended set of results}\label{Appendix:Full_results}

Our full set of results comprehends 22 HOD models. 25 realization mocks are used for each HOD model and we apply reconstruction to each mock. Then, we calculate two-point measurements before and after reconstruction in both Fourier space and configuration space. All these analyses give a total of around 2200 BAO fits used to estimate the systematics in our BAO analysis due to HOD dependence. As we are using 25 mocks for each HOD we present the results of this work by quoting the mean values of the 25 fits. Table \ref{Table:BAO_fits_PS} shows the results for the BAO fits times a factor of $10^3$. This means that our differences in the recovery of the correct value of the BAO parameters are at the sub-sub-percent level. We found that most of our shifts with respect to 1 in $\alpha_\text{iso}$ are reduced with the CV two-point measurements. We can observe that this comes together with a reduction in the dispersion $\sigma(\overline{\alpha}_\text{iso})$. Such reduction is more pronounced for mocks with less shot-noise as CV becomes more efficient with a higher number density. This is related to the factors described in Fig.~\ref{Fig:CV_reduction_factor}. We found that our $\langle\chi^2\rangle$ per DoF values are less than $\sim$1.1, which indicates that our analytical covariance matrices for the power spectrum multipoles derived from \texttt{thecov} based on the clustering of 25 mocks are appropriated for our analysis. We then use the same analytical covariance matrices for our fits to the CV two-point measurements. Hence, we observe a drop in our $\langle\chi^2\rangle$ values as the dispersion on our BAO parameters with respect to the fiducial values is reduced. We show the analogous for the configuration space analysis in Table \ref{Table:BAO_fits_CF}. We observe a similar behavior to our results in Fourier space, except for 1st-Gen where our measurements after CV increase the bias in $\alpha_\text{iso}$. This ended up driving our HOD systematics for the configuration space case. However, we also found reasonable $\langle\chi^2\rangle$ values. For BAO fits to CV two-point measurements we also observed a drop in $\langle\chi^2\rangle$, although smaller compared to the Fourier space case. This might have to do with the fact that the reduction factors due to CV are smaller in the configuration space case. This is related to CV being less efficient for configuration space analysis as the noise mitigation is less scale-dependent compared to the Fourier spaces, leading to less sensitivity at large scales.

We also show in this section the complete version of the simplified heatmaps presented in Section \ref{subsection:robustness_against_HODs}. Fig.~\ref{Fig:Heatmaps_cv_pk_alpha_iso} represents the comparison for all the permutations of pairs using our 22 HOD models. While the statistical error for $\alpha_\text{iso}$ is about 1\%, we found at most differences of around 0.08\% for HOD mocks at $z=1.1$. In the case of the HOD mocks centered at $z=0.8$, we observe that the models HMQ$_i^{(3\sigma)}$ ($i=1,2,3$) are fairly consistent with each other. A similar trend is observed for the models with complex galactic conformity ($i=4,5,6$). On the other hand, we observed mild differences between the models HMQ$_i^{(3\sigma)}$ with and without complex galactic conformity. However, these differences are of about 0.1\% at most for $\alpha_\text{iso}$. We remind the reader that we do not compare HOD models based on mocks built from simulations centered at different redshifts since we do not want to introduce simulation noise differences beyond the HOD model itself. We did not find shifts in $\alpha_\text{iso}$ above 3-$\sigma$ in Fourier space. Similar results are found for $\alpha_\text{AP}$ as shown in Fig.~\ref{Fig:Heatmaps_cv_pk_alpha_ap}. In such case, the highest significance found in $\Delta\alpha_\text{AP}$ is 2-$\sigma$ of less. We found shifts of at most 0.26\% in $\alpha_\text{AP}$, but still this is very small compared to the statistical error of DESI 2024.

The full heatmaps for $\alpha_\text{iso}$ and $\alpha_\text{AP}$ in the configuration space analysis are shown in Fig.~\ref{Fig:Heatmaps_cv_xi_alpha_iso} and Fig.~\ref{Fig:Heatmaps_cv_xi_alpha_ap}, respectively. We found higher shifts in $\alpha_\text{iso}$ between pairs of HOD models compared to the Fourier Space case. The highest shifts come from comparing 1st-Gen with other HOD models. The maximum shift found is 0.17\%, which is 1/6 of the statistical error for DESI 2024. The 1st-Gen mocks show significant differences with various HOD models at more than 3-$\sigma$. We quote the maximum shift found at more than 3-$\sigma$ as our systematic error for $\alpha_\text{iso}$. We do not find significant shifts for the models at $z=0.8$. In a similar fashion to the Fourier space analysis, we did not find any significant shift in $\alpha_\text{AP}$. Therefore, our quoted systematic error comes from the dispersion over all shifts. It is worth mentioning that we also tested the effect of the shot-noise differences in our analysis. We produced sub-sampled versions of our HOD mocks to have the same number density for all of them ($\sim 10^{-3}h^3/\text{Mpc}^{-3}$). We found that this sub-sampling leads to higher shifts in the BAO parameters and the AP parameter as well as higher significance in the shifts. However, these biases are introduced by increasing the shot-noise and reducing the efficiency of the CV technique. 

Finally, we show the measurements of the BAO feature in both Fourier space and configuration space after removing the smoothed component for each HOD model in Fig. \ref{Fig:post_recon_data}. The smoothed component is calculated after performing the BAO fits and it is then subtracted to the measurements. While we do not show explicit error bars to facilitate the comparison between measurements of various HOD models, we found that the measurements are consistent within the error bars when the covariance matrices are taken into consideration.

\afterpage{
\begin{figure*}
\begin{center}
\begin{tabular}{c}
{\includegraphics[width=\textwidth]{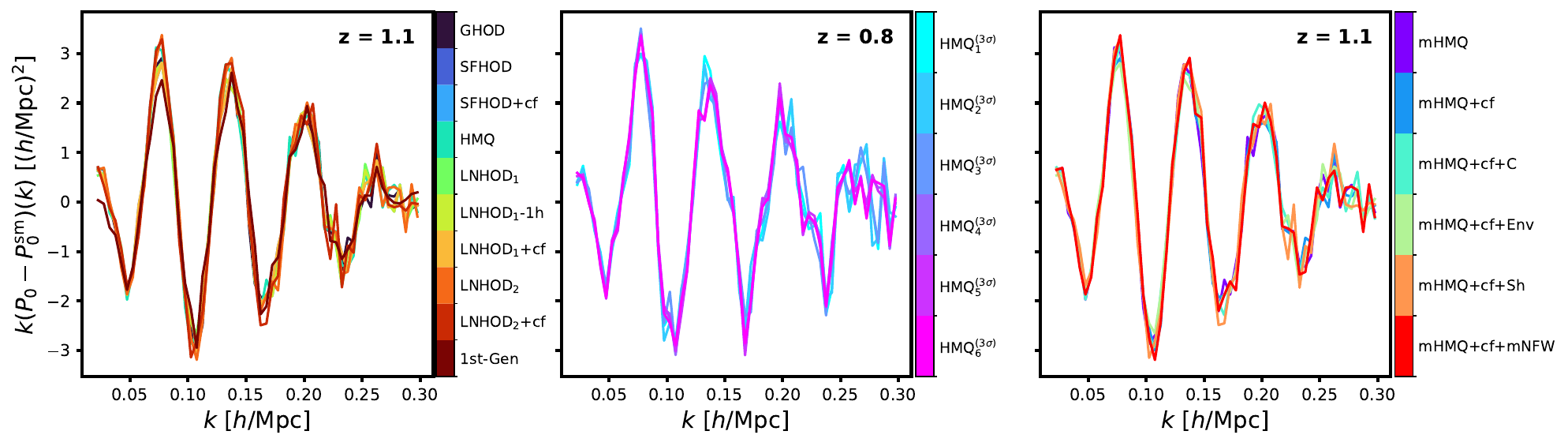}}\\
{\includegraphics[width=\textwidth]{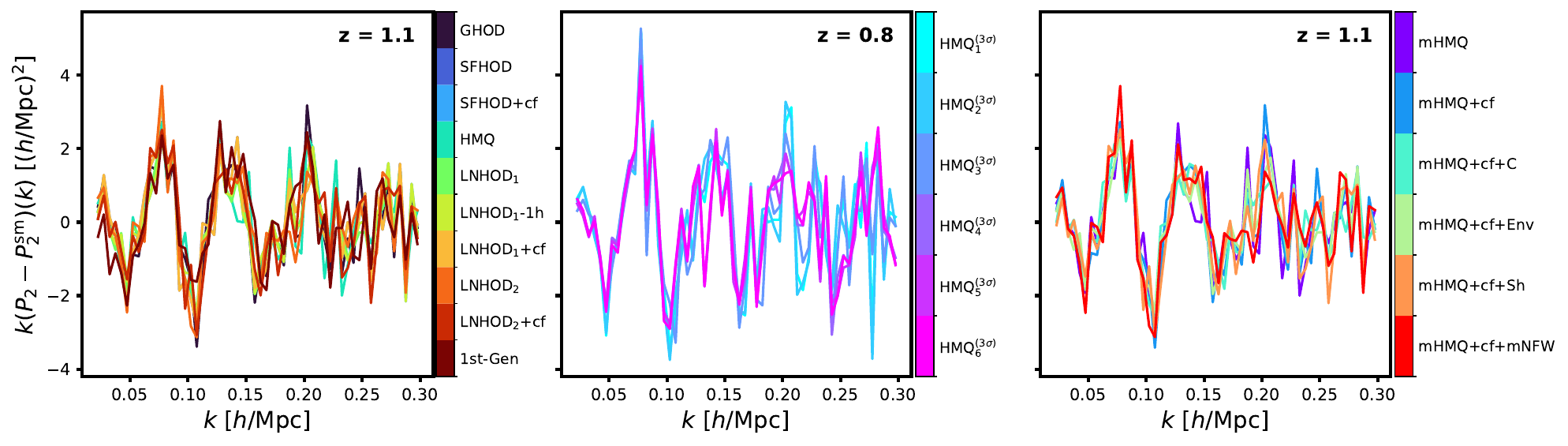}}\\
{\includegraphics[width=\textwidth]{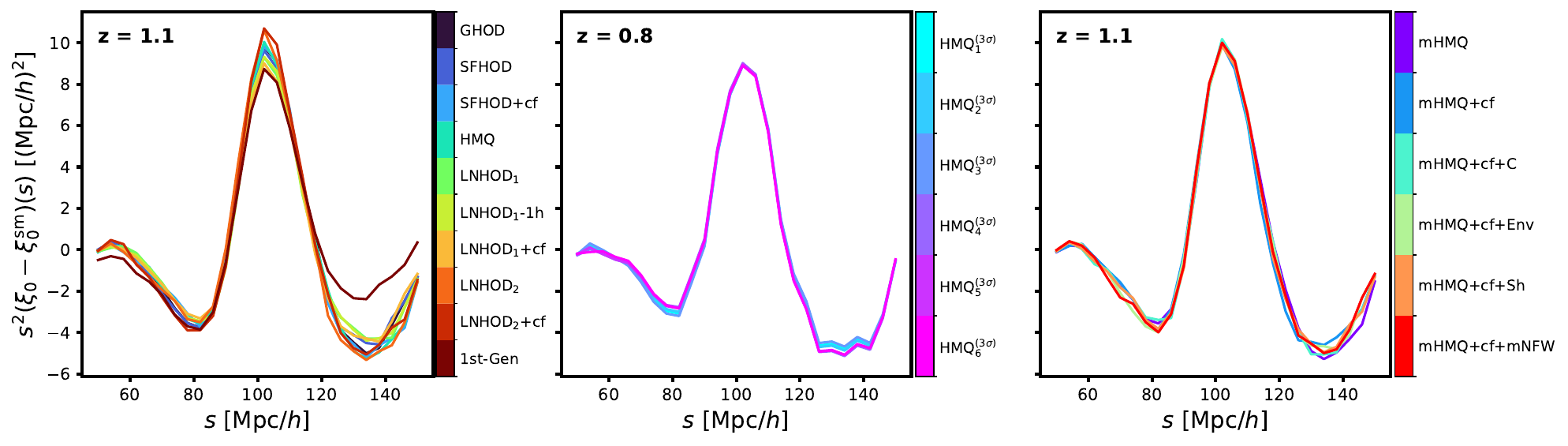}}\\
{\includegraphics[width=\textwidth]{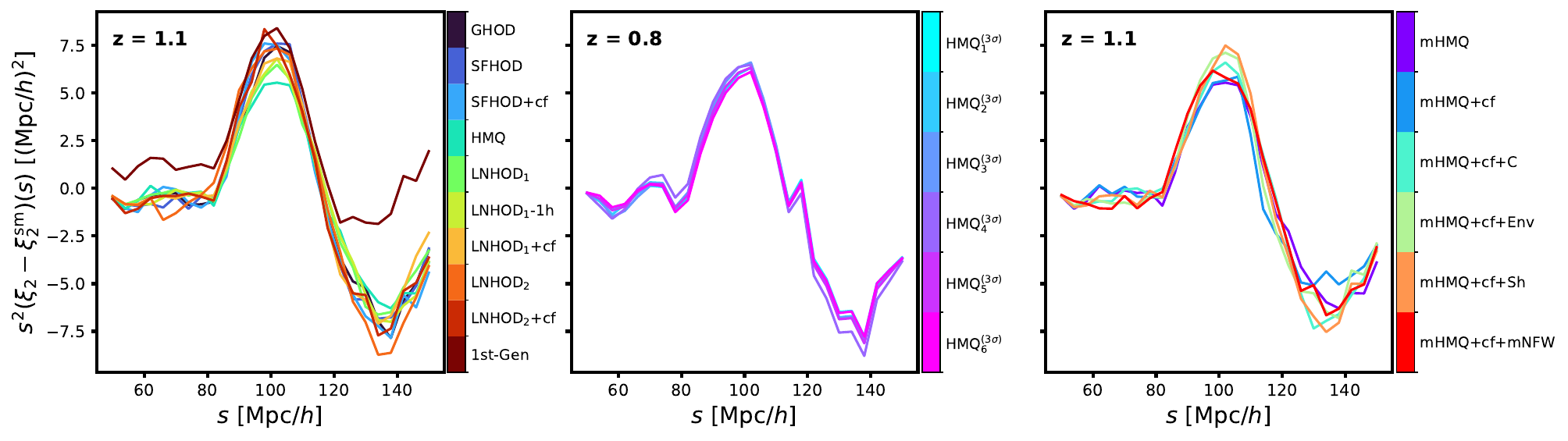}}\\
\end{tabular}
\end{center}
\caption{Two-point measurements of the BAO feature after applying rec-sym reconstruction, the CV technique, and after averaging 25 realizations for each HOD model. The measurements in the plot show the data after being subtracted with the smoothed component. We do not show the error bars for the measurements in this plot for clarity of the plot, but the BAO components of all the HOD models are consistent within the error bars. The top panels show the measurements for the monopole and quadrupole of the galaxy power spectrum and the bottom panels correspond to the monopole and quadrupole terms of the two-point correlation function.}
\label{Fig:post_recon_data}
\end{figure*}
\clearpage
}

\afterpage{
\begin{table*}[t]
\begin{center}
\setlength{\tabcolsep}{4pt} 
\resizebox{\textwidth}{!}{
{\renewcommand{\arraystretch}{1.2}
\begin{tabular}{l|c|c|c|c|c|c|c|c|c|c|c|c|c}
\hline\hline
HOD model & $\langle\Delta\alpha_\text{iso}\rangle$ &  $\langle \sigma_{\alpha_\text{iso}} \rangle$ &  $\sigma(\overline{\alpha}_\text{iso})$ &  $\langle\Delta\alpha_\text{AP}\rangle$ &  $\langle \sigma_{\alpha_\text{AP}} \rangle$ &  $\sigma(\overline{\alpha}_\text{AP})$ &  $\langle\Delta\alpha_\parallel\rangle$ &  $\langle \sigma_{\alpha_\parallel} \rangle$ &  $\sigma(\overline{\alpha}_\parallel)$ &  $\langle\Delta\alpha_\perp\rangle$ &  $\langle \sigma_{\alpha_\perp} \rangle$ &  $\sigma(\overline{\alpha}_\perp)$ & $\langle \chi^2 \rangle$ \\ \hline
\multicolumn{14}{c}{Using standard two-point measurements} \\ \hline
mHMQ+cf+mNFW &  -0.80 &  0.64 & 0.64 &0.00 & 2.30 & 2.60 &  -0.90 &1.72 &  1.92 & -0.80 & 0.96 &   1.00 &101.3 \\
GHOD &  -0.90 &0.64 &    0.70 &    -1.20 &    2.28 & 2.23 &  -1.70 &1.70 &  1.84 & -0.50 & 0.96 &   0.84 & 91.8 \\
SFHOD &  -0.70 &0.64 &    0.55 &    -1.90 &    2.30 & 2.32 &  -2.00 &1.70 &  1.80 &  0.00 & 0.96 &   0.80 & 91.4 \\
SFHOD+cf &  -1.20 &0.64 &    0.80 &    -0.50 &    2.26 & 2.19 &  -1.50 &1.68 &  1.86 & -1.00 & 0.96 &   0.90 &100.7 \\
HMQ &  -0.40 &0.64 &    0.56 &    -2.10 &    2.26 & 2.51 &  -1.80 &1.66 &  1.88 &  0.30 & 0.96 &   0.90 & 92.2 \\
mHMQ &  -0.30 &0.64 &    0.70 &    -1.20 &    2.26 & 2.25 &  -1.10 &1.66 &  1.74 &  0.20 & 0.96 &   0.94 &101.9 \\
mHMQ+cf &   0.00 &0.64 &    0.66 &    -1.70 &    2.30 & 2.54 &  -1.10 &1.70 &  1.96 &  0.60 & 0.96 &   0.94 &103.5 \\
LNHOD$_1$ &  -0.80 &0.64 &    0.58 &    -2.20 &    2.26 & 2.12 &  -2.30 &1.68 &  1.68 &  0.00 & 0.96 &   0.76 & 90.2 \\
LNHOD$_1$-1h &  -0.60 &0.62 &    0.62 &    -1.30 &    2.24 & 2.31 &  -1.50 &1.66 &  1.88 & -0.20 & 0.94 &   0.76 & 93.8 \\
LNHOD$_1$+cf &  -0.30 &0.64 &    0.64 &    -2.90 &    2.28 & 2.13 &  -2.20 &1.70 &  1.76 &  0.80 & 0.96 &   0.76 & 90.2 \\
LNHOD$_2$ &  -0.70 &0.64 &    0.70 &    -0.60 &    2.30 & 2.09 &  -1.10 &1.70 &  1.62 & -0.50 & 0.96 &   0.94 &105.8 \\
LNHOD$_2$+cf &  -0.10 &0.64 &    0.66 &    -1.20 &    2.26 & 2.54 &  -0.90 &1.68 &  1.80 &  0.40 & 0.96 &   1.10 &103.4 \\
HMQ$_1^{(3\sigma)}$ &   0.20 &0.94 &    0.96 &1.00 &    3.18 & 2.57 &   0.80 &2.28 &  2.02 & -0.10 & 1.44 &   1.24 & 98.3 \\
HMQ$_2^{(3\sigma)}$ &   0.20 &0.92 &    1.02 &1.00 &    3.12 & 3.10 &   0.80 &2.22 &  2.38 &  0.00 & 1.42 &   1.40 & 99.3 \\
HMQ$_3^{(3\sigma)}$ &   0.10 &0.96 &    0.94 &0.60 &    3.24 & 2.31 &   0.50 &2.30 &  1.84 &  0.00 & 1.48 &   1.18 & 96.1 \\
HMQ$_4^{(3\sigma)}$ &  -0.50 &0.94 &    1.23 &    -0.40 &    3.20 & 2.93 &  -0.90 &2.28 &  2.40 & -0.30 & 1.44 &   1.50 & 95.0 \\
HMQ$_5^{(3\sigma)}$ &  -0.40 &0.94 &    1.19 &0.20 &    3.20 & 2.83 &  -0.30 &2.30 &  2.22 & -0.40 & 1.44 &   1.52 & 94.5 \\
HMQ$_6^{(3\sigma)}$ &  -0.70 &0.94 &    1.25 &    -0.10 &    3.20 & 3.08 &  -0.80 &2.30 &  2.38 & -0.60 & 1.44 &   1.62 & 95.5 \\
mHMQ+cf+C &  -1.20 &0.64 &    0.61 &    -1.50 &    2.24 & 2.37 &  -2.20 &1.64 &  1.86 & -0.60 & 0.98 &   0.84 &100.3 \\
mHMQ+cf+Env &  -0.30 &0.64 &    0.60 &    -0.80 &    2.26 & 2.07 &  -0.90 &1.68 &  1.50 &  0.00 & 0.96 &   0.92 &101.4 \\
mHMQ+cf+Sh & -0.90 & 0.64 & 0.60 & -0.10 & 2.30 & 2.47 & -1.00 & 1.70 & 1.80 & -0.90 & 0.98 & 0.98 & 102.2 \\
1st-Gen & 0.00 & 0.62 & 0.69 & -0.10 & 2.28 & 2.02 & -0.10 & 1.68 & 1.68 & 0.10 & 0.94 & 0.82 & 85.1 \\
\hline
\multicolumn{14}{c}{Using CV noise-reduced two-point measurements} \\ \hline
mHMQ+cf+mNFW &  -0.20 &0.60 &    0.46 &    -0.40 &    2.16 & 1.60 &  -0.40 &1.60 &  1.08 &  0.00 & 0.90 &   0.76 & 56.7 \\
GHOD &  -0.40 &0.64 &    0.48 &    -0.90 &    2.24 & 1.51 &  -1.00 &1.64 &  1.12 &  0.00 & 0.96 &   0.68 & 40.6 \\
SFHOD &   0.20 &0.64 &    0.40 &    -2.50 &    2.28 & 1.44 &  -1.40 &1.68 &  1.08 &  1.10 & 0.96 &   0.60 & 37.7 \\
SFHOD+cf &  -0.10 &0.64 &    0.37 &    -0.90 &    2.26 & 1.60 &  -0.70 &1.66 &  1.16 &  0.20 & 0.98 &   0.64 & 39.3 \\
HMQ &   0.10 &0.62 &    0.36 &    -1.60 &    2.20 & 1.41 &  -1.00 &1.62 &  0.98 &  0.60 & 0.94 &   0.62 & 42.3 \\
mHMQ &   0.20 &0.62 &    0.42 &    -0.90 &    2.20 & 1.51 &  -0.40 &1.62 &  1.06 &  0.50 & 0.94 &   0.68 & 49.7 \\
mHMQ+cf &  -0.30 &0.66 &    0.46 &    -0.70 &    2.36 & 1.48 &  -0.80 &1.74 &  1.08 & -0.10 & 1.00 &   0.68 & 40.0 \\
LNHOD$_1$ &  -0.30 &0.62 &    0.36 &    -1.70 &    2.22 & 1.39 &  -1.40 &1.62 &  0.94 &  0.30 & 0.94 &   0.62 & 40.2 \\
LNHOD$_1$-1h &  -0.10 &0.62 &    0.37 &    -0.80 &    2.18 & 1.61 &  -0.70 &1.60 &  1.16 &  0.20 & 0.92 &   0.64 & 42.7 \\
LNHOD$_1$+cf &   0.20 &0.62 &    0.40 &    -2.50 &    2.24 & 1.43 &  -1.40 &1.64 &  1.06 &  1.10 & 0.94 &   0.60 & 39.2 \\
LNHOD$_2$ &  -0.20 &0.64 &    0.45 &    -0.20 &    2.24 & 1.62 &  -0.30 &1.66 &  1.10 &  0.00 & 0.94 &   0.76 & 53.2 \\
LNHOD$_2$+cf &   0.40 &0.62 &    0.48 &    -0.90 &    2.20 & 1.79 &  -0.20 &1.62 &  1.22 &  0.80 & 0.94 &   0.82 & 50.2 \\
HMQ$_1^{(3\sigma)}$ &   0.90 &0.92 &    0.69 &1.60 &    3.10 & 1.79 &   1.90 &2.22 &  1.30 &  0.40 & 1.42 &   0.96 & 65.2 \\
HMQ$_2^{(3\sigma)}$ &   0.90 &0.90 &    0.78 &1.40 &    3.04 & 2.47 &   1.80 &2.16 &  1.74 &  0.60 & 1.40 &   1.20 & 64.8 \\
HMQ$_3^{(3\sigma)}$ &   0.90 &0.94 &    0.72 &1.30 &    3.16 & 1.90 &   1.70 &2.24 &  1.34 &  0.50 & 1.46 &   1.04 & 62.4 \\
HMQ$_4^{(3\sigma)}$ &   0.10 &0.92 &    0.94 &0.00 &    3.14 & 2.41 &   0.00 &2.24 &  1.56 &  0.20 & 1.42 &   1.42 & 62.7 \\
HMQ$_5^{(3\sigma)}$ &   0.20 &0.92 &    0.90 &0.40 &    3.12 & 2.32 &   0.50 &2.24 &  1.42 &  0.20 & 1.42 &   1.42 & 62.0 \\
HMQ$_6^{(3\sigma)}$ &  -0.10 &0.92 &    0.96 &0.20 &    3.16 & 2.43 &   0.00 &2.26 &  1.48 & -0.10 & 1.42 &   1.50 & 63.1 \\
mHMQ+cf+C &   0.10 &0.64 &    0.36 &    -1.60 &    2.26 & 1.40 &  -1.00 &1.66 &  0.98 &  0.70 & 0.98 &   0.62 & 39.8 \\
mHMQ+cf+Env &  -0.20 &0.66 &    0.36 &    -1.50 &    2.32 & 1.40 &  -1.30 &1.70 &  0.94 &  0.30 & 0.98 &   0.64 & 36.6 \\
mHMQ+cf+Sh &   0.40 &0.62 &    0.49 &    -0.80 &    2.20 & 1.78 &  -0.10 &1.62 &  1.22 &  0.80 & 0.94 &   0.82 & 50.6 \\
1st-Gen &   0.50 &0.62 &    0.46 &0.10 &    2.24 & 1.38 &   0.50 &1.66 &  0.96 &  0.50 & 0.94 &   0.70 & 38.0 \\
\hline\hline
\end{tabular}}}
\end{center}
\caption{Post-reconstruction BAO fits in Fourier space for standard two-point measurements and CV two-point measurements. We show the constraints for both ($\alpha_\text{iso}$,$\alpha_\text{AP}$) and ($\alpha_\parallel$,$\alpha_\perp$) parameterizations for each HOD model. The $\langle\Delta\alpha_\text{iso}\rangle$ column represents the mean of the shifts in $\alpha_\text{iso}$ with respect to the fiducial. Then, while second column for $\alpha_\text{iso}$ shows the average error coming from the covariance matrix, the third column shows the standard deviation obtained from the dispersion of 25 fits for $\alpha_\text{iso}$, and so on for the rest of parameters. All values are multiplied by a $10^3$ factor, meaning that deviations from fiducial values can be of the order of the sub-sub-percent level. The last column corresponds to the average $\chi^2$ obtained from 25 realization fits with 93 DoF for all fits.}
\label{Table:BAO_fits_PS}
\end{table*}
\clearpage}

\afterpage{
\begin{table*}[t]
\begin{center}
\setlength{\tabcolsep}{4pt} 
\resizebox{\textwidth}{!}{
{\renewcommand{\arraystretch}{1.2}
\begin{tabular}{l|c|c|c|c|c|c|c|c|c|c|c|c|c}
\hline\hline
HOD model & $\langle\Delta\alpha_\text{iso}\rangle$ &  $\langle \sigma_{\alpha_\text{iso}} \rangle$ &  $\sigma(\overline{\alpha}_\text{iso})$ &  $\langle\Delta\alpha_\text{AP}\rangle$ &  $\langle \sigma_{\alpha_\text{AP}} \rangle$ &  $\sigma(\overline{\alpha}_\text{AP})$ &  $\langle\Delta\alpha_\parallel\rangle$ &  $\langle \sigma_{\alpha_\parallel} \rangle$ &  $\sigma(\overline{\alpha}_\parallel)$ &  $\langle\Delta\alpha_\perp\rangle$ &  $\langle \sigma_{\alpha_\perp} \rangle$ &  $\sigma(\overline{\alpha}_\perp)$ & $\langle \chi^2 \rangle$ \\ \hline
\multicolumn{14}{c}{Using standard two-point measurements} \\ \hline
mHMQ+cf+mNFW &  -1.10 &0.66 &    0.64 &    -0.20 &    2.34 & 2.48 &  -1.20 &1.74 &  1.80 & -1.00 & 1.00 &   1.02 & 36.0 \\
GHOD &  -1.70 &0.62 &    0.68 &    -0.90 &    2.22 & 2.30 &  -2.30 &1.64 &  1.82 & -1.30 & 0.92 &   0.90 & 32.1 \\
SFHOD &  -1.40 &0.62 &    0.58 &    -1.40 &    2.22 & 2.14 &  -2.40 &1.64 &  1.70 & -0.90 & 0.94 &   0.76 & 30.7 \\
SFHOD+cf &  -1.80 &0.66 &    0.80 &    -0.20 &    2.30 & 2.20 &  -2.00 &1.68 &  1.82 & -1.70 & 1.00 &   0.96 & 31.9 \\
HMQ &  -1.10 &0.62 &    0.54 &    -2.70 &    2.18 & 2.34 &  -2.90 &1.62 &  1.74 & -0.10 & 0.92 &   0.86 & 32.4 \\
mHMQ &  -0.90 &0.66 &    0.66 &    -1.40 &    2.30 & 2.01 &  -1.90 &1.70 &  1.60 & -0.40 & 0.98 &   0.84 & 33.9 \\
mHMQ+cf &  -0.60 &0.66 &    0.62 &    -1.20 &    2.32 & 2.35 &  -1.40 &1.70 &  1.88 & -0.20 & 1.00 &   0.80 & 35.1 \\
LNHOD$_1$ &  -1.20 &0.62 &    0.58 &    -2.30 &    2.18 & 2.13 &  -2.80 &1.62 &  1.66 & -0.40 & 0.92 &   0.80 & 31.4 \\
LNHOD$_1$-1h &  -1.20 &0.60 &    0.63 &    -1.50 &    2.14 & 2.27 &  -2.20 &1.58 &  1.84 & -0.60 & 0.90 &   0.78 & 37.4 \\
LNHOD$_1$+cf &  -0.70 &0.62 &    0.68 &    -3.30 &    2.20 & 2.03 &  -2.90 &1.64 &  1.70 &  0.50 & 0.92 &   0.80 & 30.0 \\
LNHOD$_2$ &  -1.00 &0.66 &    0.71 &    -0.80 &    2.34 & 1.94 &  -1.50 &1.72 &  1.52 & -0.70 & 1.00 &   0.92 & 35.4 \\
LNHOD$_2$+cf &  -0.70 &0.64 &    0.62 &    -1.10 &    2.28 & 2.42 &  -1.40 &1.68 &  1.70 & -0.30 & 0.98 &   1.04 & 34.2 \\
HMQ$_1^{(3\sigma)}$ &   0.20 &0.94 &    0.99 &1.50 &    3.24 & 2.43 &   1.10 &2.32 &  1.96 & -0.30 & 1.46 &   1.24 & 39.3 \\
HMQ$_2^{(3\sigma)}$ &   0.20 &0.94 &    1.03 &0.70 &    3.22 & 2.85 &   0.60 &2.30 &  2.32 &  0.10 & 1.46 &   1.26 & 37.2 \\
HMQ$_3^{(3\sigma)}$ &   0.10 &0.94 &    0.94 &0.80 &    3.22 & 2.37 &   0.60 &2.32 &  1.88 & -0.10 & 1.44 &   1.18 & 42.1 \\
HMQ$_4^{(3\sigma)}$ &  -0.20 &0.96 &    1.16 &    -1.10 &    3.32 & 3.07 &  -0.90 &2.38 &  2.44 &  0.30 & 1.50 &   1.46 & 37.4 \\
HMQ$_5^{(3\sigma)}$ &   0.00 &0.96 &    1.08 &    -0.40 &    3.32 & 2.97 &  -0.30 &2.40 &  2.32 &  0.30 & 1.50 &   1.40 & 40.4 \\
HMQ$_6^{(3\sigma)}$ &  -0.30 &0.98 &    1.17 &    -1.00 &    3.34 & 3.19 &  -1.00 &2.40 &  2.42 &  0.20 & 1.50 &   1.58 & 38.3 \\
mHMQ+cf+C &  -1.80 &0.66 &    0.61 &    -1.10 &    2.28 & 2.18 &  -2.50 &1.66 &  1.76 & -1.30 & 1.00 &   0.76 & 32.5 \\
mHMQ+cf+Env &  -1.10 &0.66 &    0.58 &    -0.80 &    2.32 & 1.84 &  -1.60 &1.70 &  1.38 & -0.80 & 1.00 &   0.84 & 33.3 \\
mHMQ+cf+Sh &  -1.50 &0.66 &    0.64 &    -0.20 &    2.32 & 2.45 &  -1.70 &1.70 &  1.82 & -1.40 & 1.00 &   0.98 & 32.9 \\
1st-Gen &   0.40 &0.58 &    0.64 &0.70 &    2.20 & 2.15 &   0.80 &1.62 &  1.76 &  0.20 & 0.90 &   0.80 & 32.2 \\ \hline
\multicolumn{14}{c}{Using CV noise-reduced two-point measurements} \\ \hline
mHMQ+cf+mNFW &   0.10 &0.74 &    0.40 &1.00 &    2.78 & 2.01 &   0.70 &1.94 &  1.36 & -0.20 & 1.22 &   0.82 & 25.6 \\
GHOD &  -0.40 &0.70 &    0.50 &    -0.30 &    2.64 & 1.65 &  -0.60 &1.86 &  1.18 & -0.30 & 1.14 &   0.76 & 24.6 \\
SFHOD &  -0.40 &0.70 &    0.39 &    -0.50 &    2.64 & 1.50 &  -0.70 &1.86 &  1.12 & -0.10 & 1.16 &   0.58 & 23.7 \\
SFHOD+cf &  -0.50 &0.74 &    0.55 &0.90 &    2.76 & 1.67 &   0.00 &1.92 &  1.30 & -0.70 & 1.22 &   0.72 & 24.9 \\
HMQ &   0.00 &0.70 &    0.35 &    -1.40 &    2.58 & 1.38 &  -1.00 &1.82 &  0.96 &  0.60 & 1.14 &   0.60 & 24.6 \\
mHMQ &   0.30 &0.74 &    0.41 &0.00 &    2.74 & 1.36 &   0.30 &1.92 &  1.00 &  0.30 & 1.20 &   0.62 & 24.5 \\
mHMQ+cf &   0.50 &0.74 &    0.53 &    -0.10 &    2.72 & 1.40 &   0.40 &1.92 &  1.08 &  0.60 & 1.22 &   0.70 & 25.9 \\
LNHOD$_1$ &   0.00 &0.70 &    0.32 &    -0.70 &    2.60 & 1.52 &  -0.50 &1.82 &  1.02 &  0.30 & 1.14 &   0.62 & 24.2 \\
LNHOD$_1$-1h &   0.00 &0.68 &    0.35 &0.10 &    2.54 & 1.66 &   0.10 &1.80 &  1.20 &  0.00 & 1.12 &   0.62 & 27.9 \\
LNHOD$_1$+cf &   0.50 &0.70 &    0.42 &    -1.80 &    2.60 & 1.45 &  -0.70 &1.84 &  1.06 &  1.20 & 1.14 &   0.64 & 22.8 \\
LNHOD$_2$ &   0.20 &0.74 &    0.46 &0.50 &    2.78 & 1.47 &   0.50 &1.96 &  0.98 &  0.10 & 1.22 &   0.74 & 26.0 \\
LNHOD$_2$+cf &   0.60 &0.74 &    0.44 &0.30 &    2.70 & 1.90 &   0.70 &1.90 &  1.30 &  0.50 & 1.20 &   0.80 & 27.4 \\
HMQ$_1^{(3\sigma)}$ &   0.70 &1.02 &    0.91 &0.60 &    3.68 & 2.31 &   1.10 &2.54 &  1.76 &  0.60 & 1.68 &   1.20 & 34.4 \\
HMQ$_2^{(3\sigma)}$ &  -0.20 &1.30 &    0.96 &2.80 &    4.72 & 2.96 &   1.40 &3.16 &  2.06 & -0.90 & 2.18 &   1.34 & 32.8 \\
HMQ$_3^{(3\sigma)}$ &   0.30 &1.00 &    0.81 &1.00 &    3.62 & 2.43 &   0.90 &2.50 &  1.62 &  0.10 & 1.66 &   1.28 & 35.5 \\
HMQ$_4^{(3\sigma)}$ &   0.20 &1.00 &    0.97 &0.70 &    3.62 & 2.51 &   0.60 &2.50 &  1.72 &  0.10 & 1.66 &   1.42 & 34.9 \\
HMQ$_5^{(3\sigma)}$ &   0.20 &1.02 &    0.94 &1.20 &    3.68 & 2.47 &   0.90 &2.54 &  1.70 & -0.10 & 1.68 &   1.38 & 33.9 \\
HMQ$_6^{(3\sigma)}$ &   0.10 &1.02 &    0.97 &1.50 &    3.70 & 2.48 &   1.00 &2.56 &  1.68 & -0.30 & 1.68 &   1.42 & 33.7 \\
mHMQ+cf+C &  -0.60 &0.74 &    0.43 &0.00 &    2.70 & 1.54 &  -0.60 &1.88 &  1.10 & -0.60 & 1.22 &   0.68 & 26.3 \\
mHMQ+cf+Env &   0.10 &0.74 &    0.44 &0.40 &    2.74 & 1.38 &   0.30 &1.92 &  0.98 &  0.00 & 1.22 &   0.66 & 25.6 \\
mHMQ+cf+Sh &  -0.40 &0.74 &    0.37 &1.10 &    2.74 & 1.70 &   0.30 &1.92 &  1.20 & -0.70 & 1.22 &   0.66 & 25.7 \\
1st-Gen &   1.10 &0.58 &    0.45 &0.00 &    2.16 & 1.66 &   1.10 &1.60 &  1.26 &  1.20 & 0.88 &   0.66 & 37.0 \\
\hline\hline
\end{tabular}}}
\end{center}
\caption{Post-reconstruction BAO fits in configuration space for standard two-point measurements and CV two-point measurements. Each row represents a different HOD model for which fits were performed. We show constraints for both ($\alpha_\text{iso}$,$\alpha_\text{AP}$) and ($\alpha_\parallel$,$\alpha_\perp$) parameterizations while values are multiplied by a $10^3$ factor, meaning that deviations from fiducial values can be of the order of the sub-sub-percent level. The last column corresponds to the average $\chi^2$ obtained from 25 realization fits with 37 DoF for all fits.  For a nomenclature used for the variables listed in this table we refer to the reader to Table~\ref{Table:BAO_parameter_description}.}
\label{Table:BAO_fits_CF}
\end{table*}
\clearpage}

\afterpage{
\begin{figure*}
  {\includegraphics[width=\textwidth]{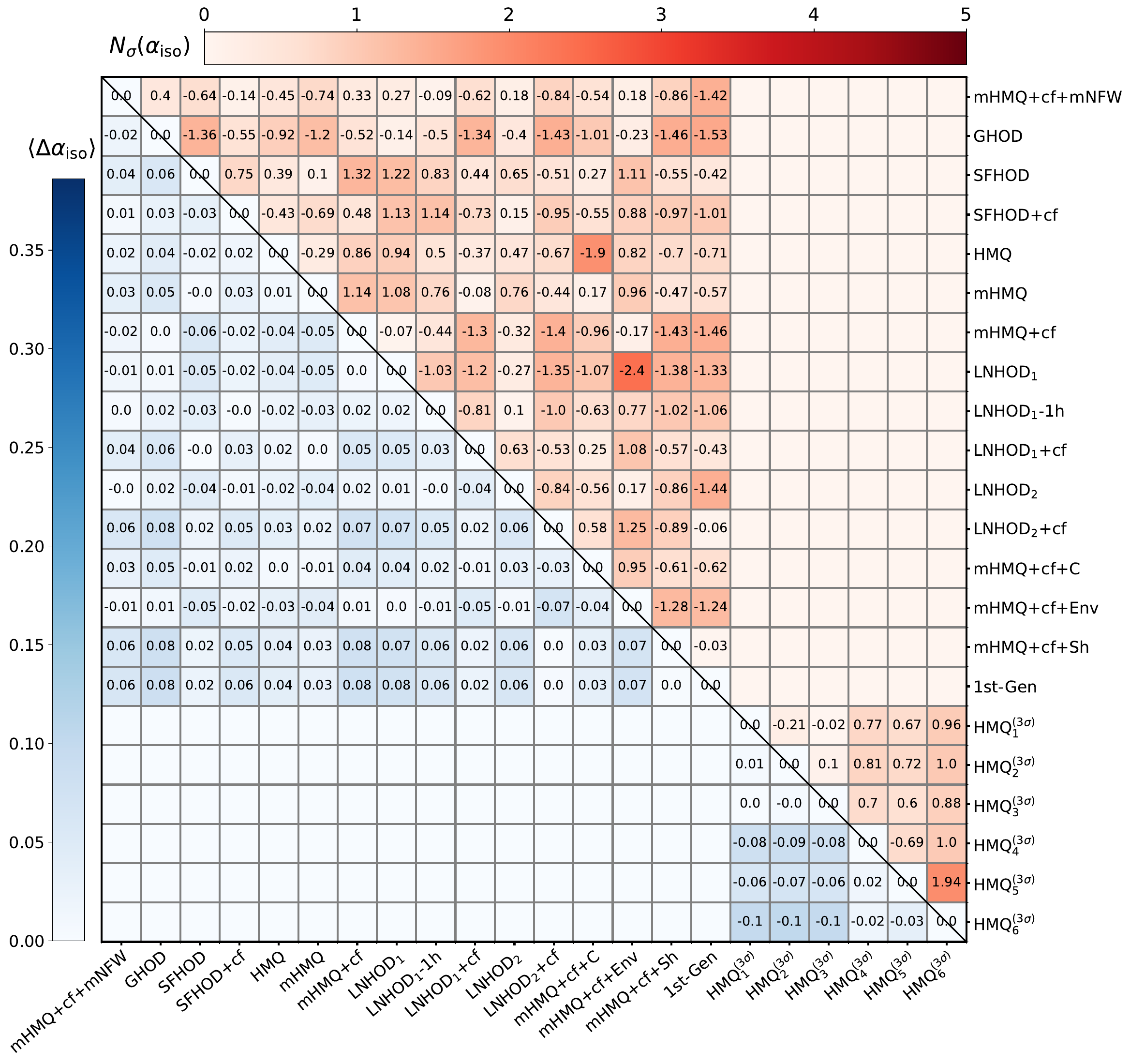}}
    \caption{Heatmap for the analysis of the shifts between pairs of HOD models in Fourier space. The heatmap shows the results for $\Delta\alpha_\text{iso}$ after the CV technique has been already applied to the post-reconstruction version of the mocks. The empty spots in the heatmap indicate that we do not compare shifts between HOD models that are not centered within the same redshift bin. The values in the lower part of the heatmap represent the shifts in $\alpha_\text{iso}$ and the values in the upper side of the heatmap show the corresponding significance of the shift.}
    \label{Fig:Heatmaps_cv_pk_alpha_iso}
\end{figure*}
\clearpage}

\afterpage{
\begin{figure*}
  {\includegraphics[width=\textwidth]{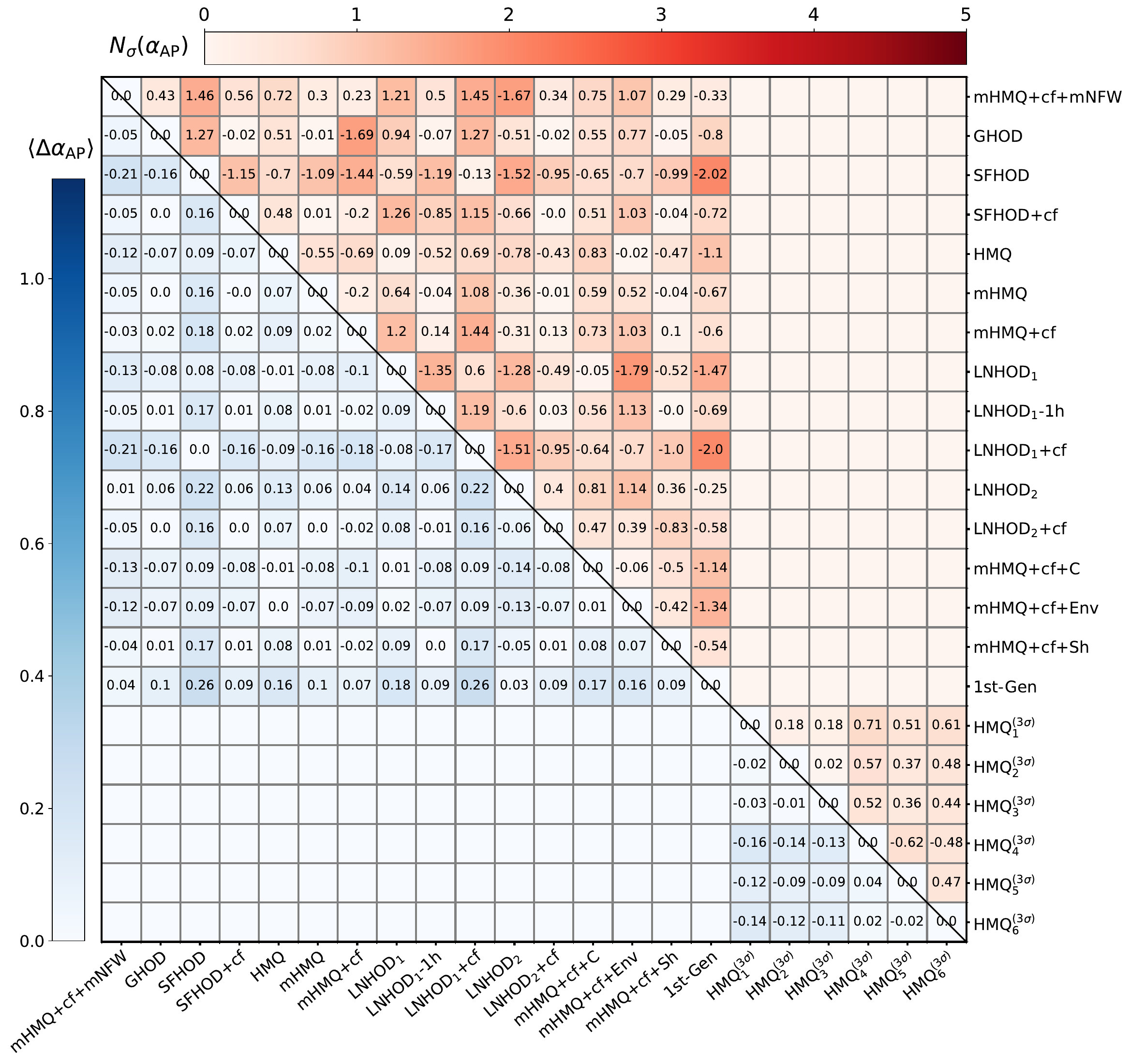}} 
    \caption{Heatmap for the comparison between pairs of HOD models when focusing on $\alpha_\text{AP}$ in Fourier space. The shifts between the measurements on $\alpha_\text{AP}$ for pairs of HOD models are shown in blue, where the blue color scale has been set to be 1/3 of the statistical error in $\alpha_\text{AP}$. The values in red show the associated value of $N_\sigma$ for each shift between a pair of HODs. The red color scale is set to peak at a 5-$\sigma$ detection level. Again, the empty spots in the heatmap indicate that we do not compare HOD models that correspond to different redshift values in order to not introduce extra noise from the simulations not due to the HOD prescription specifically.}
    \label{Fig:Heatmaps_cv_pk_alpha_ap}
\end{figure*}
\clearpage}

\afterpage{
\begin{figure*}
  {\includegraphics[width=\textwidth]{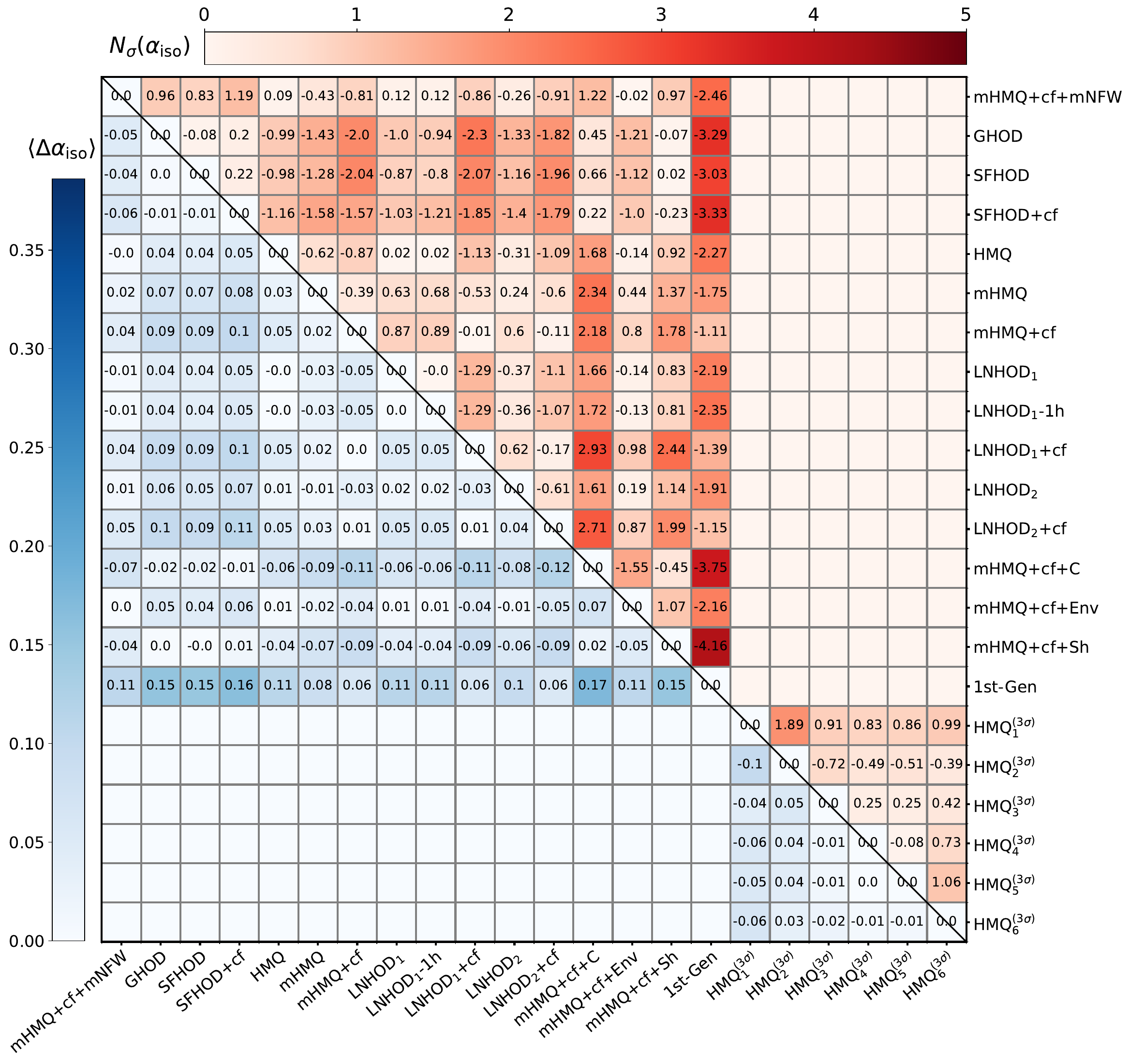}}
    \caption{Analogous heatmap to the one shown in Fig.~\ref{Fig:Heatmaps_cv_pk_alpha_iso} for the configuration space analysis. Here we also focus on comparing results from BAO fits after reconstruction and CV are applied. We notice that in the configuration space analysis case, we found biases due to HOD systematics for some pairs involving the 1st-Gen mocks. For example, a comparison of 1st-Gen with mHMQ+cf+Sh shows a systematics detection due to HOD prescription at a 4.16-$\sigma$ with a corresponding shift in $\alpha_\text{iso}$ of 0.15\%.}
    \label{Fig:Heatmaps_cv_xi_alpha_iso}
\end{figure*}
\clearpage}

\afterpage{
\begin{figure*}
  {\includegraphics[width=\textwidth]{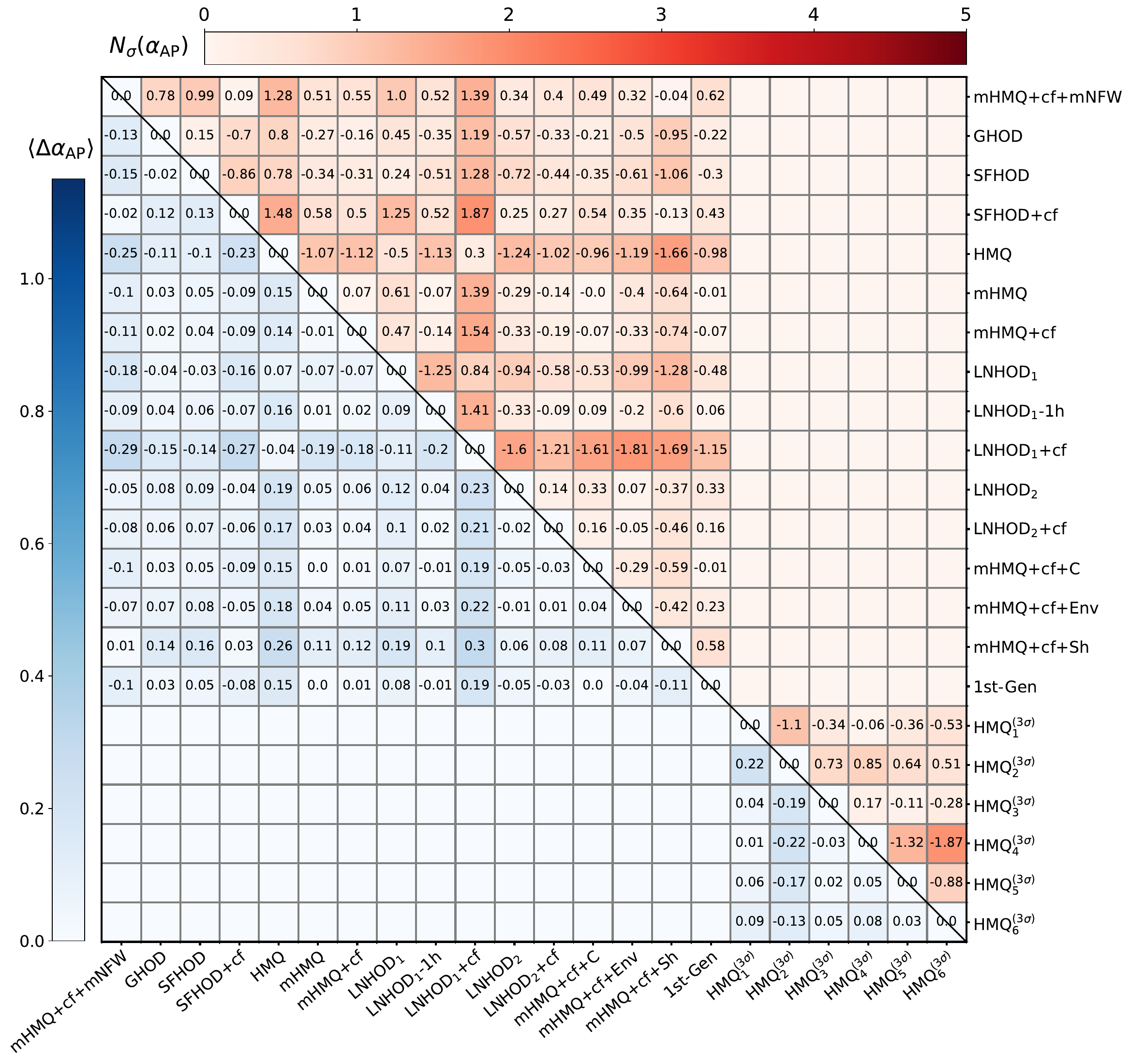}} 
    \caption{Analogous heatmap to the one described in Fig.~\ref{Fig:Heatmaps_cv_pk_alpha_ap} in configuration space. The two heatmaps produce consistent results for the shifts in $\alpha_\text{AP}$ and no systematics in the AP parameter is detected.}
    \label{Fig:Heatmaps_cv_xi_alpha_ap}
\end{figure*}
\clearpage}

%% file: DESI-2023-0285_author_list.affiliations.tex

\section{Author Affiliations}
\label{sec:affiliations}

\noindent \hangindent=.5cm $^{1}${Department of Physics, The University of Texas at Dallas, Richardson, TX 75080, USA}

\noindent \hangindent=.5cm $^{2}${Laboratoire de Physique Subatomique et de Cosmologie, 53 Avenue des Martyrs, 38000 Grenoble, France}

\noindent \hangindent=.5cm $^{3}${Ecole Polytechnique F\'{e}d\'{e}rale de Lausanne, CH-1015 Lausanne, Switzerland}

\noindent \hangindent=.5cm $^{4}${IRFU, CEA, Universit\'{e} Paris-Saclay, F-91191 Gif-sur-Yvette, France}

\noindent \hangindent=.5cm $^{5}${SLAC National Accelerator Laboratory, Menlo Park, CA 94305, USA}

\noindent \hangindent=.5cm $^{6}${Lawrence Berkeley National Laboratory, 1 Cyclotron Road, Berkeley, CA 94720, USA}

\noindent \hangindent=.5cm $^{7}${University of California, Berkeley, 110 Sproul Hall \#5800 Berkeley, CA 94720, USA}

\noindent \hangindent=.5cm $^{8}${University of Michigan, Ann Arbor, MI 48109, USA}

\noindent \hangindent=.5cm $^{9}${Center for Astrophysics $|$ Harvard \& Smithsonian, 60 Garden Street, Cambridge, MA 02138, USA}

\noindent \hangindent=.5cm $^{10}${Department of Physics \& Astronomy, Ohio University, Athens, OH 45701, USA}

\noindent \hangindent=.5cm $^{11}${Physics Department, Yale University, P.O. Box 208120, New Haven, CT 06511, USA}

\noindent \hangindent=.5cm $^{12}${Institute of Cosmology and Gravitation, University of Portsmouth, Dennis Sciama Building, Portsmouth, PO1 3FX, UK}

\noindent \hangindent=.5cm $^{13}${School of Mathematics and Physics, University of Queensland, 4072, Australia}

\noindent \hangindent=.5cm $^{14}${Center for Cosmology and AstroParticle Physics, The Ohio State University, 191 West Woodruff Avenue, Columbus, OH 43210, USA}

\noindent \hangindent=.5cm $^{15}${Department of Astronomy, The Ohio State University, 4055 McPherson Laboratory, 140 W 18th Avenue, Columbus, OH 43210, USA}

\noindent \hangindent=.5cm $^{16}${The Ohio State University, Columbus, 43210 OH, USA}

\noindent \hangindent=.5cm $^{17}${Physics Dept., Boston University, 590 Commonwealth Avenue, Boston, MA 02215, USA}

\noindent \hangindent=.5cm $^{18}${Leinweber Center for Theoretical Physics, University of Michigan, 450 Church Street, Ann Arbor, Michigan 48109-1040, USA}

\noindent \hangindent=.5cm $^{19}${Department of Physics \& Astronomy, University of Rochester, 206 Bausch and Lomb Hall, P.O. Box 270171, Rochester, NY 14627-0171, USA}

\noindent \hangindent=.5cm $^{20}${Department of Physics \& Astronomy, University College London, Gower Street, London, WC1E 6BT, UK}

\noindent \hangindent=.5cm $^{21}${Institute for Advanced Study, 1 Einstein Drive, Princeton, NJ 08540, USA}

\noindent \hangindent=.5cm $^{22}${Institute for Computational Cosmology, Department of Physics, Durham University, South Road, Durham DH1 3LE, UK}

\noindent \hangindent=.5cm $^{23}${Instituto de F\'{\i}sica, Universidad Nacional Aut\'{o}noma de M\'{e}xico,  Cd. de M\'{e}xico  C.P. 04510,  M\'{e}xico}

\noindent \hangindent=.5cm $^{24}${NSF NOIRLab, 950 N. Cherry Ave., Tucson, AZ 85719, USA}

\noindent \hangindent=.5cm $^{25}${Department of Physics \& Astronomy and Pittsburgh Particle Physics, Astrophysics, and Cosmology Center (PITT PACC), University of Pittsburgh, 3941 O'Hara Street, Pittsburgh, PA 15260, USA}

\noindent \hangindent=.5cm $^{26}${Department of Astronomy, School of Physics and Astronomy, Shanghai Jiao Tong University, Shanghai 200240, China}

\noindent \hangindent=.5cm $^{27}${Kavli Institute for Particle Astrophysics and Cosmology, Stanford University, Menlo Park, CA 94305, USA}

\noindent \hangindent=.5cm $^{28}${Departamento de F\'isica, Universidad de los Andes, Cra. 1 No. 18A-10, Edificio Ip, CP 111711, Bogot\'a, Colombia}

\noindent \hangindent=.5cm $^{29}${Observatorio Astron\'omico, Universidad de los Andes, Cra. 1 No. 18A-10, Edificio H, CP 111711 Bogot\'a, Colombia}

\noindent \hangindent=.5cm $^{30}${Institut d'Estudis Espacials de Catalunya (IEEC), 08034 Barcelona, Spain}

\noindent \hangindent=.5cm $^{31}${Institute of Space Sciences, ICE-CSIC, Campus UAB, Carrer de Can Magrans s/n, 08913 Bellaterra, Barcelona, Spain}

\noindent \hangindent=.5cm $^{32}${Departament de F\'{\i}sica Qu\`{a}ntica i Astrof\'{\i}sica, Universitat de Barcelona, Mart\'{\i} i Franqu\`{e}s 1, E08028 Barcelona, Spain}

\noindent \hangindent=.5cm $^{33}${Institut de Ci\`encies del Cosmos (ICCUB), Universitat de Barcelona (UB), c. Mart\'i i Franqu\`es, 1, 08028 Barcelona, Spain.}

\noindent \hangindent=.5cm $^{34}${Fermi National Accelerator Laboratory, PO Box 500, Batavia, IL 60510, USA}

\noindent \hangindent=.5cm $^{35}${Department of Astrophysical Sciences, Princeton University, Princeton NJ 08544, USA}

\noindent \hangindent=.5cm $^{36}${Department of Physics, The Ohio State University, 191 West Woodruff Avenue, Columbus, OH 43210, USA}

\noindent \hangindent=.5cm $^{37}${Sorbonne Universit\'{e}, CNRS/IN2P3, Laboratoire de Physique Nucl\'{e}aire et de Hautes Energies (LPNHE), FR-75005 Paris, France}

\noindent \hangindent=.5cm $^{38}${Departament de F\'{i}sica, Serra H\'{u}nter, Universitat Aut\`{o}noma de Barcelona, 08193 Bellaterra (Barcelona), Spain}

\noindent \hangindent=.5cm $^{39}${Institut de F\'{i}sica d’Altes Energies (IFAE), The Barcelona Institute of Science and Technology, Campus UAB, 08193 Bellaterra Barcelona, Spain}

\noindent \hangindent=.5cm $^{40}${Instituci\'{o} Catalana de Recerca i Estudis Avan\c{c}ats, Passeig de Llu\'{\i}s Companys, 23, 08010 Barcelona, Spain}

\noindent \hangindent=.5cm $^{41}${Department of Physics and Astronomy, Siena College, 515 Loudon Road, Loudonville, NY 12211, USA}

\noindent \hangindent=.5cm $^{42}${Department of Physics and Astronomy, University of Sussex, Brighton BN1 9QH, U.K}

\noindent \hangindent=.5cm $^{43}${Department of Physics \& Astronomy, University  of Wyoming, 1000 E. University, Dept.~3905, Laramie, WY 82071, USA}

\noindent \hangindent=.5cm $^{44}${National Astronomical Observatories, Chinese Academy of Sciences, A20 Datun Rd., Chaoyang District, Beijing, 100012, P.R. China}

\noindent \hangindent=.5cm $^{45}${Departamento de F\'{i}sica, Universidad de Guanajuato - DCI, C.P. 37150, Leon, Guanajuato, M\'{e}xico}

\noindent \hangindent=.5cm $^{46}${Instituto Avanzado de Cosmolog\'{\i}a A.~C., San Marcos 11 - Atenas 202. Magdalena Contreras, 10720. Ciudad de M\'{e}xico, M\'{e}xico}

\noindent \hangindent=.5cm $^{47}${Department of Physics and Astronomy, University of Waterloo, 200 University Ave W, Waterloo, ON N2L 3G1, Canada}

\noindent \hangindent=.5cm $^{48}${Waterloo Centre for Astrophysics, University of Waterloo, 200 University Ave W, Waterloo, ON N2L 3G1, Canada}

\noindent \hangindent=.5cm $^{49}${Perimeter Institute for Theoretical Physics, 31 Caroline St. North, Waterloo, ON N2L 2Y5, Canada}

\noindent \hangindent=.5cm $^{50}${Space Sciences Laboratory, University of California, Berkeley, 7 Gauss Way, Berkeley, CA  94720, USA}

\noindent \hangindent=.5cm $^{51}${Max Planck Institute for Extraterrestrial Physics, Gie\ss enbachstra\ss e 1, 85748 Garching, Germany}

\noindent \hangindent=.5cm $^{52}${Department of Physics and Astronomy, Sejong University, Seoul, 143-747, Korea}

\noindent \hangindent=.5cm $^{53}${Centre for Astrophysics \& Supercomputing, Swinburne University of Technology, P.O. Box 218, Hawthorn, VIC 3122, Australia}

\noindent \hangindent=.5cm $^{54}${CIEMAT, Avenida Complutense 40, E-28040 Madrid, Spain}

\noindent \hangindent=.5cm $^{55}${Department of Physics, University of Michigan, Ann Arbor, MI 48109, USA}